\documentclass[a4paper,11pt]{article}
\pdfoutput=1

\usepackage{jheppub2}
\bibstyle{JHEP}

\usepackage[english]{babel}
\usepackage[utf8]{inputenc}
\usepackage[T1]{fontenc}
\usepackage{amsfonts}
\usepackage{amsmath}         
\usepackage{amsfonts}          
\usepackage{amssymb}
\usepackage{mathtools}
\usepackage{comment}
\usepackage{tikz}
\usetikzlibrary{decorations.markings}
\usepackage{physics}
\usepackage{dsfont}
\usepackage{tabu}
\usepackage{tikz}
\usepackage{xcolor}
\usetikzlibrary{arrows.meta}

\newcommand{\overbar}[1]{\mkern 1.5mu\overline{\mkern-1.5mu#1\mkern-1.5mu}\mkern 1.5mu}

\DeclareMathOperator\vol{Vol}
\renewcommand{\tr}[1]{\text{Tr}\,{#1}}
\renewcommand{\exp}[1]{\text{exp}\,{#1}}

\title{Driven inhomogeneous CFT as a theory in curved space-time}

\author[1]{Johanna Erdmenger,}
\author[1]{Jani Kastikainen,}
\author[2]{and Tim Schuhmann}
\affiliation[1]{Institute for Theoretical Physics and Astrophysics and Würzburg-Dresden Cluster of Excellence
ct.qmat, Julius-Maximilians-Universität Würzburg, Am Hubland, 97074 Würzburg, Germany}
\affiliation[2]{Department of Physics and Astronomy, Ghent University, 9000 Ghent, Belgium}

\emailAdd{erdmenger@physik.uni-wuerzburg.de}
\emailAdd{jani.kastikainen@uni-wuerzburg.de}
\emailAdd{tim.schuhmann@ugent.be}

\abstract{For two-dimensional conformal field theories (CFTs) driven by evolving background space-time metrics in a closed universe, we present an operator formulation as a driven inhomogeneous CFT. The Hamiltonian of this theory is given by a background space-time dependent smearing of the stress tensor over the spatial slice. Emphasis is placed on the treatment of the curved-space Weyl anomaly, which we show is realized by the difference between Schrödinger and Heisenberg picture Hamiltonians once an appropriate renormalization scheme, the chirally split scheme, is chosen. As a result, the unitary evolution generated by the background metric coincides with that of a Virasoro quantum circuit. To showcase our formalism, we consider the stress tensor one-point function and the entanglement entropy of an interval in 
both operator and curved-space formulations. We find that these curved-space observables admit a state interpretation only in the chirally split scheme. 
Finally, we derive the holographic dual of the driven CFT in three-dimensional gravity, extending previous works to arbitrary driving. The holographic dictionary reproduces the stress tensor one-point function and the entanglement entropy in a diffeomorphism invariant scheme.}

\preprint{$\begin{array}{rr}
	\text{}\end{array}$
}

\begin{document}
\maketitle
\flushbottom

\newpage

\section{Introduction}

Quantum field theories (QFTs) in evolving curved space-time geometries exhibit a range of counter-intuitive effects characteristic of quantum systems subject to time-dependent driving. These effects stem from the absence of a unique ground state \cite{Fulling:1972md}, which can, for example, give rise to particle production during the expansion of space-time \cite{Parker:1968mv}. Over sufficiently long evolutions, quantum properties of the classical background driving the system are expected to become significant, ultimately leading to the breakdown of QFT as an effective field theory in expanding universes \cite{Brandenberger:2012aj,Dvali:2017eba}. Accordingly, it is important to investigate QFTs driven by evolving space-time geometries.

In this work, we investigate the operator formulation of a two-dimensional conformal field theory (CFT) in an evolving curved space-time as a driven quantum system. In general, a QFT in a curved geometry is characterized by a time-dependent Hamiltonian. Nevertheless, assuming the universe is spatially compact, the corresponding operator evolution is expected to be unitary. However, this naive expectation is violated already for free fields in dimensions higher than two, for which no unitary operator implementing the evolution in a single Fock space exists for general backgrounds (see \cite{Agullo:2015qqa} and references therein). The same is true in flat space-time when the theory is quantized along a curved foliation \cite{Torre:1998eq}. This subtlety is absent in two dimensions where the evolution on a cylindrical space-time is implementable in a single Fock space \cite{Kuchar:1989wz,Torre:1997zs}.\footnote{In \cite{Torre:1997zs}, this is proved for a massless free scalar field, but also argued to hold in the presence of mass.} Evolutions in curved space-time and along foliations are related, because a choice of a foliation in a flat background is equivalent to a choice of a metric obtained using a diffeomorphism.

In two-dimensional CFTs, it was recently shown \cite{Erdmenger:2021wzc,Erdmenger:2022lov} that the evolution along curved foliations of \cite{Torre:1997zs} realises a quantum circuit generated by the Virasoro algebra, a Virasoro circuit \cite{Caputa:2018kdj,Erdmenger:2020sup,Flory:2020eot,Flory:2020dja,Erdmenger:2024xmj}. The equivalent curved space-time interpretation was provided in \cite{deBoer:2023lrd}. In these cases, the time-dependent Hamiltonian generating the evolution is an infinite linear combination of Virasoro generators with time-dependent coefficients and the unitary operator implementing the evolution may be computed explicitly: it coincides with a representation of the Virasoro group (see \cite{Oblak:2016eij} for a review). CFTs with deformed Virasoro Hamiltonians of this type are called inhomogeneous driven CFTs, because the Hamiltonian is defined by smearing the stress tensor over the spatial slice with a spatially inhomogeneous function. The seminal example of a deformed Virasoro Hamiltonian is the sine-squared deformation \cite{Gendiar:2008udd,Hikihara:2011mtb,PhysRevA.83.052118,Katsura:2011ss} which was used in the context of one-dimensional quantum many-body systems to suppress boundary effects. Since then, inhomogeneous CFTs have received large amount of attention in field theory (for an incomplete list see \cite{Wen:2016inm,Okunishi:2016zat,Gawedzki:2017woc,Langmann:2018skr,MacCormack:2018rwq,Moosavi:2019fas,Bernard:2019mqm,Lapierre:2020ftq,Han:2020kwp,Wen:2020wee,Caputa:2020mgb,Goto:2023wai,Nozaki:2023fkx,Goto:2023yxb,Das:2023xaw,Kudler-Flam:2023ahk,Liu:2023tiq,Mao:2024cnm,Lapierre:2024lga,Bernamonti:2024fgx,Jiang:2024hgt,Das:2024lra}) and in relation to the AdS\slash CFT correspondence \cite{Caputa:2020mgb,Goto:2021sqx,Erdmenger:2021wzc,Liska:2022vrd,Kudler-Flam:2023ahk,deBoer:2023lrd,Jiang:2024hgt,Li:2025rzl}. In the time-dependent setting, the focus has been on sequences of time-independent Hamiltonians involving quenches \cite{Wen:2018vux,Liu:2023tiq} and periodically driven Floquet CFTs \cite{Wen:2018vux,Wen:2018agb,Lapierre:2020ftq,Fan:2020orx}.

Traditionally, inhomogeneous CFTs are formulated in the operator formulation by specifying a Schrödinger picture Hamiltonian as the starting point, for which \cite{Erdmenger:2021wzc,deBoer:2023lrd} provide a geometric interpretation of the resulting unitary evolution. In this work, we revisit and extend the geometric formulation \cite{deBoer:2023lrd} of an inhomogeneous CFT as a CFT in curved space-time.\footnote{The connection between curved space-times and inhomogeneous CFTs was to the knowledge of the authors first pointed in \cite{Tada:2014kza} in the context of the SSD. More recent works in this direction are \cite{Rodriguez-Laguna:2016roi,dubail2017emergence,Lapierre:2019rwj,Bernard:2019mqm}.} An important question in this context concerns the appearance of the curved space Weyl anomaly in the stress tensor operator and how it is captured by the operator formulation.

One of the main results of \cite{deBoer:2023lrd} was the discovery of the chirally split renormalization scheme adapted to CFTs driven by time-dependent Lorentzian metrics. This is a renormalization scheme in which the Weyl anomaly has been transferred to a diffeomorphism anomaly in such a way that the left- and right-moving sectors are decoupled at the quantum level as expected from the operator formulation of the theory.\footnote{An analogous scheme adapted to Euclidean signature is the holomorphically factorized scheme, see \cite{deBoer:2023lrd} for comparison.} In stress-tensor expectation values, the chirally split diffeomorphism anomaly appears as additional c-number Schwarzian terms whose derivatives act only in the spatial direction, with time treated as a parameter. This is unconventional compared to the common lore, which takes such derivatives to lie along the light-ray directions. However, as argued in \cite{deBoer:2023lrd}, and put on a firm footing here, the chirally split answer is natural from the operator, inhomogeneous CFT, point of view.

In \cite{deBoer:2023lrd}, and also in the calculations of \cite{Erdmenger:2021wzc}, the picture, either Schrödinger or Heisenberg, in which operators are expressed, was not taken into consideration. This is, however, crucial to ensure that operators and states are expressed in the same picture when building up expectation values. Here we revisit these previous results and carefully distinguish between Schrödinger and Heisenberg picture operators. We show in detail that the chirally split diffeomorphism anomaly must appear in the Heisenberg picture stress tensor operator, and therefore, also in the Heisenberg picture Hamiltonian. This implies that the anomaly in the operator formulation is beautifully accounted for by the difference between Schrödinger and Heisenberg pictures in Lorentzian signature. The location and form of the anomaly are important both for the phase factor multiplying the time-evolved CFT state in the curved background and for operator expectation values, such as the energy, as it contributes additional c-number terms.

Our results serve as a handbook for the treatment of two-dimensional CFTs in Lorentzian signature in the driven, time-dependent setting.\footnote{See also \cite{Crawford:2021adf} for work on CFTs on the flat Lorentzian cylinder.} We apply this framework to study the time-evolution of simple correlation functions of local operators. Beyond correlation functions, the entanglement entropy of a subsystem of the spatial circle is a useful diagnostic of the driven state. Its calculation in a curved background, however, is subtle. The problem was first studied in two-dimensional flat space-times in the seminal works \cite{Holzhey:1993kx,Holzhey:1994we}, and subsequently extended to curved geometries in \cite{Fiola:1994ir}. More recently, the holographic duality and specifically the Ryu--Takayanagi formula \cite{Ryu:2006bv,Ryu:2006ef} have been used to compute entanglement entropy in a family of static background metrics \cite{Miyata:2024gvr,Li:2025rzl}. For inhomogeneous CFTs driven by a sequence of time-independent Hamiltonians, entropy has been analyzed, for example, in \cite{Wen:2018agb,Fan:2020orx,Lapierre:2020ftq,Bai:2024azk,Mao:2024cnm} using twist-operator methods \cite{Calabrese:2004eu,Calabrese:2009qy}. Nevertheless, a complete analysis relating entanglement entropy on curved backgrounds to that in inhomogeneous driven CFTs is still lacking.

In this work, we fill this gap using Euclidean replica path integral methods recently developed in \cite{Estienne:2025uhh}. Since entanglement entropy is a UV divergent quantity, its finite terms are renormalization-scheme dependent, and the method of \cite{Estienne:2025uhh} allows us to track this scheme dependence in a satisfactory way. In the diffeomorphism invariant scheme, we reproduce previous results \cite{Holzhey:1993kx,Holzhey:1994we,Fiola:1994ir} as special cases, but we argue that it also contains terms incompatible with a state interpretation, i.e., terms not encoded in the unitarily time-evolved state. We show that a state-compatible result is obtained in the chirally split renormalization scheme.

Universal properties of holographic two-dimensional CFTs are captured by three-dimen\-sional Einstein gravity via the AdS\slash CFT correspondence. The gravity solution dual to a CFT driven by an evolving curved space-time was obtained in the slow-driving limit in \cite{deBoer:2023lrd}, while the static limit was analyzed in detail in \cite{Li:2025rzl}. We extend these solutions to arbitrary Virasoro driving of the CFT and use the holographic dictionary to compute the observables above, comparing them with CFT results. We find that field theory results match with gravity calculations, but only in the diffeomorphism invariant renormalization scheme. In particular, the Ryu--Takayanagi formula yields contributions to the entanglement entropy that are not encoded in the time-evolved state. The chirally split results are recovered by adding a boundary term to the Einstein action.

\begin{figure}[t]
\centering
\begin{tikzpicture}[
    every node/.style={font=\sffamily},
    doublearrow/.style={Latex-Latex, thick},
    arrowlabel/.style={midway, fill=white, inner sep=2pt, font=\sffamily\color{black}}
]

\node (top) at (0, 3) {CFT$_{1+1}$ on an evolving curved space-time {\color{black}(Sec.~\ref{sec:CFTCurvedSpacetime})}};
\node (left) at (-4, 0) {Unitary driving with a deformed Virasoro Hamiltonian};
\node (right) at (4, 0) {AdS$_3$ perspective};

\draw[doublearrow] (left) -- node[arrowlabel] {(Sec.~\ref{sec:CCrevisited})} (-1.1, 2.5);
\draw[doublearrow] (1.1, 2.5) -- node[arrowlabel] {(Sec.~\ref{sec:holography})} (right);

\end{tikzpicture}
\caption{Outline of the article. Section \ref{sec:CFTCurvedSpacetime} introduces CFT$_{1+1}$ on evolving curved space-time. In Section \ref{sec:CCrevisited}, we then provide the equivalent perspective of unitary evolution with a deformed Virasoro Hamiltonian and the holographic perspective is given in Section \ref{sec:holography}.}
\label{fig:planofpaper}
\end{figure}
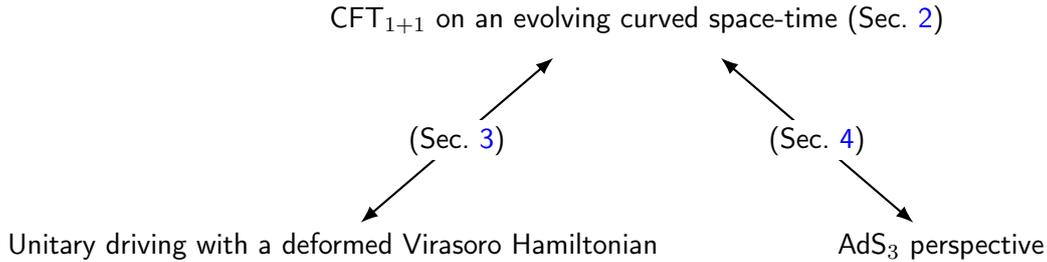

The article is organized as follows. In Section \ref{sec:CFTCurvedSpacetime}, we introduce the theory studied in this work: a Lorentzian two-dimensional CFT on an evolving curved space-time. The roles of the diffeomorphism invariant renormalization scheme and the chirally split scheme in the anomalous Ward identities of the CFT are discussed in Sections \ref{subsec:diffeo_inv_scheme} and \ref{subsec:chirally_split_scheme}, respectively. In Section \ref{sec:CCrevisited}, we rewrite the CFT on an evolving curved space-time in its operator formulation as a driven inhomogeneous CFT and then compute observables. Details of the rewriting can be found in Section \ref{subsec:unitary_evolution} while simple correlation functions are computed in Section \ref{subsec:corr_fn} and entanglement entropy in Section \ref{subsec:EE}. The holographic perspective on this general boundary theory is presented in Section \ref{sec:holography}. The dual geometry is described in Section \ref{subsec:hologrpahy_setup}, and in Section \ref{subsec:hologrpahy_observables}, we compute previously considered field theory observables using the holographic dictionary. Conclusions, discussions and future directions are provided in Section \ref{sec:conclusion}. The outline of the article is also summarized in Figure \ref{fig:planofpaper}.

\subsection{Summary of results}

Let us summarize the new results of this article. We study two-dimensional CFTs driven by general time-dependent Lorentzian background metrics on the cylinder $S^1\times \mathbb{R}$ with coordinates $(\phi,t)$. We parametrize the metric as
\begin{equation}
\label{eq:summofres_metric}
    g_{ab}(x)\,dx^adx^b  = e^{2\omega}\,(d\phi+\nu\,dt)(d\phi+\overbar{\nu}\,dt) = e^{2\sigma}\,dx^-dx^+
\end{equation}
in terms of three independent functions $\omega(\phi,t)$, $\nu(\phi,t)$ and $\overbar{\nu}(\phi,t)$ that are $2\pi$-periodic in $\phi$. Here $x^- = F_t(\phi)$ and $x^+ = \overbar{F}_t(\phi)$ are light-ray coordinates in which the metric takes the conformally flat form with a scale factor $\sigma$ that depends on $\omega$, $F_t$ and $\overbar{F}_t$.

\paragraph{Stress tensor in chirally split renormalization scheme.} The renormalized stress tensor operator of any two-dimensional conformal field theory satisfies the anomalous $\text{Diff}\ltimes \text{Weyl}$ Ward identities. We show that they are solved in the chirally split renormalization scheme in the light-ray coordinate system $x^- = F_t(\phi)$ and $x^+ = \overbar{F}_t(\phi)$ by
\begin{align}
    T_{--}^{\text{F}}(x^-,x^+) &= T_{--}(F_t(\phi))  -\frac{c}{24\pi}\frac{1}{F_t'(\phi)^{2}}\{F_t(\phi),\phi\}\,,\nonumber\\
    T_{++}^{\text{F}}(x^-,x^+) &= T_{++}(\overbar{F}_t(\phi))  -\frac{c}{24\pi}\frac{1}{\overbar{F}_t'(\phi)^{2}}\{\overbar{F}_t(\phi),\phi\}\,,\nonumber\\
    T_{-+}^{\text{F}}(x^-,x^+) &= 0\,,
    \label{eq:sumofres_chirally_split_stress_tensor}
\end{align}
where $\{F_t(\phi),\phi\}$ is the Schwarzian derivative, $T_{--}(\phi) = T(\phi)\otimes \mathbf{1}$ and $T_{++}(\phi) = \mathbf{1}\otimes T(-\phi)$ are operators that satisfy the Virasoro algebra. This extends \cite{deBoer:2023lrd} which only considered stress tensor expectation values.

For a free canonically quantized massless scalar field $\Phi$ in two dimensions, we show that the stress tensor in the chirally split scheme is obtained by point-splitting in the spatial $\phi$-coordinate alone. More precisely, we show that the renormalized stress tensor operator in the chirally split scheme is obtained as the limit
\begin{equation}
    F_t'(\phi)^2\,T_{--}^{\text{F}}(\phi,t) \equiv \lim_{\epsilon\rightarrow 0}\biggl[\frac{1}{2}\,F_t'(\phi+\epsilon)\,F_t'(\phi)\,[\Pi_-(F_t(\phi+\epsilon)),\Pi_-(F_t(\phi))]_++\frac{1}{4\pi\epsilon^{2}}\biggr]\,,
\label{eq:point_splitting_intro}
\end{equation}
where $[\cdot,\cdot]_+$ denotes the anti-commutator and $\Pi_{\pm} = \frac{\partial}{\partial x^\pm} \Phi_{\text{H}}$ are chiral conjugate momentum operators defined in terms of the Heisenberg picture scalar field operator $\Phi_{\text{H}}$. This is a purely Lorentzian point-splitting prescription and differs from those of previous literature \cite{Christensen:1976vb,davies1977quantum,Christensen:1978yd,Wald:1978pj,Polchinski:1986qf,Novotny:1998bw}. In particular, point-splitting in $\phi$ breaks diffeomorphism symmetry, but is Weyl invariant, reflecting the symmetries of the chirally split renormalization scheme.

\paragraph{Equivalence with inhomogeneous driven CFT.} Time-evolution in the background metric \eqref{eq:summofres_metric} can equivalently be expressed as unitary time-evolution with a time-dependent Schrödinger picture Hamiltonian $H_{\text{S}}(t)$. We show as a core calculation of our article, taking into account the anomaly contributions precisely, that the Schrödinger picture Hamiltonian associated to a CFT on the background \eqref{eq:summofres_metric} in the chirally split renormalization scheme is given by
\begin{equation}
    H_{\text{S}}(t) = -\int_{0}^{2\pi}d\phi\,\nu(\phi,t)\,T_{--}(\phi)+\int_{0}^{2\pi}d\phi\,\overbar{\nu}(\phi,t)\,T_{++}(\phi)\,.
\end{equation}
This is precisely what usually serves as the starting point in the literature of driven inhomogeneous CFTs with deformed Virasoro Hamiltonians. The associated unitary evolution operator $U(t)$ is given by
\begin{equation}
    U(t) = \overleftarrow{\mathcal{T}}\exp{\biggl(-i\int_{0}^{t} ds\,H_{\text{S}}(s)\biggr)} = V_{f_t}\otimes \overbar{V}_{\overbar{f}_t}\,,
\end{equation}
where $f_t = F_t^{-1}$ is the inverse function with $t$ treated as a parameter. $U(t)$ is a tensor product of unitary projective representations $V_{f_t}$ of curves $f_t$ on $\widetilde{\text{Diff}}_+S^1$ and therefore precisely implements a Virasoro quantum circuit.

It follows that the Heisenberg picture stress tensor operator in the background metric is given by
\begin{equation}
    T_{\pm\pm}^{\text{H}}(\phi,t) = U(t)^\dagger\,T_{\pm\pm}^{\text{S}}(\phi)\,U(t)\,,
\end{equation}
where $T_{\pm\pm}^{\text{S}}(\phi) = T_{\pm\pm}^{\text{H}}(\phi,0)$ is the time-independent Schrödinger picture operator. Because the curved-space diffeomorphism Ward identity in the chirally split renormalization scheme coincides with the Heisenberg equation for the Heisenberg picture stress tensor, we obtain the equality
\begin{equation}
			T_{--}^{\text{H}}(\phi,t) = F_t'(\phi)^2\,T_{--}^{\text{F}}(x^-,x^+)\,,\quad T_{++}^{\text{H}}(\phi,t) = \overbar{F}_t'(\phi)^2\,T_{++}^{\text{F}}(x^-,x^+)\,.
\end{equation}
This identifies the solution \eqref{eq:sumofres_chirally_split_stress_tensor} of the curved-space Ward identities with a Heisenberg picture operator. We highlight as a key insight that the Schwarzian derivatives in $T_{\pm\pm}^{\text{F}}$ are consistent with naturally stemming from the Schwarzian transformation of the stress tensor under the adjoint action of the Virasoro group element $V_{f_t}$. In other words, the curved-space $\text{Diff}\ltimes \text{Weyl}$ anomaly in the chirally split scheme is accounted for by the difference of operators in Schrödinger and Heisenberg pictures. This is a special and desired property of the chirally split scheme and not true for other generic renormalization schemes. Our analysis fills a missing gap in \cite{Erdmenger:2021wzc, deBoer:2023lrd} where the difference between Schrödinger and Heisenberg pictures and the associated appearance of Schwarzian terms was not treated in detail.

\paragraph{Evolution of correlation functions.} Rewriting of the curved-space CFT as an inhomogeneous CFT allows us to compute correlation functions in its operator formulation directly. We consider the time-evolution of Lorentzian time-ordered correlation functions when at time $t = 0$ the CFT is initialized to the $SL(2,\mathbb{R})$ invariant vacuum $\ket{0}$. Assuming the operators are spacelike separated, the correlation functions take the form
\begin{equation}
    \langle A(\phi_1,t_1)\,\cdots\,A(\phi_n,t_n)\rangle = \bra{0} A_{\text{H}}(\phi_1,t_1)\,\cdots\,A_{\text{H}}(\phi_n,t_n) \ket{0}.
\end{equation}
The one-point function of the stress tensor components in the vacuum state evolve as
\begin{equation}
    \langle T_{--}(\phi,t)\rangle = -\frac{c}{48\pi}F_t'(\phi)^2-\frac{c}{24\pi}\{F_t(\phi),\phi\}\,,\,\,  \langle T_{++}(\phi,t)\rangle = -\frac{c}{48\pi}\overbar{F}_t'(\phi)^2-\frac{c}{24\pi}\{\overbar{F}_t(\phi),\phi\}\,,
\end{equation}
We see that the presence of the background metric modifies the Casimir energy, which we highlight arises from the time-dependent state $U(t)\ket{0}$.

For spacelike separated insertions, the two-point function of a scalar field, which transforms with weight $\Delta$ under Weyl transformations, is given by
\begin{equation}
    \langle \mathcal{O}(\phi,t)\,\mathcal{O}(\theta,s) \rangle = \frac{b_{\mathcal{O}}}{(2\pi)^4}\Biggl[\frac{e^{-2\,(\omega(\phi,t)+\omega(\theta,s))}\,F_{t}'(\phi)\,\overbar{F}_{t}'(\phi)\,F_{s}'(\theta)\,\overbar{F}_{s}'(\theta)}{16\sin^2{\bigl(\frac{F_t(\phi)-F_s(\theta)}{2}\bigr)}\sin^2{\bigl(\frac{\overbar{F}_t(\phi)-\overbar{F}_s(\theta)}{2}\bigr)}}\Biggr]^{\Delta\slash 2}\,.
    \label{eq:scalar_2_point_intro}
\end{equation}
This is the expected answer obtained by a standard Weyl transformation of the two-point function in flat space, however, our operator approach allows to give an interpretation for various terms appearing in the expression: the dependence on $F_t,\overbar{F}_t$ arises from the state $U(t)\ket{0}$ while the $\omega$-dependence arises from explicit metric dependence of the field itself.

\paragraph{Evolution of entanglement entropy.} We study the entanglement entropy $S(t)$ of a finite spatial interval $(\phi_1,\phi_2)\subset S^1$ in the instantaneous pure state $U(t)\ket{0}$ generated by the driving with the background metric. We employ curved-space methods to evaluate the entropy and examine its consistency with a purely operator-algebraic (inhomogeneous CFT) interpretation.

More precisely, we perform the computation using the replica trick where $\tr{\rho(t)^n}$ is expressed in terms of a Euclidean path integral over a replica manifold. This requires an analytic continuation of the time-dependent Lorentzian metric to a Euclidean metric which is possible in the light-ray coordinates $x^\pm$. In the diffeomorphism invariant scheme, the path integrals are completely fixed by the Weyl anomaly and controlled by the Liouville action. However, because the replica manifold contains conical singularities at the entangling points $\phi_{1,2}$, the Liouville action must be regularized by cutting holes around them. This may be done in a diffeomorphism invariant manner by cutting out geodesic disks. We show that the resulting entropy obtained in the diffeomorphism invariant scheme is given by
\begin{equation}
    S_{\text{D}}(t) = \frac{c}{6}\,(\omega(\phi_1,t) + \omega(\phi_2,t))+\frac{c}{12}\Biggl[\log{\Biggl(\frac{4\sin^2{\bigl(\frac{F_t(\phi_1)-F_t(\phi_2)}{2}\bigr)}}{\varepsilon_1\varepsilon_2\, F_{t}'(\phi_1)\,F_{t}'(\phi_2)}\Biggr)}+(F_t\leftrightarrow \overbar{F}_t)\Biggr]\,,
\end{equation}
where $\varepsilon_{1,2}$ are the radii of two geodesic disks in the flat metric $\eta$ that are cut-out around the entangling points to regulate the Euclidean path integrals. This result follows completely from the transformation properties of the path integrals under Weyl transformations, controlled by the Liouville action, and under diffeomorphisms.

This formula is problematic for the reason that it depends on the Weyl factor of the metric. This is in tension with the fact that the Hamiltonian of the CFT does not depend on $\omega$ due to Weyl invariance so that the time-evolved state $U(t)\ket{0}$ must also be independent of $\omega$. Because entanglement entropy is a property of the state alone, it cannot depend on $\omega$. We show that the solution to this puzzle is to calculate the path integrals in the chirally split scheme, in which the result is given by
\begin{equation}
    S(t) = \frac{c}{12}\log{\Biggl(\frac{4\sin^2{\bigl(\frac{F_t(\phi_1)-F_t(\phi_2)}{2}\bigr)}}{\varepsilon_1\varepsilon_2\, F_{t}'(\phi_1)\,F_{t}'(\phi_2)}\Biggr)}+(F_t\leftrightarrow \overbar{F}_t)\,,
\end{equation}
which is independent of the Weyl factor and hence amenable to a state interpretation. This reinforces the article’s central tenet: scheme-dependent quantities in curved spacetime admit an operator-algebraic interpretation only in the chirally split scheme.

\paragraph{Holographic dual.} We write down the holographic dual of the driven CFT, extending the results of \cite{deBoer:2023lrd} beyond the slow-driving regime and those of \cite{Li:2025rzl} beyond the static limit, to encompass arbitrary driving. The bulk solution dual to the background metric of the CFT is given by
\begin{equation}
	\widetilde{G}_{\mu\nu}(X)\,dX^{\mu}dX^{\nu} = 	\ell^{2}\,\biggl[\frac{1}{z^{2}}\,(dz^{2} + e^{2\sigma}\,dx^-dx^+ ) + g_{(2)ab}\,dx^adx^b + z^2g_{(4)ab}\,dx^adx^b\biggr]\,,
    \label{eq:Gtilde_solution_intro}
\end{equation}
where the components are explicitly
\begin{gather}
	g_{(2)\pm\pm} = \partial_\pm^{2}\sigma -(\partial_\pm\sigma)^{2}-\frac{1}{4}\,,\quad g_{(2)-+} = \partial_-\partial_{+}\sigma\,,\\
    g_{(4)\pm\pm} = e^{-2\sigma}\,g_{(2)-+}\,g_{(2)\pm\pm}\,,\quad g_{(4)-+} = \frac{1}{2}\,e^{-2\sigma}\,\bigl(g_{(2)--}\,g_{(2)++}+g_{(2)-+}^2\bigr)\,.
    \label{eq:g2_g4_sol_intro}
\end{gather}
This geometry can be obtained from Poincaré AdS$_3$ by a diffeomorphism that we specify in the body of the article. We use this geometry to compute the aforementioned correlation functions now using the holographic dictionary, including the RT formula for the calculation of the entanglement entropy, and compare with field theory results. Starting from the Einstein action, we find perfect agreement with the field theory calculations in the diffeomorphism invariant renormalization scheme. In order to reproduce observables in the chirally split scheme, the Einstein action has to be supplemented with an additional boundary term.

\section{The renormalized stress tensor in evolving curved space-time}
\label{sec:CFTCurvedSpacetime}

We begin by studying properties of two-dimensional CFTs on background metrics in Lorentzian signature. We devote detailed attention to the stress tensor and its renormalization scheme dependence. Concerning that, we introduce the \textit{diffeomorphism invariant scheme} in Section \ref{subsec:diffeo_inv_scheme} and the \textit{chirally split scheme} in Section \ref{subsec:chirally_split_scheme}.

\subsection{Parametrization of the background metric}

Consider the cylinder $(S^{1}\times \mathbb{R},g)$ equipped with a Lorentzian metric $g$ and parametrized by coordinates $(\phi,t)$ where $\phi \sim \phi + 2\pi$ is the periodic spatial coordinate and $t \in \mathbb{R}$ is the time coordinate. Any two-dimensional Lorentzian metric can be written as
\begin{equation}
    g_{ab}(x)\,dx^adx^b  = e^{2\omega}\,(d\phi+\nu\,dt)(d\phi+\overbar{\nu}\,dt)\,,
    \label{eq:curvedg}
\end{equation}
with three arbitrary functions $\omega,\nu,\overbar{\nu}$ of the two coordinates $(\phi,t)$. Since the metric is Lorentzian, we have $g_{tt} = e^{2\omega}\nu\overbar{\nu} < 0$ so that we take $\nu < 0$ and $\overbar{\nu} > 0$.

There is a useful parametrization of the metric in terms of elements of the Lorentzian conformal group $\widetilde{\text{Diff}}_+S^{1}\times \widetilde{\text{Diff}}_+S^{1}$ where
\begin{equation}
    \widetilde{\text{Diff}}_+S^{1} \equiv \{F\colon \mathbb{R}\rightarrow \mathbb{R}\,\vert\, F(\phi+2\pi) = F(\phi)+2\pi\,,\, F'(\phi)>0\}
\end{equation}
is the universal cover of the group $\text{Diff}_+S^{1}$ of orientation-preserving diffeomorphisms of the circle. Consider now a curve $(F_t,\overbar{F}_t)\in \widetilde{\text{Diff}}_+S^{1}\times \widetilde{\text{Diff}}_+S^{1}$ parametrized by $t >0$. Then without loss of generality, we can write the metric components as \cite{deBoer:2023lrd}
\begin{equation}
	\nu(\phi,t) = \frac{\dot{F}_{t}(\phi)}{F_{t}'(\phi)}\,,\quad \overbar{\nu}(\phi,t) = \frac{\dot{\overbar{F}}_t(\phi)}{\overbar{F}_t'(\phi)}\,.
 \label{eq:varphinunubar}
\end{equation}
Therefore up to the Weyl factor, there is a one-to-one correspondence between 2D metrics and curves on $\widetilde{\text{Diff}}_+S^{1}\times \widetilde{\text{Diff}}_+S^{1}$. Defining the light-ray coordinates\footnote{The conditions $F_t'(\phi)>0,\overbar{F}_t'(\phi)>0$ imply that the maps $\phi\mapsto x^-$, $\phi\mapsto x^+$ are bijective and orientation-preserving.}
\begin{equation}
	x^{-} = F_{t}(\phi)\,,\quad x^{+} = \overbar{F}_t(\phi)\,,
	\label{eq:lightraycoords}
\end{equation}
the metric takes the conformally flat form
\begin{equation}
    g_{ab}(x)\,dx^adx^b  = e^{2\omega+2\varphi}\,dx^-dx^+\,,\quad e^{-2\varphi} = F_t'(\phi)\,\overbar{F}_t'(\phi)\,.
    \label{eq:conformally_flat_g}
\end{equation}
Requiring the transformation \eqref{eq:lightraycoords} to be orientation-preserving (amounting to a positive Jacobian determinant for the diffeomorphism \eqref{eq:lightraycoords}) imposes the condition $\overbar{\nu}> \nu$.

The parametrization \eqref{eq:varphinunubar} of the metric is subject an ambiguity: $\nu$ and $\overbar{\nu}$ are invariant under the left-action $F_t\rightarrow h\circ F_t$ and $\overbar{F}_t\rightarrow \overbar{h}\circ \overbar{F}_t$ with time-independent elements $h,\overbar{h}\in \widetilde{\text{Diff}}_+S^{1} $ which is equivalent to a conformal transformation of the light-ray coordinates $x^-\rightarrow h^{-1}(x^-)$ and $x^+\rightarrow \overbar{h}^{-1}(x^+)$. We fix this ambiguity by imposing  initial conditions $F_{t=0} \equiv h$ and $\overbar{F}_{t=0} \equiv \overbar{h}$ for the curve $(F_t,\overbar{F}_t)$.\footnote{When we relate CFTs on evolving curved space-time, as discussed in this section, to driven inhomogeneous CFTs with deformed Virasoro Hamiltonians in Section \ref{sec:CCrevisited}, this ambiguity will manifest itself as the freedom of choice of the initial state of the time-evolution. The simple choice $h = \overbar{h} = \text{id}$ amounts to choosing the initial state to be the vacuum state as will be explained in Section \ref{sec:CCrevisited}.}

\paragraph{Static space-time.} Various static cases studied in previous works are different special cases of this evolving curved space-time. We will now comment on some of them. The general static case is obtained for time-independent functions $\partial_t\nu = \partial_t\overbar{\nu} = 0$ which we parametrize as
\begin{equation}
    \nu(\phi,t) = -\frac{1}{p'(\phi)}\,,\quad \overbar{\nu}(\phi,t) = \frac{1}{\overbar{p}'(\phi)}\,.
    \label{eq:static_nu_nubar}
\end{equation}
Here $p$ is determined by $\nu$ up to a single integration constant which is fixed by the boundary condition $p(\phi+2\pi) = p(\phi) + 2\pi$ (and similarly for $\overbar{p}$ in terms of $\overbar{\nu}$). The static case \eqref{eq:static_nu_nubar} corresponds to the choice of the curve
\begin{equation}
    F_t(\phi) = (h\circ p^{-1})(p(\phi)-t)\,,\quad \overbar{F}_t(\phi) = (\overbar{h}\circ \overbar{p}^{-1})(\overbar{p}(\phi)+t)\,,
    \label{eq:static_curve}
\end{equation}
which satisfy the above initial condition at $t = 0$. With this in place, we can now explicitly connect to other literature. The setup that corresponds to a metric which is conformally flat in the original $(\phi,t)$ coordinates
\begin{equation}
    g_{ab}(x)\,dx^adx^b  = e^{2\omega}\,(-dt^2+d\phi^2)\,.
    \label{eq:g_conformally_flat}
\end{equation}
amounts to the choice $p = \overbar{p} = \text{id}$ for which $\nu = -1$,  $\overbar{\nu} = 1 $. This is considered for example in \cite{Cotler:2022weg}. Notice that the choice $p = \overbar{p} = \text{id}$ imposes a simple time-dependence on the curve $(F_t,\overbar{F}_t)$, namely $F_t(\phi)= h(\phi-t)$ and $\overbar{F}_t(\phi)= h(\phi+t)$. Another static setup which has been of recent interest in \cite{Miyata:2024gvr,Li:2025rzl} is the case in which $\omega = 0$ and $p = \overbar{p}$. In this case, the background metric takes the form
\begin{equation}
    g_{ab}(x)\,dx^adx^b  = -\frac{1}{p'(\phi)^2}\,dt^2+d\phi^2\,.
    \label{eq:staticg}
\end{equation}
The investigations we carry out in this article contain, but are not limited to, these special cases.

\subsection{Renormalized Ward identities in the diffeomorphism invariant scheme}
\label{subsec:diffeo_inv_scheme}

In this section, by solving the renormalized Ward identities for diffeomorphism and Weyl transformations, we derive the Heisenberg picture\footnote{The relation between Heisenberg and Schrödinger pictures will be worked out in Section \ref{sec:CCrevisited}.} stress tensor operator in the metric we just introduced in \eqref{eq:curvedg}.

A CFT on a Riemannian manifold $(M,g)$ is described by an action $I[\Phi,g;M]$ which satisfies $I[\psi\Phi,\psi g;M]  = I[\Phi,g;D(M)] $ under an element $\psi = (D,\chi) \in \text{Diff}\ltimes \text{Weyl}$ of the group of diffeomorphisms and Weyl transformations which acts on the metric and on the fundamental field as
\begin{align}
		(\psi g)_{ab}(x) &= e^{2\chi(D(x))}\,\frac{\partial D^{c}}{\partial x^{a}}\frac{\partial D^{d}}{\partial x^{b}}\,g_{cd}(D(x))\,,\\
        (\psi \Phi)_{a_1\ldots a_n}(x) &=  e^{-\Delta_{\Phi}\chi(D(x))}\,\frac{\partial D^{b_1}}{\partial x^{a_1}}\cdots\frac{\partial D^{b_n}}{\partial x^{a_n}}\,\Phi_{b_1\ldots b_n}(D(x))\,,
        \label{eq:psi_action}
\end{align}
where $\Delta_\Phi$ is the Weyl dimension of $\Phi$. Under diffeomorphisms that preserve the underlying manifold $D(M) = M$, the action is completely $\text{Diff}\ltimes \text{Weyl}$ invariant $I[\psi\Phi,\psi g;M]  = I[\Phi,g;M]$. This distinction is important when $M$ has boundaries which can be moved by the diffeomorphism. In this section, $M$ does not play a role so we simply denote the action by $I[\Phi,g]$.

The stress tensor of the CFT in the metric $g$ is then defined as
\begin{equation}
	T_{ab}^{\text{cl}}(x) \equiv -\frac{2}{\sqrt{-g}}\frac{\delta I[\Phi,g]}{\delta g^{ab}(x)}\,.
 \label{eq:defTab}
\end{equation}
The $\text{Diff}\ltimes \text{Weyl}$ transformation property of the action implies that the stress tensor satisfies the classical Ward identities (see Appendix \ref{app:ward_identities_derivation})
\begin{equation}
    \nabla^{b}T_{ab}^{\text{cl}} = E_\alpha(\Phi)\,\pounds_a\Phi^\alpha\,,\quad g^{ab}\,T_{ab}^{\text{cl}} = \Delta_{\Phi}\, E_\alpha(\Phi)\,\Phi^\alpha\,,
    \label{eq:diffweylward}
\end{equation}
where $E_{\alpha}(\Phi) = 0$ are the equations of motion for $\Phi$ coming from the variation of the action $I[\Phi,g]$ with respect to $\Phi$ and $\pounds_a$ denotes the Lie derivative with the vector field $(\xi_a)^b = \delta^b_a$.

Upon quantization, the classical fundamental field $\Phi(\phi,t)$ is promoted to a Heisenberg picture field operator $\Phi_{\text{H}}(\phi,t)$ that satisfies the equations of motion $E_{\alpha}(\Phi_{\text{H}}) = 0$. At first glance, after quantization the right-hand side of the Ward identities \eqref{eq:diffweylward} should vanish due to $E_{\alpha}(\Phi_{\text{H}}) = 0$, however, this is not the case due to UV divergences in the operator product $E(\Phi_{\text{H}}(x))\,\Phi_{\text{H}}(x')$ in the coincidence limit $x'\rightarrow x$. This divergence must be regulated and subtracted (renormalized) which leaves a finite term that violates the classical Ward identities: the $\text{Diff}\ltimes \text{Weyl}$ anomaly. We illustrate this for a free conformal scalar field in two and four dimensions using point-splitting and heat kernel regularization in Appendix \ref{app:heatkernel}.

As is well known, the $\text{Diff}\ltimes \text{Weyl}$ anomaly is more general and applies to any possibly strongly-coupled CFT. General arguments based on Wess--Zumino consistency conditions and diffeomorphism invariance imply that in two dimensions it takes the form \cite{Osborn:1991gm,Deser:1993yx,Boulanger:2007st}
\begin{equation}
	\nabla^{b}T_{ab}^{\text{D}} = 0\,,\quad  g^{ab}\,T_{ab}^{\text{D}} = -\frac{c}{24\pi}\,R\,,
 \label{eq:weylanomaly2}
\end{equation}
where $T_{ab}^{\text{D}} $ denotes the renormalized stress tensor operator, $R$ is the Ricci scalar of the background metric $g$ and $c$ is the central charge of the CFT.\footnote{The factor in front of the Ricci scalar is independent of the normalization of the classical action of the CFT, however, it is sensitive to the factors in the definition \eqref{eq:defTab} of the stress tensor as a variation of the action. The factor $-\frac{c}{24\pi}$ here is for the standard convention with $-2$ in \eqref{eq:defTab} and for a free massless scalar field, $c = 1$. See Appendix \ref{app:heatkernel} for details.} In this scheme, the $\text{Diff}\ltimes \text{Weyl}$ anomaly appears as a pure Weyl anomaly and we refer to this scheme therefore as the \textit{diffeomorphism invariant scheme}. The superscript in $T_{ab}^{\text{D}} $ denotes precisely this.

These equations should be understood as operator valued\footnote{Usually, the renormalized Ward identities \eqref{eq:weylanomaly2} are written for the expectation value of the stress tensor, but they are valid more generally as operator valued equations.} equations where $T_{ab}^{\text{D}}(\phi,t) $ is a composite Heisenberg picture operator constructed from the fundamental field $\Phi_{\text{H}}(\phi,t)$. The first equation in \eqref{eq:weylanomaly2} is identified with the Heisenberg equation describing the time-evolution of $T_{ab}^{\text{D}}(\phi,t) $, however, there is a subtlety regarding the renormalization scheme in which the additive c-number terms arising from the anomaly are correctly accounted for. This point is clarified in the next section.

It is useful to separate terms responsible for the $\text{Diff}\ltimes \text{Weyl}$ anomaly in stress tensor operator as
\begin{equation}
    T_{ab}^{\text{D}}= T_{ab} + C_{ab}^{\text{D}}\,,
    \label{eq:TDW}
\end{equation}
where $T_{ab}$ is an \textit{operator} that satisfies the $\text{Diff}\ltimes \text{Weyl} $ Ward identities without anomalies (the classical on-shell identities)
\begin{equation}
    \nabla^{b}T_{ab}= 0\,,\quad  g^{ab}\,T_{ab} = 0\,,
    \label{eq:operatorWard}
\end{equation}
while $C_{ab}^{\text{D}}$ are c-numbers that solve the anomalous Ward identities
\begin{equation}
	\nabla^{b}C_{ab}^{\text{D}} = 0\,,\quad  g^{ab}\,C_{ab}^{\text{D}} = -\frac{c}{24\pi}\,R\,.
    \label{eq:C_D_Wards}
\end{equation}
It is well known that the solution to \eqref{eq:C_D_Wards} can be written as a functional derivative
\begin{equation}
	C_{ab}^{\text{D}} = -\frac{2}{\sqrt{-g}}\frac{\delta A_{\text{D}}[g]}{\delta g^{ab}}
    \label{eq:CD_ab}
\end{equation}
of the Polyakov action \cite{Polyakov:1981rd}
\begin{equation}
	A_{\text{D}}[g]  = -\frac{c}{96\pi}\int d^{2}x\sqrt{-g}\,R\,\frac{1}{\nabla^2}\,R\,.
    \label{eq:Polyakov_text}
\end{equation}
The Polyakov action is diffeomorphism invariant $A_{\text{D}}[\widetilde{g}] =A_{\text{D}}[g]$, leading to the first equation in \eqref{eq:C_D_Wards}. Under Weyl transformations it transforms as $A_{\text{D}}[e^{2\chi}g] =A_{\text{D}}[g] + I_{\text{Lio}}[\chi,g]$ where the Liouville action is defined as
\begin{equation}
    I_{\text{Lio}}[\chi,g] \equiv\frac{c}{24\pi}\int d^2x\sqrt{-g}\,\bigl( g^{ab}\,\partial_a \chi\,\partial_b\chi +\chi R \bigr)\,.
    \label{eq:Liouville_action_text}
\end{equation}
From this transformation property of the Polyakov action it follows that \eqref{eq:CD_ab} is explicitly (see Appendix \ref{subapp:diff_invariant_scheme} for a derivation)
\begin{equation}
    C_{ab}^{\text{D}} = \frac{c}{12\pi}\,\biggl[\nabla_{a}\sigma\,\nabla_b\,\sigma+\nabla_a\nabla_b\,\sigma-g_{ab}\,\biggl(\nabla^2 \sigma+\frac{1}{2}\, \nabla^{c}\sigma\,\nabla_c\sigma\biggr)\biggr]\,.
    \label{eq:CD_stress_tensor}
\end{equation}
The classical action $I[\Phi,g]$ of the CFT is UV divergent after quantization and must be regularized and renormalized. After renormalization in the diffeomorphism invariant scheme (for example using dimensional regularization), the $I[\Phi,g]$ is promoted to the renormalized action
\begin{equation}
    I_{\text{D}}[\Phi_{\text{H}},g] = \mathcal{N}_{\text{D}}\{I[\Phi_{\text{H}},g]\}+A_{\text{D}}[g]\,,
    \label{eq:renaction_diff_scheme}
\end{equation}
where $\mathcal{N}_{\text{D}}\{I[\Phi_{\text{H}},g]\}$ is the operator valued and normal-ordered classical action of the CFT involving the Heisenberg picture field operator $\Phi_{\text{H}}$ and $\mathcal{N}_{\text{D}}\{\,\cdot\,\}$ denotes normal-ordering in the diffeomorphism invariant scheme. The couplings in the normal-ordered action are renormalized and differ from the bare couplings. The Polyakov action $A_{\text{D}}[g]$ is the finite c-number left after renormalization. The variation of \eqref{eq:renaction_diff_scheme} with respect to $g$ produces the renormalized stress tensor operator $T_{ab}^{\text{D}}$: the first term in \eqref{eq:renaction_diff_scheme} gives the normal-ordered operator part $T_{ab}$ while the Polyakov action produces the anomalous part $C_{ab}^{\text{D}}$ \eqref{eq:CD_stress_tensor}.

\subsection{Chirally split renormalization scheme}\label{subsec:chirally_split_scheme}

To be able to later connect to standard calculations in the operator formulation, we will now introduce another renormalization scheme for the stress tensor. In the light-ray coordinates \eqref{eq:lightraycoords} where the metric takes the conformally flat form \eqref{eq:conformally_flat_g}, the non-anomalous Ward identities \eqref{eq:operatorWard} for the operator part of the stress tensor are solved by
\begin{equation}
    T_{\pm\pm}(x^-,x^+) = T_{\pm\pm}(x^{\pm})\,,\quad T_{-+}(x^{-},x^{+}) = 0\,.
\end{equation}
This implies that the two-dimensional operator $T_{ab}$ ``splits'' (or factorizes) into two mutually commuting one-dimensional operators $T_{--}(x^-) = T_{--}(F_t(\phi))$ and $T_{++}(x^+) = T_{++}(\overbar{F}_t(\phi))$ which are chiral in the sense of depending on $(\phi,t)$ only through $F_t$ and $\overbar{F}_t$ respectively. For the full renormalized operator, this is not true in the diffeomorphism invariant scheme, because $T_{--}^{\text{D}}$ depends on both $(F_t,\overbar{F}_t)$ due to the anomalous term $C_{--}^{\text{D}}$ that mixes them (similarly for $T_{++}^{\text{D}}$). This is a detail that has led to past confusion and is often overlooked in the literature.

To remedy this, we will introduce another renormalization scheme that we call the \textit{chirally split scheme}. In this scheme, also the renormalized stress tensor is chirally split. The stress tensor in the chirally split scheme is denoted by
\begin{equation}
    T_{ab}^{\text{F}} = T_{ab} + C_{ab}^{\text{F}}\,,
\end{equation}
where $C_{ab}^{\text{F}}$ is related to $C_{ab}^{\text{D}}$ by the addition of finite counterterms local in the background metric and the superscript stands for ``factorized''. To find these counterterms, we work at the level of the actions. Using the fact that under Weyl transformations the Polyakov action shifts by the Liouville action, we obtain for the metric \eqref{eq:curvedg} that
\begin{equation}
    A_{\text{D}}[g] = I_{\text{Lio}}[\omega,\hat{g}]+A_{\text{D}}[\hat{g}] = I_{\text{Lio}}[\varphi,\eta]+I_{\text{Lio}}[\omega,\hat{g}]+A_{\text{D}}[\eta]\,,
    \label{eq:AD_g_transformation}
\end{equation}
where we have defined the metric $\hat{g}_{ab} \equiv e^{-2\omega}\,g_{ab}$. In \cite{deBoer:2023lrd} and Appendix \ref{subapp:chirally_split_scheme} it is shown that
\begin{equation}
    I_{\text{Lio}}[\varphi,\eta] = \Gamma[\nu]+\overbar{\Gamma}[\overbar{\nu}] +K[\nu,\overbar{\nu}]\,,
    \label{eq:Liouville_varphi}
\end{equation}
where the functionals are given by\footnote{These expressions correct factors found in \cite{deBoer:2023lrd}; see Appendix \ref{subapp:chirally_split_scheme} for details.}
\begin{gather}
	\Gamma[\nu]  = \frac{c}{48\pi}\int d\phi dt\, \nu\,\partial_\phi^{2}\log{F_t'(\phi)}\,, \quad \overbar{\Gamma}[\overbar{\nu}] = -\frac{c}{48\pi}\int d\phi dt\, \overbar{\nu}\,\partial_\phi^{2}\log{\overbar{F}_t'(\phi)}\,,\nonumber\\ K[\nu,\overbar{\nu}] = \frac{c}{48\pi}\int d\phi dt\,\frac{1}{\overbar{\nu}-\nu}\,[(\partial_\phi \nu)+(\partial_\phi \overbar{\nu})]^{2}\,.
	\label{appdecomposition}
\end{gather}
Since $A_{\text{D}}[\eta] = 0$, we obtain the decomposition
\begin{equation}
	A_{\text{D}}[g] = \Gamma[\nu]+\overbar{\Gamma}[\overbar{\nu}] +K[\nu,\overbar{\nu}]+I_{\text{Lio}}[\omega,\hat{g}]\,.
    \label{eq:Polyakov_decomposition_text}
\end{equation}
We see that the action \eqref{eq:Polyakov_decomposition_text} has chirally split and non-split parts. The anomaly action in the chirally split scheme is obtained as
\begin{equation}
	A_{\text{F}}[g] \equiv \Gamma[\nu]+\overbar{\Gamma}[\overbar{\nu}]  = A_{\text{D}}[g] -(K[\nu,\overbar{\nu}]+I_{\text{Lio}}[\omega,\hat{g}])\,,
    \label{eq:chirally_split_scheme}
\end{equation}
which is factorized meaning it decomposes into a sum of terms that each depend on $F_t$ and $\overbar{F}_t$ respectively. We see that $K[\nu,\overbar{\nu}]+I_{\text{Lio}}[\omega,\hat{g}]$ is the counterterm local in the metric components $(\omega,\nu,\overbar{\nu})$ that allows to move between the diffeomorphism invariant and chirally split schemes at the level of the effective actions. Using \eqref{eq:chirally_split_scheme}, we define
\begin{equation}
	C_{ab}^{\text{F}} = -\frac{2}{\sqrt{-g}}\frac{\delta A_{\text{F}}[g]}{\delta g^{ab}}\,.
	\label{eq:Chatdef}
\end{equation}
As shown in Appendix \ref{subapp:chirally_split_scheme}, its components are explicitly
\begin{equation}
	C_{--}^{\text{F}} = -\frac{c}{24\pi}\frac{1}{F_t'(\phi)^{2}}\{F_t(\phi),\phi\}\,,\quad C_{++}^{\text{F}} = -\frac{c}{24\pi}\frac{1}{\overbar{F}_t'(\phi)^{2}}\{\overbar{F}_t(\phi),\phi\}\,,\quad C_{-+}^{\text{F}} = 0\,,
\end{equation}
where the Schwarzian derivative is given by
\begin{equation}
    \{F(\phi),\phi\} = \biggl(\frac{F''(\phi)}{F'(\phi)}\biggr)'-\frac{1}{2}\biggl(\frac{F''(\phi)}{F'(\phi)}\biggr)^2\,.
\end{equation}
Therefore the renormalized stress tensor operator in the chirally split scheme has components
\begin{align}
    T_{--}^{\text{F}}(x^{-},x^{+}) &= T_{--}(F_t(\phi))  -\frac{c}{24\pi}\frac{1}{F_t'(\phi)^{2}}\{F_t(\phi),\phi\}\,,\nonumber\\
    T_{++}^{\text{F}}(x^{-},x^{+}) &= T_{++}(\overbar{F}_t(\phi))  -\frac{c}{24\pi}\frac{1}{\overbar{F}_t'(\phi)^{2}}\{\overbar{F}_t(\phi),\phi\}\,,\nonumber\\
    T_{-+}^{\text{F}}(x^{-},x^{+}) &= 0\,.
    \label{eq:chirally_split_stress_tensor}
\end{align}
We see that in the chirally split scheme, the renormalized stress tensor also splits into two parts that depend on $F_t$ and $\overbar{F}_t$ respectively. Using the identity $F_t'(\phi)^{-2}\{F_t(\phi),\phi\} = -\{f_t(F_t(\phi)),F_t(\phi)\}$ with $f_t = F_t^{-1}$ and \eqref{eq:lightraycoords}, they can also be written as
\begin{align}
    T_{--}^{\text{F}}(x^{-},x^{+}) &= T_{--}(x^-)  +\frac{c}{24\pi}\{f_t(x^-),x^-\}\,,\nonumber\\
    T_{++}^{\text{F}}(x^{-},x^{+}) &= T_{++}(x^+)  +\frac{c}{24\pi}\{\overbar{f}_t(x^+),x^+\}\,,\nonumber\\
    T_{-+}^{\text{F}}(x^{-},x^{+}) &= 0\,.
    \label{eq:chirally_split_stress_tensor_simple}
\end{align}
The Ward identities satisfied by the stress tensor in the chirally split scheme, and also in the diffeomorphism invariant scheme, are derived in Appendix \ref{app:actions_anomalies}. A summarizing comparison of the identities in the two renormalization schemes is provided in Figure \ref{fig:scheme comparison}.

\begin{figure}[t]
\centering
{\tabulinesep=1.4mm
\begin{tabu} {|c|c|}
    \hline
    Scheme  & $\text{Diff}\ltimes \text{Weyl}$ Ward identities  \\
    \hline 
    Diff. invariant&\hspace{0.5cm}
    $\vcenter{\hbox{$
        \begin{array}{rcl}
        (\partial_t - \nu\, \partial_\phi-2\partial_\phi \nu)\, (F_t'^{2}T_{--}^{\text{D}})-\dfrac{c}{96\pi}\,\,e^{2\omega}\,(\partial_t - \overbar{\nu}\, \partial_\phi)\, R  & = & 0        \vspace{0.15cm}\\
        (\partial_t - \overbar{\nu}\, \partial_\phi-2\partial_\phi \overbar{\nu})(\overbar{F}_t'^{2} T_{++}^{\text{D}}) - \dfrac{c}{96\pi}\,e^{2\omega}\,(\partial_t - \nu\, \partial_\phi)\, R & = & 0\vspace{0.15cm}\\
        F_t'\overbar{F}_t'\,T_{-+}^{\text{D}} + \dfrac{c}{96\pi}\,e^{2\omega}R & = & 0
        \end{array}\hspace{0.5cm}
    $}}$  \\
    \hline
    Chirally split & 
    $\vcenter{\hbox{$
        \begin{array}{rcl}
        (\partial_t -\nu\,\partial_\phi - 2\partial_\phi\nu)\,(F_t'^{2} T_{--}^{\text{F}}) + \dfrac{c}{24\pi}\,\partial_\phi^{3} \nu & = & 0        \vspace{0.15cm}\\
        (\partial_t -\overbar{\nu}\,\partial_\phi - 2\partial_\phi\overbar{\nu})\,(\overbar{F}_t'^{2}T_{++}^{\text{F}})  +\dfrac{c}{24\pi}\,\partial_\phi^{3} \overbar{\nu} & = & 0\vspace{0.15cm}\\
        F_t'\overbar{F}_t'\,T_{-+}^{\text{F}} & = & 0
        \end{array}
    $}}$ \\
    \hline
\end{tabu}
}
\caption{The renormalized $\text{Diff}\ltimes \text{Weyl}$ Ward identities in two different renormalization schemes: diffeomorphism invariant and chirally split scheme. The derivation of this form of the Ward identities can be found in Appendix \ref{app:actions_anomalies}.}
\label{fig:scheme comparison}
\end{figure}

\paragraph{The connection to the Virasoro algebra.} Let us now point out the strength of the chirally split scheme by thinking about the connection to the Virasoro algebra.
Let us denote the generator of the Virasoro algebra by $T(\phi)$ which generates the algebra as \cite{Besken:2020snx,Erdmenger:2024xmj}
\begin{equation}
	[T(\phi_1),T(\phi_2)] = -i\,\bigl(T(\phi_1)+T(\phi_2)\bigr)\,\delta'_{2\pi}(\phi_1-\phi_2)+\frac{ic}{24\pi}\,\delta'''_{2\pi}(\phi_1-\phi_2)\,,
 \label{eq:TTbarcommutator}
\end{equation}
where $\delta_{2\pi}(\phi) \equiv \frac{1}{2\pi}\sum_{n=-\infty}^\infty e^{in\phi}$ is the $2\pi$-periodic delta function. Expanding in Fourier modes, the Virasoro algebra takes the form
\begin{equation}
    T(\phi) = \frac{1}{2\pi}\sum_{n=-\infty}^\infty L_n\,e^{in\phi}\,,\quad [L_n,L_m] = (n-m)\,L_{n+m}+\frac{c}{12}\,n^{3}\,\delta_{n,-m}\,.
    \label{eq:stresstensorexp}
\end{equation}
When going to the quantum regime in two-dimensional CFTs, the chiral light-ray components $T_{\pm\pm}(x^\pm)$ of the stress tensor are promoted to a pair of commuting operators as\footnote{Strictly speaking, the quantization \eqref{eq:quantizedTs} is valid for the flat space stress tensor, but because the stress tensor with indices down is Weyl invariant, it does not make a difference.}
\begin{equation}
	T_{--}(\phi) = T(\phi)\otimes \mathbf{1}\,,\quad T_{++}(\phi) = \mathbf{1}\otimes \overbar{T}(\phi)\,,
	\label{eq:quantizedTs}
\end{equation}
where we have defined $\overbar{T}(\phi) \equiv T(-\phi)$. The use of $\overbar{T}(\phi)$ and the corresponding barred sector is needed as will be explained in Section \ref{subsec:unitary_evolution} (see footnote \ref{note:conjugate} below).

The Virasoro algebra is the Lie algebra of the Virasoro group which is the central extension of the Lie group $\widetilde{\text{Diff}}_+S^1$ (see \cite{Oblak:2016eij} for review). Let us now focus on the behavior of the stress tensor under the adjoint action of the Virasoro group. This is formally implemented by the unitary projective representations $V_{f_t}$ of curves $f_t$ on $\widetilde{\text{Diff}}_+S^1$ as defined in \cite{deBoer:2023lrd,Erdmenger:2024xmj} under which $T(\phi)$ transforms as \cite{Fewster:2004nj,Fewster:2018srj}
\begin{equation}
	V_{F_t}\,T(\phi)\,V_{F_t}^\dagger = F'_t(\phi)^{2}\,T(F_t(\phi)) - \frac{c}{24\pi}\,\{F_t(\phi),\phi\}\,.
	\label{eq:Ttransformation_text_sec2}
\end{equation}
Given this equation, we can now highlight the main strength of the chirally split scheme. Comparing \eqref{eq:Ttransformation_text_sec2} to the stress tensor of a CFT on curved driven space-time in the chirally split scheme \eqref{eq:chirally_split_stress_tensor}, we find
\begin{align}
    \big(V_{F_t}\otimes \overbar{V}_{\overbar{F}_t}\big)\,T_{--}(\phi)\,\big(V_{F_t}^\dagger\otimes \overbar{V}^\dagger_{\overbar{F}_t}\big) =  F'_t(\phi)^{2} \, T_{--}^{\text{F}}(x^{-},x^{+})
    \label{eq:chirally_split_Virasoro}
\end{align}
whereas this is not satisfied for the diffeomorphism invariant scheme (see Section \ref{subsec:unitary_evolution} for details)
\begin{align}
    \big(V_{F_t}\otimes \overbar{V}_{\overbar{F}_t}\big)\,T_{--}(\phi)\,\big(V_{F_t}^\dagger\otimes \overbar{V}^\dagger_{\overbar{F}_t}\big) \neq  F'_t(\phi)^{2} \, T_{--}^{\text{D}}(x^{-},x^{+})\,.
\end{align}
We note that only in the chirally split scheme, the c-number terms due to the $\text{Diff}\ltimes \text{Weyl}$ anomaly are compatible with the central term of the Virasoro algebra \eqref{eq:TTbarcommutator}. This is one key difference between the chirally split scheme and the diffeomorphism invariant scheme that we would like to conclude this section with: in the chirally split scheme, the connection to Virasoro symmetry is direct and natural, as the stress tensor in this scheme precisely coincides with the generator of the Virasoro algebra under adjoint action of $V_{f_t}$. We will discuss this connection again in more detail in Section \ref{subsec:unitary_evolution} and conclude this section with an example. 

\paragraph{Example: free massless scalar field.} We will now show that there exists a point-splitting prescription for a free massless scalar field theory in the background metric \eqref{eq:curvedg} that directly produces the chirally split stress tensor operator \eqref{eq:chirally_split_stress_tensor} after canonical quantization. To this end, let us consider the action
\begin{equation}
	I[\Phi,g] = \int d^{2}x\sqrt{-g}\,\mathcal{L} = -\frac{1}{2}\int d^{2}x\sqrt{-g}\,g^{ab}\,\partial_{a}\Phi\,\partial_b \Phi\,.
    \label{eq:free_massless_2d}
\end{equation}
In light-ray coordinates, the classical stress tensor has components
\begin{equation}
	 T_{\pm\pm}^{\text{cl}} = (\partial_{\pm}\Phi)^{2}\,,\quad T_{-+}^{\text{cl}} =0\,.
     \label{eq:classical_stress_scalar}
\end{equation}
The conjugate momentum in the slicing provided by the coordinates $(\phi,t)$ is given by (see Appendix \ref{app:point_splitting_space} for the detailed definition)
\begin{equation}
    \Pi(\phi,t) = \frac{\partial (\sqrt{-g}\,\mathcal{L})}{\partial \dot{\Phi}(\phi,t)} = -F_t'(\phi)\,\partial_-\Phi +\overbar{F}_t'(\phi)\,\partial_+\Phi\,,
    \label{eq:conjugate_momentum}
\end{equation}
where $\partial_{\pm} \equiv \frac{\partial}{\partial x^\pm}$. The theory is quantized by promoting $\Phi$ and $\Pi$ into Heisenberg picture operators $\Phi_{\text{H}}$ and $\Pi_{\text{H}}$ which satisfy the canonical equal-time commutation relations
\begin{equation}
    [\Phi_{\text{H}}(\phi_1,t),\Pi_{\text{H}}(\phi_2,t)] = i\,\delta_{2\pi}(\phi_1-\phi_2)\,.
    \label{eq:canonical_commutation_relation_text}
\end{equation}
As usual, the scalar field operator $\Phi_{\text{H}}$ satisfies the equations of motion $\partial_-\partial_+ \Phi_{\text{H}} = 0$ which implies
\begin{equation}
    \Phi_{\text{H}}(\phi,t) = \Phi_-(x^-) + \Phi_+(x^+) = \Phi_-(F_t(\phi)) + \Phi_+(\overbar{F}_t(\phi))\,,
\end{equation}
where $\Phi_\pm$ are chiral operators. Defining chiral momentum operators $\Pi_{\pm}(x^\pm) = \partial_{\pm}\Phi_{\pm}(x^\pm)$, the Heisenberg picture conjugate momentum operator takes the form
\begin{equation}
    \Pi_{\text{H}}(\phi,t) = -F_t'(\phi)\,\Pi_-(F_t(\phi))+\overbar{F}_t'(\phi)\,\Pi_+(\overbar{F}_t(\phi))\,.\label{eq:conjugate_momentum_operator_text}
\end{equation}
These formulae have also appeared in \cite{Torre:1997zs} in the context of evolution along non-trivial foliations of flat space.

In Appendix \ref{app:point_splitting_space}, we show that the equal-time commutation relation \eqref{eq:canonical_commutation_relation_text} implies
\begin{equation}
    \Pi_-(\phi) = -\frac{1}{\sqrt{2}}\,(J(\phi)\otimes \mathbf{1})\,,\quad \Pi_+(\phi) = \frac{1}{\sqrt{2}}\,(\mathbf{1}\otimes \overbar{J}(\phi))\,,
    \label{eq:Pi_J_relation_text}
\end{equation}
where $\overbar{J}(\phi) = J(-\phi)$ and $J(\phi)$ satisfies the $U(1)$ current algebra
\begin{equation}
    [J(\phi_1),J(\phi_2)] = -i\,\delta_{2\pi}'(\phi_1-\phi_2)\,.
    \label{eq:U1_current_algebra_text}
\end{equation}
The introduction of the operator $\overbar{J}(\phi)$ is necessary due to the relative minus sign in \eqref{eq:conjugate_momentum_operator_text}.

On-shell, the classical stress tensor is $T_{\pm\pm}^{\text{cl}} = (\partial_\pm\Phi_\pm^{\text{cl}})^2$ which in terms of the classical conjugate momentum gives
\begin{equation}
		T_{--}^{\text{cl}}(\phi,t) = \Pi_-^{\text{cl}}(F_t(\phi))^2\,,\quad T_{++}^{\text{cl}}(\phi,t) = \Pi_+^{\text{cl}}(\overbar{F}_t(\phi))^2\,.
        \label{eq:classical_stress_tensor_Pi_text}
\end{equation}
Promoting to operators $\Pi^{\text{cl}}_{\pm}\rightarrow \Pi_{\pm}$, the \eqref{eq:classical_stress_scalar} is promoted to a Heisenberg picture operator $T_{ab}^{\text{H}}$ that naively satisfies the diffeomorphism Ward identity $\partial_{\mp}T_{\pm\pm}^{\text{H}} = 0$ (see Section \ref{subsec:unitary_evolution} below for a careful treatment). However, this is not the case due to UV divergences appearing in the coincidence limit of the product of two currents $J$. As shown in Appendix \ref{app:point_splitting_space}, the divergence appear in the anti-commutator as
\begin{equation}
    \frac{1}{4}\,[J(\phi_1),J(\phi_2)]_+ = -\frac{1}{4\pi}\frac{1}{4\sin^{2}(\frac{\phi_1-\phi_2}{2})} + T(\phi_1)+\frac{1}{2\pi}\frac{1}{24} + \mathcal{O}(\phi_1-\phi_2)\,,\quad \phi_2\rightarrow \phi_1\,,
    \label{eq:anti_commutator_expansion_text}
\end{equation}
where $T(\phi)$ is a stress tensor operator that satisfies the Virasoro algebra \eqref{eq:TTbarcommutator} with central charge $c = 1$ and the anti-commutator is defined as $[A(\phi_1),B(\phi_2)]_+ \equiv A(\phi_1)\,B(\phi_2)+B(\phi_2)\,A(\phi_1)$. The first term in \eqref{eq:anti_commutator_expansion_text} should be understood as a derivative of the Cauchy principal value distribution as explained in Appendix \ref{app:point_splitting_space}. Because of this divergence, we define the stress tensor operator by point-splitting in space: we displace one of the conjugate momentum factors in \eqref{eq:classical_stress_tensor_Pi_text} in $\phi$ and replace the product with $1\slash 2$ times their anti-commutator. Then we take the coincidence limit and subtract the divergence. More precisely, we define the renormalized operator
\begin{equation}
    F_t'(\phi)^2\,T_{--}^{\text{F}}(\phi,t) \equiv \lim_{\epsilon\rightarrow 0}\biggl[\frac{1}{2}\,F_t'(\phi+\epsilon)\,F_t'(\phi)\,[\Pi_-(F_t(\phi+\epsilon)),\Pi_-(F_t(\phi))]_++\frac{1}{4\pi\epsilon^{2}}\biggr]\,,
\label{eq:point_splitting_text}
\end{equation}
where the last term subtracts the divergence coming from the $\epsilon\rightarrow 0$ limit. The use of the anti-commutator ensures that the resulting operator is Hermitian. The $T_{++}^{\text{F}}$ operator is defined similarly in terms $\overbar{F}_t$ and $\Pi_+$. The anti-commutator in \eqref{eq:point_splitting_text} is controlled by \eqref{eq:anti_commutator_expansion_text} and the $\epsilon\rightarrow 0$ limit yields exactly the solution \eqref{eq:chirally_split_stress_tensor} of the renormalized Ward identities in the chirally split scheme with $c = 1$ (see Appendix \ref{app:point_splitting_space}). Thus we have demonstrated, that for a free massless scalar field, the chirally split scheme is equivalent to a point-splitting scheme with minimal subtraction of the divergence coming from the coincidence limit.

\section{CFTs driven by background metrics: operators and observables}
\label{sec:CCrevisited}

It is known that Lorentzian CFTs on evolving curved space-times can also be viewed through the looking-glass of driven inhomogeneous CFTs with deformed Virasoro Hamiltonians \cite{deBoer:2023lrd} (see \cite{Tada:2014kza,dubail2017emergence,Lapierre:2019rwj,Bernard:2019mqm} for earlier work). In Section \ref{subsec:unitary_evolution}, we revisit this equivalence to resolve subtleties concerning renormalization scheme dependence and the treatment of operators in the Schrödinger and Heisenberg pictures. In this way, we precisely connect the contents of Section \ref{sec:CFTCurvedSpacetime} to the standard operator approach of investigating driven inhomogeneous CFTs, namely Hamiltonian time-evolution with time-dependent deformed Virasoro Hamiltonians, and at the same time highlight assumptions underlying this approach. After establishing this equivalent viewpoint, we discuss simple correlation functions explicitly in Section \ref{subsec:corr_fn} and provide a derivation of the entanglement entropy of an interval based on recent work \cite{Estienne:2025uhh} in Section \ref{subsec:EE}, again highlighting the influence of the renormalization scheme.

\subsection{CFT Hamiltonian and unitary time-evolution}\label{subsec:unitary_evolution}

Let us begin by investigating the Hamiltonian associated to a CFT in an evolving curved space-time, starting classically. Our goal is to show that after quantization the Hamiltonian operator in the Schrödinger picture matches exactly with the Hamiltonian of a driven inhomogeneous CFT. The classical Hamiltonian is defined as a Hamilton--Jacobi variation of the action as
\begin{equation}
	H_{\text{cl}}(t) \equiv -\tfrac{d}{ds}I[\Phi_s,g;M_s]\Big\vert_{s = t}\,,
	\label{eq:HamiltonianHJ_text}
\end{equation}
where $M_s$ is a finite section of the cylinder $S^1\times (0,s)$, $\Phi_s$ is a solution of the classical equations of motion with boundary conditions imposed at the two space-like boundary components of $\partial M_s$. Note the $s$-derivative acts on the solution $\Phi_s$ and on the integration domain $M_s$, but not on the background metric $g$. Equation \eqref{eq:HamiltonianHJ_text} is a field theory generalization of the formula for the classical Hamiltonian of a free point-particle.\footnote{For old work on Hamilton--Jacobi formulation of classical field theory and its quantization, see \cite{freistadt1955classical,freistadt1956quantized}.} In the usual context, the on-shell action $I[\Phi_s,g;M_s]$ is called the Hamilton's principal function which can be thought of as a solution of the equation \eqref{eq:HamiltonianHJ_text}, the Hamilton--Jacobi equation, given a predetermined classical Hamiltonian on the left-hand side.

As shown using diffeomorphism invariance in Appendix \ref{app:Hamiltonian_derivation} explicitly, the classical Hamiltonian obtained in this way is equivalently given by a more familiar expression
\begin{equation}
	H_{\text{cl}}(t)  = -\int_{0}^{2\pi} d\phi\sqrt{\gamma}\,\zeta^{a}n^{b}\,T_{ab}^{\text{cl}}\,,
 \label{eq:Noetherchargeformula}
\end{equation}
where $T_{ab}^{\text{cl}}$ is the classical Einstein--Hilbert stress tensor \eqref{eq:defTab} in the metric $g$, $\gamma$ is the induced metric of the constant-$t$ slice and
\begin{equation}
	\zeta^{a} = \delta^{a}_{t},\quad n_{a} = \frac{\partial_a t}{\sqrt{-g^{ab}\,\partial_a t\,\partial_b t}}\,.
    \label{eq:n_down_def}
\end{equation} 
Here $n^{a} = g^{ab}\,n_b$ is the timelike past-pointing unit normal vector of the constant-$t$ slices. When the background metric is independent of $t$ (the static case), $\zeta^{a}$ is a Killing vector and the Hamiltonian is time-independent on-shell due to the conservation equation of the stress tensor $T_{ab}^{\text{cl}}$.

The formula \eqref{eq:Noetherchargeformula} is the Hamiltonian of the theory generating $t$-translations along a fixed $(\phi,t)$-foliation in curved space-time. Usually in the literature, the same Hamiltonian appears as the generator of translations along a non-trivial curved foliation of flat space-time. In that case, the $n^a$ is replaced with the unit normal of the foliation and $T_{ab}^{\text{cl}}$ with the flat space stress tensor which may also be obtained from the Noether procedure. In a two-dimensional CFT, the two Hamiltonians are equivalent by diffeomorphism invariance of the integrand and by Weyl invariance of $T_{ab}^{\text{cl}}$. For a two-dimensional QFT, $T_{ab}^{\text{cl}}$ is no longer Weyl invariant so that the equivalence breaks down: a QFT couples to the Weyl factor of the metric which cannot be encoded by a non-trivial foliation. The same holds for the Hamiltonian in higher dimensions, where a higher-dimensional metric contains more information than a choice of a foliation (diffeomorphism). We have not found a direct reference for \eqref{eq:Noetherchargeformula} which is why we have dedicated Appendix \ref{app:Hamiltonian_derivation} for its derivation.

In the background metric \eqref{eq:curvedg}, the classical Hamiltonian \eqref{eq:Noetherchargeformula} is explicitly (see also \cite{deBoer:2023lrd})
\begin{align}
	H_{\text{cl}}(t) = &-\int_{0}^{2\pi}d\phi\,\dot{F}_t(\phi)\,F_t'(\phi)\,T_{--}^{\text{cl}}(x^-,x^+)+\int_{0}^{2\pi}d\phi\,\dot{\overbar{F}}_t(\phi)\,\overbar{F}_t'(\phi)\,T_{++}^{\text{cl}}(x^-,x^+)\nonumber\\
	&+\int_{0}^{2\pi}d\phi\,(\dot{F}_t(\phi)\,\overbar{F}_t'(\phi) -\dot{\overbar{F}}_t(\phi)\,F_t'(\phi))\,T_{-+}^{\text{cl}}(x^-,x^+)\,,
\label{eq:freeNoethercharge}
\end{align}
where $x^\pm$ are given in terms of $(\phi,t)$ via \eqref{eq:lightraycoords}. At the classical level, the last term vanishes on-shell due to $T_{-+}^{\text{cl}} = 0$. Due to Weyl invariance of the classical action of the CFT, the Hamiltonian is independent of the Weyl factor $\omega$.

\paragraph{Quantization.} We will now move on to the quantum formulation of the CFT. Upon quantization, the classical Hamiltonian \eqref{eq:Noetherchargeformula} (in our case explicitly \eqref{eq:freeNoethercharge}) is promoted to a Hamiltonian operator. However, this is subtle and raises the following three questions that we need to answer to perform quantization properly.
\begin{enumerate}
    \item[\textbf{Q1}] Is the resulting Hamiltonian operator in the Schrödinger or the Heisenberg picture? This distinction is important because standard unitary state evolution is generated by the Schrödinger Hamiltonian. If the Hamiltonian obtained by quantization is in the Heisenberg picture, we need to convert it to the Schrödinger picture first to specify state evolution.
    \item[\textbf{Q2}] The Hamiltonian contains the stress tensor which after quantization is a composite operator whose UV divergences must be renormalized. How does the choice of renormalization scheme impact the Hamiltonian?
    \item[\textbf{Q3}] Which renormalization scheme produces a Schrödinger picture Hamiltonian which matches with the deformed Virasoro Hamiltonian of an inhomogeneous CFT? We will see that this is the scheme which underlies the operator formulation of the theory.
\end{enumerate}
We now answer these questions. Let the Schrödinger picture Hamiltonian of a CFT on a constant-$t$ slice in the background metric $g$ be given by $H_{\text{S}}(t)$. The Hamiltonian has explicit time-dependence, because the metric acts as a time-dependent classical background field driving the CFT. The state evolution of the CFT is generated by $H_{\text{S}}(t)$ via the Schrödinger equation
\begin{equation}
	\partial_t\ket{\Psi(t)} = -i\,H_{\text{S}}(t)\ket{\Psi(t)}.
\end{equation}
We choose the initial state to be $\ket{\Psi(0)} = \ket{0}\otimes \ket{0} $ where the vacuum state is defined by $L_0\ket{0} = -\frac{c}{24}\ket{0}$ and $L_{n>0}\ket{0} = 0$. From now on, we will abuse notation and denote $\ket{0}\otimes \ket{0}\equiv \ket{0}$. As a result, the Schrödinger picture state at time $t$ is given by $\ket{\Psi(t)} = U(t)\ket{0}$ where the unitary operator
\begin{equation}
	U(t) \equiv\overleftarrow{\mathcal{T}}\exp{\biggl(-i\int_{0}^{t} ds\,H_{\text{S}}(s)\biggr)} = \overrightarrow{\mathcal{T}}\exp{\biggl(-i\int_{0}^{t} ds\,H_{\text{H}}(s)\biggr)}\,,
	\label{eq:schrodingerunitary}
\end{equation}
and the time-ordering operator $\overleftarrow{\mathcal{T}}$ orders larger values of $s$ to the left (while $\overrightarrow{\mathcal{T}}$ orders them to the right).\footnote{The second equality in \eqref{eq:schrodingerunitary} containing the Heisenberg Hamiltonian \eqref{eq:HH_HS} is a mathematical identity that involves a reversal of the time-ordering; see for example \cite{Erdmenger:2024xmj} for a proof.} The corresponding Heisenberg picture Hamiltonian is given by
\begin{equation}
    H_{\text{H}}(t) \equiv U(t)^{\dagger}\,H_{\text{S}}(t)\,U(t)\,.
    \label{eq:HH_HS}
\end{equation}
With these definitions set up, we can now answer \textbf{Q1}: \textit{the classical Hamiltonian $H_{\text{cl}}(t)$ is promoted to the Heisenberg picture Hamiltonian $H_{\text{H}}(t)$ upon quantization.} To see this, note that the formula \eqref{eq:Noetherchargeformula} involves the stress tensor \eqref{eq:defTab} that satisfies classical $\text{Diff}\ltimes \text{Weyl}$ Ward identities when the equations of motion are satisfied. Upon quantization, it is replaced by a renormalized stress tensor operator which satisfies the renormalized Ward identities \eqref{eq:weylanomaly2} as discussed in Sections \ref{subsec:diffeo_inv_scheme} and \ref{subsec:chirally_split_scheme}. The picture of the stress tensor is fixed by the fact that in quantum field theory equations of motion are always satisfied by Heisenberg picture operators.\footnote{This is simply the statement that classical dynamics are encoded into Heisenberg equations in the quantum theory.} Thus it satisfies the renormalized Ward identities as a Heisenberg picture operator for which the conservation equation is its Heisenberg equation. In contrast, the Schrödinger picture stress tensor is time-independent and cannot be a solution of the on-shell Ward identities. This will be seen more explicitly below. Now, because we know that solutions of the Ward identities are Heisenberg operators, it follows that the Hamiltonian \eqref{eq:Noetherchargeformula} (\eqref{eq:freeNoethercharge} in our model specifically) becomes the Heisenberg picture Hamiltonian upon quantization.

To tackle \textbf{Q2}, we look at the formula \eqref{eq:Noetherchargeformula} which defines the classical Hamiltonian in terms of the stress tensor. Quantization in different schemes corresponds to replacing $T_{ab}^{\text{cl}}$ with the renormalized stress tensor operator of the corresponding scheme, for example, by $T_{ab}^{\text{D}}$ or $T_{ab}^{\text{F}}$ in diffeomorphism invariant and chirally split schemes respectively. The stress tensors in different renormalization schemes differ (only) by c-number terms, a standard property of renormalization. To answer \textbf{Q2} precisely, \textit{the choice of renormalization scheme modifies the additive c-number terms of the Heisenberg picture Hamiltonian.} In the unitary operator \eqref{eq:schrodingerunitary}, these c-numbers appear only as different phases, but, as often overlooked, they contribute significantly to one- and higher-point correlation functions of the stress tensor, which we will see explicitly below.\footnote{Note that the relation between Schrödinger and Heisenberg picture operators is always scheme independent because the additional phases of the unitary $U(t)$ cancel in the adjoint action \eqref{eq:HH_HS}.} To take this c-number seriously is a key aspect of this article.

This analysis, which highlights the influence of the choice of the renormalization scheme on the c-number of the Heisenberg Hamiltonian, motivates us to investigate \textbf{Q3}: \textit{we will now show that the answer is the chirally split scheme.} To this end, by plugging the stress tensor \eqref{eq:chirally_split_stress_tensor} in \eqref{eq:Noetherchargeformula}, the Heisenberg picture Hamiltonian operator in the chirally split scheme takes the form
\begin{align}
    H_{\text{H}}(t) = &-\int_{0}^{2\pi}d\phi\,\nu(\phi,t)\,F_t'(\phi)^2\,T_{--}(F_t(\phi))+\int_{0}^{2\pi}d\phi\,\overbar{\nu}(\phi,t)\,\overbar{F}_t'(\phi)^2\,T_{++}(\overbar{F}_t(\phi))\nonumber\\
    &+\frac{c}{24\pi}\int_{0}^{2\pi}d\phi\,\nu(\phi,t)\,\{F_t(\phi),\phi\}-\frac{c}{24\pi}\int_{0}^{2\pi}d\phi\,\overbar{\nu}(\phi,t)\,\{\overbar{F}_t(\phi),\phi\}\,,
    \label{eq:HF}
\end{align}
where the operators $T_{\pm\pm}(x^\pm)$ are given in terms of generators of the Virasoro algebra in \eqref{eq:quantizedTs}. Note that, as stated before, the Heisenberg picture Hamiltonian contains the additional c-number contribution.

We claim that when the Heisenberg picture Hamiltonian is given by \eqref{eq:HF}, the corresponding Schrödinger picture Hamiltonian, upon imposing the initial condition $F_0 = \overbar{F}_0 = \text{id}$ (general case is discussed below), is given by
\begin{equation}
    H_{\text{S}}(t) = -\int_{0}^{2\pi}d\phi\,\nu(\phi,t)\,T_{--}(\phi)+\int_{0}^{2\pi}d\phi\,\overbar{\nu}(\phi,t)\,T_{++}(\phi)\,,
    \label{eq:HS_split}
\end{equation}
where all time-dependence is in the metric components $\nu,\overbar{\nu}$. This is precisely the Hamiltonian that is studied in the literature of driven inhomogeneous CFTs with deformed Virasoro Hamiltonians.

Before we discuss the connection to inhomogeneous CFTs in more detail, let us prove that \eqref{eq:HS_split} arises from \eqref{eq:HF} by going from Heisenberg to Schrödinger picture via the relation \eqref{eq:HH_HS} when time-evolution is generated by the Schrödinger Hamiltonian \eqref{eq:HS_split}. To see how this works, we can substitute \eqref{eq:HS_split} to the definition of the unitary operator \eqref{eq:schrodingerunitary} to obtain
\begin{equation}
    U(t) = \overleftarrow{\mathcal{T}}\exp{\biggl(i\int_{0}^{t} ds\int_{0}^{2\pi}d\phi\,\nu(\phi,t)\,T(\phi)\biggr)}\otimes \overleftarrow{\mathcal{T}}\exp{\biggl(-i\int_{0}^{t} ds\int_{0}^{2\pi}d\phi\,\overbar{\nu}(\phi,t)\,\overbar{T}(\phi)\biggr)}\,.
    \label{eq:U_t_factorization}
\end{equation}
By introducing the inverse $f_t = F_t^{-1}$, we may write
\begin{equation}
    \nu(\phi,t) = -(\dot{f}_t\circ f_t^{-1})(\phi)\,,\quad \overbar{\nu}(\phi,t) = -(\dot{\overbar{f}}_t\circ \overbar{f}_t^{-1})(\phi)\,.
\end{equation}
Therefore up to a minus sign,  $\nu,\overbar{\nu}$ are exactly the tangent vectors of the curves $f_t,\overbar{f}_t$ on the Lie group $\widetilde{\text{Diff}}_+S^1$, respectively. Since we have assumed that $f_0 = \overbar{f}_0 = \text{id}$, the unitary factors in \eqref{eq:U_t_factorization} coincide with unitary projective representations of $\widetilde{\text{Diff}}_+S^1$ as defined in \cite{deBoer:2023lrd,Erdmenger:2024xmj}. This is a key aspect of this proof because it enables relating the renormalization scheme dependent c-number appearing in the Heisenberg Hamiltonian to the fixed c-number in the central term of the Virasoro algebra as discussed already at the end of Section \ref{subsec:chirally_split_scheme}. Explicitly, we get\footnote{The appearance of the relative minus sign in the classical Hamiltonian \eqref{eq:freeNoethercharge} explains why the left-movers $x^+$ evolve under the complex conjugate representation $\overbar{V}_{\overbar{f}_t}$.\label{note:conjugate}}
\begin{equation}
    U(t) = V_{f_t}\otimes \overbar{V}_{\overbar{f}_t}\,.
    \label{eq:unitary_operator}
\end{equation}
Using $V_{f_t}^\dagger = V_{F_t}$ together with the transformation law of $T(\phi)$ under adjoint action of $V_{F_t}$ given in \eqref{eq:Ttransformation_text_sec2},\footnote{The adjoint action of the complex conjugate representation $\overbar{V}_{\overbar{F}_t}$ on $\overbar{T}(\phi)$ is given by the same formula \cite{Erdmenger:2024xmj}.} we see that \eqref{eq:HF} is indeed obtained from \eqref{eq:HS_split} via the relation between the pictures \eqref{eq:HH_HS} and thereby complete the proof. To provide further detail, $V_{f}$ and $\overbar{V}_{f}$ satisfy the composition laws \cite{Erdmenger:2024xmj}
\begin{equation}
    V_{f_t}\,V_{h_t} = e^{iB(f_t,h_t)}\,V_{f_t\circ h_t}\,,\quad \overbar{V}_{f_t}\,\overbar{V}_{h_t} = e^{-iB(f_t,h_t)}\,\overbar{V}_{f_t\circ h_t}\,.
    \label{eq:composition_law}
\end{equation}
Here the phase factor is explicitly
\begin{equation}
    B(f_t,h_t) = b(f_t,h_t)+\frac{c}{48\pi}\int_{0}^{2\pi}d \phi\,\frac{h''_t(\phi)}{h'_t( \phi)}\,\log{f'_t(h_t(\phi))}\,,
    \label{eq:2cocycle}
\end{equation}
where the second term is the Thurston--Bott two-cocycle and $b(f_t,h_t) = a(f_t\circ h_t) - a(f_t)-a(h_t)$ is a coboundary with \cite{deBoer:2023lrd}
\begin{equation}
    a(f_t) = \frac{c}{48\pi}\int_0^t ds\int_{0}^{2\pi}d \phi\,\frac{\dot{F}_s(\phi)}{F_s'(\phi)}\biggl(\frac{F_s''(\phi)}{F_s'(\phi)}\biggr)'\,.
\end{equation}
Let us now discuss the connection to previous work on driven inhomogeneous CFTs with deformed Virasoro Hamiltonians \cite{Wen:2016inm,Okunishi:2016zat,Gawedzki:2017woc,Langmann:2018skr,MacCormack:2018rwq,Moosavi:2019fas,Bernard:2019mqm,Lapierre:2020ftq,Han:2020kwp,Wen:2020wee,Caputa:2020mgb,Goto:2023wai,Nozaki:2023fkx,Goto:2023yxb,Das:2023xaw,Kudler-Flam:2023ahk,Liu:2023tiq,Mao:2024cnm,Lapierre:2024lga,Bernamonti:2024fgx,Jiang:2024hgt,Das:2024lra}. In this literature, the starting point is usually a statement of the Hamiltonian of interest, precisely in the form of \eqref{eq:HS_split}, with specified smearing profiles $\nu$ and $\overbar{\nu}$. What we want to emphasize is that this coincides exactly with the Schrödinger picture Hamiltonian of the curved space CFT in the case where we employ the chirally split renormalization scheme. This provides an answer to \textbf{Q3}. It is important to highlight that, in order to make this connection, it is essential to work in the chirally split scheme. Only in this scheme, as already discussed in detail in Section \ref{subsec:chirally_split_scheme} around equation \eqref{eq:chirally_split_Virasoro}, does the stress tensor $T^{\text{F}}_{ab}$ that solves the Ward identities coincide with the transformation of the generator of the Virasoro algebra $T(\phi)$ under the adjoint action of the Virasoro group. Hence, employing the chirally split scheme is central to bridge the study of driven inhomogeneous CFTs with deformed Virasoro Hamiltonians and Lorentzian two-dimensional CFTs on evolving curved space-time.

Up to the Schwarzian derivatives, the above Schrödinger \eqref{eq:HS_split} and Heisenberg \eqref{eq:HF} picture Hamiltonians have appeared in the case of a free massless scalar field quantized along non-trivial foliations of flat space in \cite{Torre:1997zs} (see also \cite{Kuchar:1989wz}). In our case, the Schwarzian derivatives appear in the Heisenberg Hamiltonian while in \cite{Torre:1997zs} they appear in the Schrödinger Hamiltonian. As we have seen, the fact that they appear in the Heisenberg Hamiltonian is required from the curved space $\text{Diff}\ltimes \text{Weyl}$ anomaly point of view, marking a significant difference to the considerations in \cite{Torre:1997zs}.

Let us also discuss the impact of the choice of the initial state in this approach. In the above derivation, we assumed that $F_0 = \overbar{F}_0 = \text{id}$ which implies that $U(0) = 1$ as required by the boundary condition $\ket{\Psi(0)} = \ket{0}$ for the Schrödinger equation. This is a standard choice in quantum circuit literature but we could also instead choose a different initial state. If we choose the initial state to be the Virasoro state $\ket{\Psi(0)}\equiv (V_{h^{-1}}\otimes \overbar{V}_{\overbar{h}^{-1}}) \ket{0}$, then the state at time $t$ is given by
\begin{equation}
    \ket{\Psi(t)} = e^{iB(f_t,h^{-1}) - iB(\overbar{f}_t,\overbar{h}^{-1})}\,\bigl(V_{f_t\circ h^{-1}}\otimes \overbar{V}_{\overbar{f}_t\circ \overbar{h}^{-1}}\bigr)\ket{0},
\end{equation}
where we have used the composition law \eqref{eq:composition_law}. Up to the overall phase factor, this is the same unitary evolution starting from the vacuum state as above, but with $(F_t,\overbar{F}_t)$ replaced by $(h\circ F_t,\overbar{h}\circ F_t)$. Therefore the ambiguity in the definition of $(F_t,\overbar{F}_t)$ that we have encountered before and that keeps the metric \eqref{eq:curvedg} invariant, amounts to an ambiguity in the choice of the initial state of the evolution. In our construction here, the initial condition $F_0 = h$ and $ \overbar{F}_0 = \overbar{h}$ amounts to the choice of the initial state $(V_{h^{-1}}\otimes \overbar{V}_{\overbar{h}^{-1}})\ket{0}$ up to phase factors.

\paragraph{Connection to quantum circuits.} The Schrödinger Hamiltonian \eqref{eq:HS_split} matches with the generator of a Virasoro (or conformal) circuit defined in \cite{Flory:2020eot,Flory:2020dja,Erdmenger:2024xmj}. Thus Schrödinger picture unitary time-evolution $\ket{\Psi(t)} = V_{f_t}\otimes \overbar{V}_{\overbar{f}_t}\ket{0}$ in the background metric \eqref{eq:curvedg} generates a Virasoro circuit. The fact that unitary evolution in the background metric \eqref{eq:curvedg} gives rise to a Virasoro circuit was proved in \cite{deBoer:2023lrd}. However, the difference to \cite{deBoer:2023lrd}, in addition to phase factors, is that our circuit \eqref{eq:unitary_operator} is associated to the inverse diffeomorphisms $(f_t,\overbar{f}_t)$ instead of $(F_t,\overbar{F}_t)$. The origin of the difference is the detailed treatment of the Schrödinger and Heisenberg pictures done here. Furthermore, comparing to and building on top of \cite{Erdmenger:2021wzc}, in our treatment we have now specified that the generating Hamiltonian of the circuit in the Heisenberg picture, not in the Schrödinger picture, is identified with the Hamiltonian of the CFT in curved spacetime that arises from \eqref{eq:Noetherchargeformula} under quantization.

\paragraph{Static background metric.} In the static case \eqref{eq:static_nu_nubar}, the Schrödinger Hamiltonian \eqref{eq:HS_split} is time-independent and given by
\begin{equation}
    H_{\text{S}}(t) = \int_{0}^{2\pi}d\phi\,\frac{T_{--}(\phi)}{p'(\phi)}+\int_{0}^{2\pi}d\phi\,\frac{T_{++}(\phi)}{\overbar{p}'(\phi)}\,.
\end{equation}
Using the transformation law \eqref{eq:Ttransformation_text_sec2} and a change of integration variables, this can be written as
\begin{equation}
    H_{\text{S}}(t) =(V_P\otimes \overbar{V}_{\overbar{P}})\,H_0\,(V_P\otimes \overbar{V}_{\overbar{P}})^\dagger + \frac{c}{24\pi}\int_0^{2\pi}d\phi\,(\{P(\phi),\phi\}+ \{\overbar{P}(\phi),\phi\})\,,
    \label{eq:static_HS}
\end{equation}
where we have defined the Hamiltonian in the flat metric
\begin{equation}
    H_0 \equiv \int_0^{2\pi} d\phi\,(T_{--}(\phi) + T_{++}(\phi)) = L_0\otimes \mathbf{1} + \mathbf{1}\otimes L_0
    \label{eq:H0_def}
\end{equation}
and $P \equiv p^{-1}$, $\overbar{P} \equiv \overbar{p}^{-1}$. It follows that the unitary operator \eqref{eq:schrodingerunitary} generating the evolution is given by
\begin{equation}
    U(t) = e^{-\frac{ict}{24\pi}\int_0^{2\pi}d\phi\,(\{P(\phi),\phi\}+ \{\overbar{P}(\phi),\phi\})}\,(V_P\otimes \overbar{V}_{\overbar{P}})\,e^{-iH_0 t}\,(V_P\otimes \overbar{V}_{\overbar{P}})^\dagger\,.
\end{equation}
In general, the time-evolved state $\ket{\Psi(t)}$ is time-dependent, because $H_0$ does not commute with the Virasoro unitaries $V_P,\overbar{V}_{\overbar{P}}$. However, when the initial state is a Virasoro state is chosen to be $\ket{\Psi(0)} = (V_{p^{-1}}\otimes \overbar{V}_{\overbar{p}^{-1}})\ket{0}$, corresponding to $F_0 = h \equiv p$ and $\overbar{F}_0 = \overbar{h} \equiv \overbar{p}$, the state remains the same up to a time-dependent phase
\begin{equation}
    \ket{\Psi(t)} = e^{-\frac{ict}{24\pi}\int_0^{2\pi}d\phi\,(\{P(\phi),\phi\}+ \{\overbar{P}(\phi),\phi\}-1)}\ket{\Psi(0)}.
    \label{eq:time_ind_state}
\end{equation}
In this special case, correlation functions and entanglement entropies will be completely time-independent as we will see below.

\paragraph{Heisenberg picture stress tensor.} Above, we derived the Heisenberg picture Hamiltonian based on the assumption that the stress tensor appearing in the classical Hamiltonian \eqref{eq:Noetherchargeformula} is promoted to a Heisenberg picture operator upon quantization. The assumption is based on the fact that the stress tensor satisfies the classical Ward identities which are only true on-shell and which at the quantum level can only be satisfied by Heisenberg operators. As alluded to in Section \ref{subsec:chirally_split_scheme}, we now show this assumption is consistent: given the Schrödinger picture Hamiltonian operator derived above based on this assumption, the solution \eqref{eq:chirally_split_stress_tensor} of the Ward identities in the chirally split scheme is a Heisenberg picture operator.

The fact that \eqref{eq:chirally_split_stress_tensor} should be understood as a Heisenberg picture operator is the most clear in flat space $x^- = F_t(\phi) = \phi -t$ and $x^+ = F_t(\phi) = \phi + t$ for which the Schwarzian derivatives vanish. The time-evolution is generated by the time-independent Hamiltonian $H_0 = L_0\otimes \mathbf{1}+\mathbf{1}\otimes L_0$ \eqref{eq:H0_def} and we see that
\begin{equation}
    T_{\pm\pm}^{\text{F}}(x^-,x^+) = T_{\pm\pm}(\phi\pm t) = e^{i t H_0}\,T_{\pm\pm}(\phi)\,e^{-i t H_0}\,,
\end{equation}
where $T_{\pm\pm}(\phi)$ is the Schrödinger picture operator in this case. In general, the Heisenberg picture stress tensor is given by
\begin{equation}
    T_{\pm\pm}^{\text{H}}(\phi,t) = U(t)^\dagger\,T_{\pm\pm}^{\text{S}}(\phi)\,U(t)\,,\quad T_{-+}^{\text{H}}(\phi,t) = 0\,,
\end{equation}
where $U(t)$ is defined in \eqref{eq:unitary_operator} and the Schrödinger picture stress tensor $T_{\pm\pm}^{\text{S}}(\phi) = T_{\pm\pm}^{\text{H}}(\phi,0)$ is time-independent. The Schrödinger picture operator may contain explicit time-dependence through the background metric, but here we require its complete time-independence as an extra condition to fix the Heisenberg operator uniquely.\footnote{This choice has no impact on the form of the Schrödinger and Heisenberg picture Hamiltonians.}

Choosing the Schrödinger stress tensor as
\begin{equation}
    T_{--}^{\text{S}}(\phi) = T(\phi)\otimes \mathbf{1}\,,\quad T_{++}^{\text{S}}(\phi) = \mathbf{1}\otimes \overbar{T}(\phi)\,,\quad T_{-+}^{\text{S}}(\phi) = 0\,,
\end{equation}
and use the transformation law \eqref{eq:Ttransformation_text_sec2}, we see that the Heisenberg picture stress tensor
\begin{equation}
    T_{--}^{\text{H}}(\phi,t) = F_t'(\phi)^2\,T_{--}^{\text{F}}(x^-,x^+)\,,\quad T_{++}^{\text{H}}(\phi,t) = \overbar{F}_t'(\phi)^2\,T_{++}^{\text{F}}(x^-,x^+)
    \label{eq:Heisenberg_stress_tensor}
\end{equation}
is given in terms of the chirally split stress tensor \eqref{eq:chirally_split_stress_tensor}. Therefore the Schwarzian derivatives appearing in the chirally split stress tensor are exactly accounted for by the difference between Schrödinger and Heisenberg picture operators. Put another way, the operator formulation of the CFT matches with the curved space CFT in the chirally split scheme. This is not true in other schemes, such as in the diffeomorphism invariant scheme, where the c-number terms are different.

The diffeomorphism Ward identity satisfied by the chirally split stress tensor is derived in Appendix \ref{app:actions_anomalies} and shown in Figure \ref{fig:scheme comparison}. For the Heisenberg picture stress tensor \eqref{eq:Heisenberg_stress_tensor}, it amounts to
\begin{equation}
	\partial_t T_{--}^{\text{H}} = (\nu\,\partial_\phi + 2\partial_\phi\nu)\,T_{--}^{\text{H}} -  \frac{c}{24\pi}\,\partial_\phi^{3} \nu\,,\quad \partial_t T_{++}^{\text{H}} = (\overbar{\nu}\,\partial_\phi + 2\partial_\phi\overbar{\nu})\,T_{++}^{\text{H}} -\frac{c}{24\pi}\,\partial_\phi^{3} \overbar{\nu}\,,
\end{equation}
which can be identified with the Heisenberg equations $\partial_t T_{\pm\pm}^{\text{H}}(\phi,t) = -i\,[T_{\pm\pm}^{\text{H}}(\phi,t),H_{\text{H}}(t)]$ as advertized above. Note that the Heisenberg picture stress tensor includes the factors of $F_t'^2$ and $\overbar{F}_t'^2$ and the Schrödinger picture stress tensor is completely time-independent.

Lastly, let us consider the massless free scalar field introduced at the end of Section \ref{subsec:chirally_split_scheme}. In that case, the $U(1)$ current transforms as
\begin{equation}
    V_{F_t}\,J(\phi) \,V_{F_t}^\dagger = F'_t(\phi)\,J(F_t(\phi))\,.
    \label{eq:conjugate_momentum_transformation}
\end{equation}
Thus the conjugate momentum operator \eqref{eq:conjugate_momentum_operator_text} in terms of the current \eqref{eq:Pi_J_relation_text} is indeed in the Heisenberg picture $\Pi_{\text{H}}(\phi,t) = U(t)^{\dagger}\,\Pi_{S}(\phi)\,U(t)$ given the Schrödinger picture operator
\begin{equation}
    \Pi_{S}(\phi) = -\Pi_-(\phi) + \Pi_+(\phi)\,.
\end{equation}
With this we can also define the Schrödinger picture stress tensor directly via point-splitting\footnote{See also the formula \eqref{eq:T_from_point_splitting} in Appendix \ref{app:point_splitting_space}.}
\begin{equation}
    T_{\pm\pm}^{\text{S}}(\phi) \equiv \lim_{\epsilon\rightarrow 0}\biggl(\frac{1}{2}\,[\Pi_{\pm}(\phi+\epsilon),\Pi_{\pm}(\phi)]_++\frac{1}{4\pi\epsilon^{2}}\biggr)\,.
\label{eq:point_splitting_schrodinger}
\end{equation}
The Heisenberg picture stress tensor is obtained by using \eqref{eq:conjugate_momentum_transformation} which indeed gives the formula \eqref{eq:point_splitting_text}.

In Appendix \ref{app:Heisenberg_tensors}, we give a treatment of the quantization of a general tensor field $\mathcal{O}_{a_1\ldots a_n}$ with Weyl weight $\Delta_{\text{w}}$ and write down the corresponding of Schrödinger and Heisenberg picture operators. The analysis contains the stress tensor and the chiral-half of the conjugate momentum of a free scalar field above as special cases.

\subsection{Time-evolution of simple correlation functions}
\label{subsec:corr_fn}

In the previous section, we have described a CFT in curved space-time using its operator formulation, which we can now use to compute observables. To this end, let us consider vacuum time-ordered correlation functions of the type
\begin{equation}
    \langle A(\phi_1,t_1)\,\cdots\,A(\phi_n,t_n)\rangle \equiv \bra{0} \mathcal{T}\{A_{\text{H}}(\phi_1,t_1)\,\cdots\,A_{\text{H}}(\phi_n,t_n)\} \ket{0},
    \label{eq:time_ordered_correlator}
\end{equation}
where $A_{\text{H}}(\phi,t)$ is a Heisenberg picture operator, none of the operators lie on the same time-slice $t_1 \neq t_2\neq \ldots \neq t_n$ and the vacuum is defined as $\ket{0} \equiv \ket{0}\otimes \ket{0}$. A one-point function reduces to an \textit{expectation value} of the Schrödinger picture operator $A_{\text{S}}(\phi,t) = U(t)\,A_{\text{H}}(\phi,t)\,U(t)^\dagger$ in the instantaneous state $\ket{\Psi(t)} = U(t)\ket{0}$ of the evolution\footnote{Note that the Schrödinger picture operator can still have explicit time-dependence.}
\begin{equation}
    \langle A(\phi,t)\rangle = \bra{\Psi(t)} A_{\text{S}}(\phi,t) \ket{\Psi(t)} .
\end{equation}
Of course, this is equivalent to considering the expectation value of the Heisenberg picture operator $A_H(\phi,t)$ in the state $\ket{0}$. Let us now assume for simplicity that the operators are space-like separated in the background metric $g$ which implies that the Heisenberg picture operators commute with each other $[A_{\text{H}}(\phi_i,t_i),A_{\text{H}}(\phi_j,t_j)] = 0$ for $i,j = 1,\ldots, n$. Then the time-ordering can be removed and \eqref{eq:time_ordered_correlator} reduces to
\begin{equation}
    \langle A(\phi_1,t_1)\,\cdots\,A(\phi_n,t_n)\rangle = \bra{0} A_{\text{H}}(\phi_1,t_1)\,\cdots\,A_{\text{H}}(\phi_n,t_n) \ket{0}.
\end{equation}
We will now compute several expectation values of the operators introduced above.

\paragraph{The stress tensor.} Let us first consider the expectation value of the stress tensor operator in the chirally split scheme. In the Heisenberg picture it is given in terms of the solution \eqref{eq:chirally_split_stress_tensor_simple} of the renormalized Ward identities by \eqref{eq:Heisenberg_stress_tensor}. It follows that
\begin{equation}
    \langle T_{--}(\phi,t)\rangle = -\frac{c}{48\pi}F_t'(\phi)^2-\frac{c}{24\pi}\{F_t(\phi),\phi\}\,,\,\,  \langle T_{++}(\phi,t)\rangle = -\frac{c}{48\pi}\overbar{F}_t'(\phi)^2-\frac{c}{24\pi}\{\overbar{F}_t(\phi),\phi\}\,,
    \label{eq:stress_1_point}
\end{equation}
where we have used that the vacuum state satisfies
\begin{equation}
    \bra{0}L_n\ket{0} = -\frac{c}{24} \,\delta_{n,0}\,.
    \label{eq:L_n_vacuum}
\end{equation}
Let us answer what the interpretation of these expectation values is. Here, the left-hand side is the vacuum expectation value of the time-dependent Heisenberg picture operator in the vacuum state $\ket{0}$, or equivalently, it is the expectation value of the time-independent Schrödinger picture operator in the instantaneous state $\ket{\Psi(t)} = U(t)\ket{0}$. The right-hand side of equation \eqref{eq:stress_1_point} also appears in \cite{deBoer:2023lrd}, but there both the state and the operator on the left-hand side are time-dependent. In addition, the stress tensor one-point function in curved backgrounds has previously been studied in \cite{Isler:1989ty,Capri:1992tb,Massacand:1993cj,Lamb:1994bz,Capri:1995iw} using various regularization schemes, and in particular, using the diffeomorphism invariant scheme in \cite{Page:1996ax}. Our result \eqref{eq:stress_1_point} differs from these earlier works.

Using the Heisenberg Hamiltonian \eqref{eq:HF}, the one-point function of the Hamiltonian (or average energy) becomes
\begin{equation}
    \langle H(t)\rangle = \frac{c}{24\pi}\int_{0}^{2\pi}d\phi\,\nu(\phi,t)\,\biggl(\frac{1}{2}\,F_t'(\phi)^2+\{F_t(\phi),\phi\}\biggr)-(F_t\leftrightarrow \overbar{F}_t)\,.
    \label{eq:energy_1_point}
\end{equation}
Let us consider as a example the static metric \eqref{eq:static_nu_nubar} with the initial conditions $F_0 = p$ and $\overbar{F}_0 = \overbar{p}$ for which
\begin{equation}
    F_t(\phi) = p(\phi)-t\,,\quad \overbar{F}_t(\phi) = \overbar{p}(\phi)+t\,.
\end{equation}
In this case, the one-point function \eqref{eq:energy_1_point} is time-independent and given by
\begin{equation}
    \langle H(t)\rangle  = -\frac{c}{12}+\frac{c}{24\pi}\int_0^{2\pi}d\phi\,(\{P(\phi),\phi\}+ \{\overbar{P}(\phi),\phi\})\,,
    \label{eq:time_independent_energy}
\end{equation}
where we have performed a change of integration variables $\phi \rightarrow P(\phi)$ and used the identity $p'(P(\phi))^{-2}\{p(P(\phi)),P(\phi)\} = -\{P(\phi),\phi\}$. The choice $F_0 = p$, $\overbar{F}_0 = \overbar{p}$ amounts to choosing the initial state to be a Virasoro state $\ket{\Psi(0)} = (V_{p^{-1}}\otimes \overbar{V}_{\overbar{p}^{-1}})\ket{0}$ as explained in Section \ref{subsec:unitary_evolution}. Indeed, this same expression is obtained as the expectation value of the time-independent Schrödinger Hamiltonian \eqref{eq:static_HS} in the time-independent Virasoro state \eqref{eq:time_ind_state}. Using \eqref{eq:static_nu_nubar}, we can write \eqref{eq:time_independent_energy} explicitly in components of the metric as
\begin{equation}
    \langle H(t)\rangle  = -\frac{c}{12}+\frac{c}{48\pi}\int_0^{2\pi}d\phi\,\biggl(\frac{\nu'(\phi)^2}{\nu(\phi)}-\frac{\overbar{\nu}'(\phi)^2}{\overbar{\nu}(\phi)}\biggr)\,.
\end{equation}
Therefore in a static background with a compatible choice of an initial state, the average total energy is a local functional of the background metric.

\paragraph{Scalar primary field.} Next, we consider a scalar field $\mathcal{O}$ which transforms under the $\psi = (\chi,D)\in \text{Diff}\ltimes \text{Weyl}$ group as
\begin{equation}
    (\psi\mathcal{O})(x) = e^{-\Delta \chi(D(x)))}\,\mathcal{O}(D(x))\,.
    \label{eq:psi_scalar}
\end{equation}
The treatment of the quantization of general tensor fields can be found in Appendix \ref{app:Heisenberg_tensors} containing the scalar field as a special case. In the background metric \eqref{eq:curvedg}, the Heisenberg picture scalar field operator is given by \eqref{eq:Heisenberg_tensor_field} with $n = 0$ and $h = \overbar{h} = \Delta\slash 2$:
\begin{equation}
    \mathcal{O}_{\text{H}}(\phi,t) = e^{-\Delta\,\omega(\phi,t)}\,F_t'(\phi)^{\Delta\slash 2}\,\overbar{F}_t'(\phi)^{\Delta\slash 2}\,\mathcal{O}_{\Delta\slash 2}(F_t(\phi))\otimes \overbar{\mathcal{O}}_{\Delta\slash 2}(\overbar{F}_t(\phi))\,,
    \label{eq:2D_scalar_operator}
\end{equation}
where the chiral primary field $\mathcal{O}_{h}(\phi)$ is defined by \eqref{eq:primary_transformation_law} and $\overbar{\mathcal{O}}_{h}(\phi) \equiv \mathcal{O}_{h}(-\phi)$. Note that $\mathcal{O}_{\text{H}}$ carries an explicit dependence on $\omega(\phi,t)$ due to its non-trivial transformation law \eqref{eq:psi_scalar} under Weyl scalings.

First, we see that the one-point function of the scalar field in the background metric vanishes $\langle \mathcal{O}(\phi,t)\rangle = 0$ due to $\bra{0}\mathcal{O}_h(\phi)\ket{0} = 0$ which follows from the $SL(2,\mathbb{R})$ invariance of the vacuum state and the transformation law \eqref{eq:primary_transformation_law} of the primary under the adjoint action of the Virasoro group. Consider then the two-point function of two scalar fields at $(\phi,t)$ and $(\theta,s)$ which are spacelike separated $-(t-s)^2+(\phi-\theta)^2>0$. In this case, we have
\begin{equation}
    \langle \mathcal{O}(\phi,t)\,\mathcal{O}(\theta,s)\rangle = \bra{0} \mathcal{O}_{\text{H}}(\phi,t)\,\mathcal{O}_{\text{H}}(\theta,s) \ket{0}.
    \label{eq:scalar_2_point}
\end{equation}
By substituting \eqref{eq:2D_scalar_operator}, this reduces to a product of vacuum two-point functions of chiral primaries. Together with the transformation law \eqref{eq:primary_transformation_law}, the vacuum two-point function of a chiral primary operator is fixed up to a constant $b_{\mathcal{O}}$ by $SL(2,\mathbb{R})$ invariance of the vacuum state in the standard way as (see for instance \cite{Erdmenger:2024xmj})
\begin{equation}
    \bra{0}\mathcal{O}_{\Delta\slash 2}(\phi_1)\,\mathcal{O}_{\Delta\slash 2}(\phi_2)\ket{0} =  \lim_{\varepsilon\rightarrow 0^+}\frac{1}{(2\pi)^2}\frac{b_{\mathcal{O}}}{\bigl[2i\sin{\bigl(\frac{\phi_1-\phi_{2}+i\varepsilon}{2}\bigr)}\bigr]^{\Delta}}\,.
    \label{eq:primary2pointfunction_text}
\end{equation}
Here flipping the sign of the $i\varepsilon$ amounts to exchanging the order of the operators on the left-hand side assuming $\Delta $ is a positive integer. This two-point function implies that the commutator $[\mathcal{O}_{\Delta\slash 2}(\phi_1),\mathcal{O}_{\Delta\slash 2}(\phi_2)]$ vanishes when $\phi_1\neq \phi_2$ and in the point-limit $\phi_1 = \phi_2$ it contains only singularities involving the Dirac delta function and its derivatives at least when $\Delta$ is an integer (see \cite{Erdmenger:2024xmj} for more details). The vanishing of the commutator for $\phi_1\neq \phi_2$ ensures that the scalar operator \eqref{eq:2D_scalar_operator} respects two-dimensional causality and commutes with itself at spacelike separation in the metric \eqref{eq:curvedg}. Using \eqref{eq:primary2pointfunction_text} in \eqref{eq:scalar_2_point}, 
we obtain
\begin{equation}
    \langle \mathcal{O}(\phi,t)\,\mathcal{O}(\theta,s) \rangle = \frac{b_{\mathcal{O}}}{(2\pi)^4}\Biggl[\frac{e^{-2\,(\omega(\phi,t)+\omega(\theta,s))}\,F_{t}'(\phi)\,\overbar{F}_{t}'(\phi)\,F_{s}'(\theta)\,\overbar{F}_{s}'(\theta)}{-16\sin^2{\bigl(\frac{F_t(\phi)-F_s(\theta)}{2}\bigr)}\sin^2{\bigl(\frac{\overbar{F}_t(\phi)-\overbar{F}_s(\theta)}{2}\bigr)}}\Biggr]^{\Delta\slash 2}\,,
    \label{eq:scalar_2_point_final}
\end{equation}
where the singularities at $(\phi_1,t_1) = (\phi_2,t_2)$ are of the Cauchy principal value type analogous to \eqref{eq:Fourier_expansion_absolute_n} of Appendix \ref{app:point_splitting_space} (see \cite{Erdmenger:2024xmj} for details).

\subsection{Entanglement entropy}
\label{subsec:EE}

Before we move towards holography, we calculate the entanglement entropy of an interval in the evolving curved space-time. We perform the calculation using the replica trick and a recent formulation \cite{Estienne:2025uhh} of the Liouville action on a replica manifold. The novelty of our derivation in this setting is that it does not rely on twist operators which makes the regularization scheme of the Euclidean path integrals completely transparent. At the end, we attempt to give a state interpretation for the resulting entropy and find that it exists when the path integral is computed in the chirally split scheme.

To this end, consider an interval $\mathcal{A}(t) = (\phi_1,\phi_2)\times \{t\}\subset (0, 2 \pi)\times \mathbb{R}$ on a constant-$t$ slice of the cylinder $S^1\times \mathbb{R}$ and let $\mathcal{A}(t)^c$ be its complement. The reduced density matrix of $\mathcal{A}(t)$ in the global state $\ket{\Psi(t)} = U(t)\ket{0}$ generated by the unitary evolution \eqref{eq:unitary_operator} in the curved space-time background \eqref{eq:curvedg} is given by $\rho(t) = \tr{_{\mathcal{A}(t)^c}\ket{\Psi(t)}\bra{\Psi(t)}}$. Then the entanglement entropy of the interval in the curved space-time is defined as
\begin{equation}
    S(t) \equiv -\tr{(\rho(t)\log{\rho(t)})} = -\lim_{n\rightarrow 1^+}\partial_n\tr{\rho(t)^n}\,.
    \label{eq:EE}
\end{equation}
We will compute this using the replica trick \cite{Calabrese:2004eu,Calabrese:2009qy} by performing the calculation in Euclidean signature and eventually analytically continuing to Lorentzian signature.\footnote{Analytic continuation has been used to compute time-evolution of entanglement entropy for example in \cite{Hartman:2013qma,Sivaramakrishnan:2017yng,Malvimat:2018cfe}.}

The Euclidean continuation of the time-dependent background \eqref{eq:curvedg} cannot be performed directly in the $t$-coordinate, because this would give a complex metric. Instead, we must perform the continuation in light-ray coordinates, which we write as $x^{\pm} = \theta \pm s$, where $\theta\in (0,2\pi)$ and $s\in\mathbb{R}$ are different from $(\phi,t)$. In these coordinates, the Lorentzian background metric \eqref{eq:curvedg} is explicitly $g_{ab}\,dx^adx^b = e^{2\sigma(x^-,x^+)}\,dx^-dx^+$ where $\sigma\colon \mathbb{R}\times \mathbb{R}\rightarrow \mathbb{R}$ is given by $\sigma = \omega + \varphi$ as defined in \eqref{eq:conformally_flat_g}. Then we start from a Euclidean metric $g_{\text{e}}$ of the form
\begin{equation}
    (g_{\text{e}})_{ab}\,dx^adx^b \equiv e^{2\sigma_{\text{e}}(w,\overbar{w})}\,dwd\overbar{w}\,,
    \label{eq:Euclidean_metric}
\end{equation}
where $w,\overbar{w}\in (0,2\pi)\times \mathbb{R}\subset \mathbb{C}$ are complex coordinates such that $\overbar{w} = w^*$ and we have defined a biholomorphic function $\sigma_{\text{e}}\colon \mathbb{C}\times \mathbb{C} \rightarrow \mathbb{R}$ which satisfies
\begin{equation}
    \sigma_{\text{e}}(x^-,x^+) = \sigma(x^-,x^+)\,,
\end{equation}
when the two independent arguments are restricted to the real axis $x^\pm\in\mathbb{R}$. Writing $w = \theta + i\tau$, $\overbar{w} = \theta -i\tau$, we see that the Lorentzian metric $g$ is obtained by analytic continuation of $\tau\rightarrow -is$ to the negative imaginary axis such that $w\rightarrow x^-$ and $\overbar{w}\rightarrow x^+$. For the replica trick, we therefore consider the Euclidean manifold $(S^1\times \mathbb{R},g_{\text{e}})$ in what follows.

Let us define the replica manifold \cite{Calabrese:2004eu,Calabrese:2009qy} which is an $n$-sheeted cover of the cylinder $(S^1\times \mathbb{R},g_{\text{e}})$ branched at the points $(w_i,\overbar{w}_i)$ with $i = 1,2$ which after analytic continuation coincide with the location of the entangling points in light-ray coordinates $w_i\rightarrow x^-_i = F_t(\phi_i)$ and $\overbar{w}_i\rightarrow x^+_i = \overbar{F}_t(\phi_i)$. There exists a ramified map\footnote{In \cite{Estienne:2025uhh}, the ramified map is defined as $\widetilde{\mathcal{P}}\colon M_n\rightarrow S^1\times \mathbb{R}$ where $M_n$ is the replica space. However, as long as there is a coordinate function $\psi\colon M_n\rightarrow S^1\times \mathbb{R}$ covering $M_n$ with a single chart (as is the case in this paper), we can write $\mathcal{P} = \widetilde{\mathcal{P}}\circ \psi^{-1}$.} $\mathcal{P}\colon S^1\times \mathbb{R}\rightarrow S^1\times \mathbb{R}$, which is an $n$-to-$1$ map from the cylinder to itself such that the replica manifold can be modeled\footnote{The metric $\mathcal{P}^*g_{\text{e}}$ has conical excesses at the branch points of the ramified map. Therefore when $(S^1\times \mathbb{R},\mathcal{P}^*g_{\text{e}})$ is embedded in higher-dimensional Euclidean space, it is an $n$-sheeted surface where each sheet is locally $(S^1\times \mathbb{R},g_{\text{e}})$ around each branch point.} as $(S^1\times \mathbb{R},\mathcal{P}^*g_{\text{e}})$. Then we define the following ratio of partition functions \cite{Estienne:2025uhh}
\begin{equation}
    \mathcal{R}_{(S^1\times \mathbb{R},g_{\text{e}})}(w_1,\overbar{w}_1;w_2,\overbar{w}_2) \equiv \frac{Z[\mathcal{P}^*g_{\text{e}};S^1\times \mathbb{R}]}{Z[g_{\text{e}};S^1\times \mathbb{R}]^n}\,.
    \label{eq:ratio}
\end{equation}
Here $Z[g_{\text{e}};M]$ denotes the Euclidean path integral over a manifold $(M,g_{\text{e}})$ defined as
\begin{equation}
    Z[g_{\text{e}};M] \equiv \int [d\Phi_{\text{e}}]\,e^{-I_{\text{E}}[\Phi_{\text{e}},g_{\text{e}};M]}\,,
    \label{eq:Z_def}
\end{equation}
where the Euclidean action of the CFT is defined as $I_{\text{E}}[\Phi_{\text{e}},g_{\text{e}};M] = -iI[\Phi,g;M]$ and $\Phi_{\text{e}}$ denotes the Euclidean fundamental field.

The definition \eqref{eq:ratio} takes as input the Euclidean cylinder $(S^1\times \mathbb{R},g_{\text{e}})$ and the ramified map $\mathcal{P}$ through which it depends on the locations of the two branch points $(w_i,\overbar{w}_i)$. Upon analytic continuation described above, the ratio \eqref{eq:ratio} becomes the Lorentzian function $\mathcal{R}_{(S^1\times \mathbb{R},g)}(x^-_1,x^+_1;x^-_2,x^+_2)$ which determines the Rényi entropy of the interval $(\phi_1,\phi_2) \in S^1$ as \cite{Calabrese:2004eu,Calabrese:2009qy}
\begin{equation}
    \tr{\rho(t)^n} = \mathcal{R}_{(S^1\times \mathbb{R},g)}(F_t(\phi_1),\overbar{F}_t(\phi_1);F_t(\phi_2),\overbar{F}_t(\phi_2))\,.
    \label{eq:Renyi_entropy}
\end{equation}
In the literature, what is often neglected, is the renormalization scheme in which the path integrals are computed: we claim that equation \eqref{eq:Renyi_entropy}, which identifies the trace with a ratio of path integrals, is valid only in the chirally split scheme as this is the scheme reproduced by the operator formulation. We will now show how to compute \eqref{eq:ratio} using results of \cite{Estienne:2025uhh} in the diffeomorphism invariant scheme from which the expression in the chirally split scheme may be obtained by adding the finite counterterm described in Section \ref{subsec:chirally_split_scheme}.

\paragraph{Diffeomorphism invariant scheme.} In the diffeomorphism invariant scheme, the Euclidean path integral \eqref{eq:Z_def} over a manifold $(M,g_{\text{e}})$ is performed with a diffeomorphism invariant integration measure and denoted by $Z_{\text{D}}[g;M]$. This implies that the transformation property of the classical action under diffeomorphisms is lifted to the path integral $Z_{\text{D}}[D^*g;D^{-1}(M)] = Z_{\text{D}}[g,M] $. We can use this to relate the replica path integral \eqref{eq:ratio} on the cylinder $S^1\times \mathbb{R}$ to a path integral on the plane $\mathbb{R}^2$. To this end, let $\mathcal{C}\colon S^1\times \mathbb{R} \rightarrow \mathbb{R}^2 $ be the diffeomorphism \cite{Besken:2020snx}
\begin{equation}
    \mathcal{C}^w(w,\overbar{w}) = C(w) \equiv \tan{\frac{w}{2}}\,,\quad \mathcal{C}^{\overbar{w}}(w,\overbar{w}) = \overbar{C}(\overbar{w}) \equiv \tan{\frac{\overbar{w}}{2}}\,,
    \label{eq:tan_map_euclidean}
\end{equation}
which maps the cylinder to the plane such that $w=\pm i\infty$ are mapped to $w = \pm i$. Then diffeomorphism invariance implies
\begin{equation}
    \mathcal{R}_{(S^1\times \mathbb{R},g_{\text{e}})}(w_1,\overbar{w}_1;w_2,\overbar{w}_2) = \mathcal{R}_{(\mathbb{R}^2,(\mathcal{C}^{-1})^*g_{\text{e}})}(\mathcal{C}(w_1,\overbar{w}_1);\mathcal{C}(w_2,\overbar{w}_2))\,,
    \label{eq:mapping_ratio}
\end{equation}
where $(\mathcal{C}^{-1})^*g_{\text{e}}$ is the pull-back of the metric on the cylinder to the plane. We will now compute $\mathcal{R}_{(\mathbb{R}^2,g_{\text{e}})}(w_1,\overbar{w}_1;w_2,\overbar{w}_2)$ for a generic metric $g_{\text{e}}$ (we denote it by the same symbol as the original metric on the cylinder for simplicity) on the plane $\mathbb{R}^2$ taking the form
\begin{equation}
    (g_{\text{e}})_{ab}\,dx^adx^b = e^{2\chi_{\text{e}}}\,dwd\overbar{w}\,.
    \label{eq:g_e_chi}
\end{equation}
The result on the cylinder is then obtained via \eqref{eq:mapping_ratio}.

First, we use the fact that on the plane, we have (our conventions are the same as in \cite{Estienne:2025uhh})
\begin{equation}
    \frac{Z_{\text{D}}[g_{\text{e}};\mathbb{R}^2]}{Z_{\text{D}}[\eta_{\text{e}}; \mathbb{R}^2]} = e^{ I_{\text{Lio}}[\chi_{\text{e}},\eta_{\text{e}};\mathbb{R}^2]}\,,
    \label{eq:normal_Liouville}
\end{equation}
where $I_{\text{Lio}}[\chi,g;M]$ is the (Euclidean) Liouville action defined in \eqref{eq:Liouville_action_text}. It describes the usual non-invariance of the path integral under Weyl transformations in the diffeomorphism invariant scheme. This equation is valid as long as the Weyl factor $\chi_{\text{e}}$ is free of singularities.

For the path integral over the replica manifold of the plane, modeled as $(\mathcal{P}^*g_{\text{e}},\mathbb{R}^2)$, the formula \eqref{eq:normal_Liouville} requires modification due to conical singularities in the pull-back metric $\mathcal{P}^*g_{\text{e}}$ at the branch points $(w_i,\overbar{w}_i)\in \mathbb{R}^2$ \cite{Estienne:2025uhh}. In the case of a plane $\mathbb{R}^2$, the ramified map $\mathcal{P}\colon M_n\rightarrow \mathbb{R}^2$ is a conformal diffeomorphism of the flat metric of the form
\begin{equation}
    \mathcal{P}^w(w,\overbar{w}) = P(w) \equiv \biggl(\frac{w-w_1}{w-w_2}\biggr)^n\,,\quad \mathcal{P}^{\overbar{w}}(w,\overbar{w}) = \overbar{P}(\overbar{w}) \equiv \biggl(\frac{\overbar{w}-\overbar{w}_1}{\overbar{w}-\overbar{w}_2}\biggr)^n\,.
    \label{eq:uniformization_map}
\end{equation}
Thus the pull-back of \eqref{eq:g_e_chi} is explicitly
\begin{equation}
    (\mathcal{P}^*g_{\text{e}})_{ab}(x)\,dx^adx^b = e^{2\widetilde{\chi}_{\text{e}}(w,\overbar{w})}\,dwd\overbar{w}\,,
\end{equation}
where the Weyl factor is given by
\begin{gather}
\widetilde{\chi}_{\text{e}}(w,\overbar{w}) = \kappa_{\text{e}}(w,\overbar{w})+ \Sigma_{\text{e}}(w,\overbar{w})\,,\label{eq:conical_liouville_field}\\
    \kappa_{\text{e}}(w,\overbar{w}) \equiv \frac{1}{2}\log{(P'(w)\,\overbar{P}'(\overbar{w}))}\,,\quad \Sigma_{\text{e}}(w,\overbar{w}) \equiv \chi_{\text{e}}(P(w),\overbar{P}(\overbar{w}))\label{eq:kappae}\,.  
\end{gather}
By substituting \eqref{eq:uniformization_map}, we see that the Weyl factor contains logarithmic singularities of the form
\begin{equation}
    \widetilde{\chi}_{\text{e}}(w,\overbar{w}) = \frac{\gamma_i}{2} \log{\vert w-w_i\vert^2} +\mathcal{O}(1)\,,\quad w\rightarrow w_i\,,
\end{equation}
where $\gamma_i \equiv n-1$. The logarithmic singularities in the Weyl factor manifest in the metric as conical singularities with angular excesses $2\pi \gamma_i $ (see Appendix \ref{app:regularized_liouville}). Therefore the replica path integral $Z_{\text{D}}[\mathcal{P}^*g_{\text{e}};\mathbb{R}^2]$ must be regularized appropriately. Motivated by \cite{Estienne:2025uhh}, we regularize by cutting holes around the singularities $\mathbb{R}^2\rightarrow \mathbb{R}^2\,\backslash\,\mathcal{B}$ and consider the path integral
\begin{equation}
    Z_{\text{D}}^{\text{reg}}[\mathcal{P}^*g_{\text{e}};\mathbb{R}^2] \equiv Z_{\text{D}}[\mathcal{P}^*g_{\text{e}};\mathbb{R}^2\,\backslash\, \mathcal{B}]\,.
\end{equation}
Here $\mathcal{B} = \cup_{i=1}^2 B_{\tilde{\delta}_i}(w_i;\mathcal{P}^*g_{\text{e}})$ is the union of disks centered at $(w_i,\overbar{w}_i)$ of geodesic radius $\tilde{\delta}_i$ measured in the conically singular metric $\mathcal{P}^*g_{\text{e}} $ \cite{Estienne:2025uhh}. The use of geodesic disks ensures that the integration domain is defined in a diffeomorphism invariant manner and we have also chosen the radius $\tilde{\delta}_i$ to be different for each singular point. For non-singular Weyl factors, we can take the limit $\tilde{\delta}_i\rightarrow 0$ in which $\mathbb{R}^2\backslash\, \mathcal{B} \rightarrow \mathbb{R}^2 $. Thus the regularized Liouville action reduces to the ordinary one in this limit.

With this regulation in mind, the correct replica path integral for the plane is given by
\begin{equation}
    \mathcal{R}_{(\mathbb{R}^2,g_{\text{e}})}^{\tilde{\delta}_1,\tilde{\delta}_2}(w_1,\overbar{w}_1;w_2,\overbar{w}_2)\equiv \frac{Z_{\text{D}}^{\text{reg}}[\mathcal{P}^*g_{\text{e}};\mathbb{R}^2]}{Z_{\text{D}}^{\text{reg}}[g_{\text{e}};\mathbb{R}^2]^n}\,,
\end{equation}
which replaces the original definition \eqref{eq:ratio}. Taking $\tilde{\delta}_i\rightarrow 0$ produces divergences which map to divergences of the entanglement entropy.

Locally, the Liouville action still controls the anomaly away from the branch points. Therefore ratios of replica path integrals transform as \cite{Estienne:2025uhh}
\begin{equation}
    \frac{Z_{\text{D}}^{\text{reg}}[\mathcal{P}^*g_{\text{e}};\mathbb{R}^2]}{Z_{\text{D}}^{\text{reg}}[\eta_{\text{e}};\mathbb{R}^2]} = e^{ I_{\text{Lio}}^{\text{reg}}[\widetilde{\chi}_{\text{e}},\eta_{\text{e}};\mathbb{R}^2]}\,.
    \label{eq:regularized_Z_transformation}
\end{equation}
As shown in \cite{Estienne:2025uhh}, and reviewed in detail in Appendix \ref{app:regularized_liouville}, the regularized replica path integral transforms under Weyl transformations as
\begin{equation}
    \mathcal{R}_{(\mathbb{R}^2,g_{\text{e}})}^{\tilde{\delta}_1,\tilde{\delta}_2}(w_1,\overbar{w}_1;w_2,\overbar{w}_2) = e^{-\Delta_n( \chi_{\text{e}}(w_1,\overbar{w}_1)+ \chi_{\text{e}}(w_2,\overbar{w}_2))}\;\mathcal{R}_{(\mathbb{R}^2,\eta_{\text{e}})}^{\tilde{\delta}_1,\tilde{\delta}_2}(w_1,\overbar{w}_1;w_2,\overbar{w}_2)\,,
    \label{eq:R_transformation_diff}
\end{equation}
where we have defined
\begin{equation}
    \Delta_n \equiv \frac{c}{12}\frac{\gamma_i\,(\gamma_i+2)}{\gamma_i+1} = \frac{c}{12}\biggl(n-\frac{1}{n}\biggr)\,.
    \label{eq:twist_scaling_dimension}
\end{equation}
Equation \eqref{eq:R_transformation_diff} indicates that $\mathcal{R}_{(\mathbb{R}^2,\eta_{\text{e}})}^{\tilde{\delta}_1,\tilde{\delta}_2}$ transforms under Weyl transformations of $\eta_{\text{e}}$ as a product of two local fields of Weyl weight $\Delta_n$ inserted at $w_1$ and $w_2$. These local fields can be identified with twist fields \cite{Calabrese:2004eu,Calabrese:2009qy}.

Now we will apply these to the case $g_{\text{e}}\rightarrow (\mathcal{C}^{-1})^*g_{\text{e}}$ which is a pull-back of the metric on the cylinder. Given the metric \eqref{eq:Euclidean_metric} on the cylinder, we see that $(\mathcal{C}^{-1})^*g_{\text{e}} = e^{2\chi_{\text{e}}}\,\eta_{\text{e}}$ where the Weyl factor
\begin{equation}
	\chi_{\text{e}}(w,\overbar{w}) = \sigma_{\text{e}}(C^{-1}(w),\overbar{C}^{-1}(\overbar{w})) +\frac{1}{2}\log{(C^{-1})'(w)}+\frac{1}{2}\log{(\overbar{C}^{-1})'(\overbar{w})}\,.
    \label{eq:chi_e}
\end{equation}
Substituting to \eqref{eq:R_transformation_diff} and using \eqref{eq:mapping_ratio}, we obtain
\begin{align}
	&\mathcal{R}_{(S^1\times \mathbb{R},g_{\text{e}})}^{\tilde{\delta}_1,\tilde{\delta}_2}(w_1,\overbar{w}_1;w_2,\overbar{w}_2)\\
	&= \mathcal{R}_{(\mathbb{R}^2,\eta_{\text{e}})}^{\tilde{\delta}_1,\tilde{\delta}_2}(C(w_1), \overbar{C}(\overbar{w}_1);C(w_2), \overbar{C}(\overbar{w}_2))\prod_{i=1}^2\,\bigl[e^{-2 \sigma_{\text{e}}(w_i,\overbar{w}_i)}\,C'(w_i)\, \overbar{C}'(\overbar{w}_i)\bigr]^{\Delta_n\slash 2}\,.
	\label{eq:cylinder_R_flat}
\end{align}
This formula relates the replica path integral on the curved cylinder $(S^1\times\mathbb{R},g_{\text{e}})$ to the path integral on the flat plane $(\mathbb{R}^2,\eta_{\text{e}})$. In the diffeomorphism invariant scheme, the flat space result is completely fixed by the $SL(2,\mathbb{C})$ Ward identity up to a single constant $b(n)$, see Appendix \ref{subapp:Liouville_wards}. The result is
\begin{equation}
    \mathcal{R}_{(\mathbb{R}^2,\eta_{\text{e}})}^{\tilde{\delta}_1,\tilde{\delta}_2}(w_1,\overbar{w}_1;w_2,\overbar{w}_2) = b(n)\,\biggl[\frac{\tilde{\delta}_1 \tilde{\delta}_2}{(w_1-w_2)(\overbar{w}_1-\overbar{w}_2)}\biggr]^{\Delta_n}\,,
    \label{eq:flatplane_result}
\end{equation}
which is valid to leading order in $\tilde{\delta}_{1,2}\rightarrow 0$. Thus substituting to \eqref{eq:cylinder_R_flat}, we obtain
\begin{equation}
    \mathcal{R}_{(S^1\times \mathbb{R},g_{\text{e}})}^{\tilde{\delta}_1,\tilde{\delta}_2}(w_1,\overbar{w}_1;w_2,\overbar{w}_2) = b(n)\,\biggl[\frac{\prod_{i=1}^2\tilde{\delta}_i^2 \,e^{-2\sigma_{\text{e}}(w_i,\overbar{w}_i)}\,C'(w_i)\,\overbar{C}'(w_i)}{(C(w_1)-C(w_2))^2(\overbar{C}(\overbar{w}_1)-\overbar{C}(\overbar{w}_2))^2}\biggr]^{\Delta_n\slash 2}\,.
    \label{eq:cylinder_ratio_final}
\end{equation}
Substituting the map \eqref{eq:tan_map_euclidean} and using $\sigma = \varphi + \omega$ as defined in \eqref{eq:conformally_flat_g}, we obtain the Rényi entropy
\begin{equation}
    \tr{\rho(t)^n}\vert_{\text{D}} = b(n)\,\Biggl[\frac{\tilde{\delta}_1^2\tilde{\delta}_2^2\,e^{-2\,(\omega(\phi_1,t)+\omega(\phi_2,t))}\,F_{t}'(\phi_1)\,\overbar{F}_{t}'(\phi_1)\,F_{t}'(\phi_2)\,\overbar{F}_{t}'(\phi_2)}{16\,\sin^2{\bigl(\frac{F_t(\phi_1)-F_t(\phi_2)}{2}\bigr)}\sin^2{\bigl(\frac{\overbar{F}_t(\phi_1)-\overbar{F}_t(\phi_2)}{2}\bigr)}}\Biggr]^{\Delta_n\slash 2}\,,
    \label{eq:trace_rho_n_diff_inv}
\end{equation}
where $\vert_{\text{D}}$ indicates that the result is valid in the diffeomorphism invariant scheme. Normalizability of the density matrix $\tr{\rho(t)} = 1$ requires $b(n=1) = 1$. Notice that the geodesic cut-offs $\tilde{\delta}_i$ in the metric $\mathcal{P}^*g = e^{2\sigma\circ \mathcal{P}}\mathcal{P}^*\eta$ are given in terms of the cut-offs $\tilde{\varepsilon}_i$ in the pull-back of the flat metric $\mathcal{P}^*\eta$ by $\tilde{\delta}_i = e^{\sigma(w_i,\overbar{w}_i)}\,\tilde{\varepsilon}_i$. Using $\sigma = \omega - \frac{1}{2}\log{F_t'} - \frac{1}{2}\log{\overbar{F}_t'}$, we see that \eqref{eq:trace_rho_n_diff_inv} reduces to
\begin{equation}
    \tr{\rho(t)^n}\vert_{\text{D}} = b(n)\,\Biggl[\frac{\tilde{\varepsilon}_1^2\tilde{\varepsilon}_2^2}{16\,\sin^2{\bigl(\frac{F_t(\phi_1)-F_t(\phi_2)}{2}\bigr)}\sin^2{\bigl(\frac{\overbar{F}_t(\phi_1)-\overbar{F}_t(\phi_2)}{2}\bigr)}}\Biggr]^{\Delta_n\slash 2}\,,
    \label{eq:trace_rho_n_diff_inv_2}
\end{equation}
which is equal to the Rényi entropy in flat space obtained by setting $\sigma = 0$ in \eqref{eq:trace_rho_n_diff_inv}. Therefore the curved space result \eqref{eq:EE_1} in terms of the curved space cut-offs $\tilde{\delta}_{i}$ is equal to the flat space expression \eqref{eq:trace_rho_n_diff_inv_2} in terms of the flat space cut-offs $\tilde{\varepsilon}_i$. By writing $\tr{\rho(t)^n}\vert_{\text{D}} \equiv \mathcal{R}_{\text{D}}(g;\tilde{\delta}_i)$, this result can be summarized as $\mathcal{R}_{\text{D}}(g;\tilde{\delta}_i) = \mathcal{R}_{\text{D}}(e^{2\sigma}\eta;e^{\sigma}\tilde{\varepsilon}_i) = \mathcal{R}_{\text{D}}(\eta;\tilde{\varepsilon}_i)$. It follows that the curved space Rényi entropy in terms of the flat space cut-offs is obtained by a simple Weyl rescaling of the cut-offs in the flat space Rényi entropy $\mathcal{R}_{\text{D}}(g;\tilde{\varepsilon}_i) = \mathcal{R}_{\text{D}}(\eta;e^{-\sigma}\tilde{\varepsilon}_i) $.\footnote{This fact is often used in the literature, but here we have provided an explicit derivation starting from the Liouville action. We emphasize that this can be done only in the diffeomorphism invariant scheme.} Therefore replacing $\tilde{\varepsilon}_i\rightarrow e^{-\sigma}\tilde{\varepsilon}_i$ in \eqref{eq:trace_rho_n_diff_inv_2}, the entanglement entropy \eqref{eq:EE} of the interval $(\phi_1,\phi_2)\subset S^1$ in the metric $g$ with flat space cut-offs becomes
\begin{equation}
    S_{\text{D}}(t) = \frac{c}{6}\,(\omega(\phi_1,t) + \omega(\phi_2,t))+\frac{c}{12}\Biggl[\log{\Biggl(\frac{4\sin^2{\bigl(\frac{F_t(\phi_1)-F_t(\phi_2)}{2}\bigr)}}{\varepsilon_1\varepsilon_2\, F_{t}'(\phi_1)\,F_{t}'(\phi_2)}\Biggr)}+(F_t\leftrightarrow \overbar{F}_t)\Biggr] -b'(1)\,,
    \label{eq:EE_1}
\end{equation}
where we have used $\partial_n\Delta_n\vert_{n=1} = \frac{c}{6}$ and defined the geodesic cut-offs $\varepsilon_{i} = \tilde{\varepsilon}_{i}\vert_{n = 1} $ in the flat metric $\eta$. In Section \ref{sec:holography}, we derive the same expression using the Ryu--Takayanagi formula.

Consider the flat metric $g = \eta$ obtained by setting $\sigma  = 0$.\footnote{In this case, the coordinates $(\phi,t)$ define a non-trivial slicing of the flat cylinder.} The entanglement entropy \eqref{eq:EE_1} reduces to
\begin{equation}
    S_{\text{D}}(t)\vert_{\text{flat}} = \frac{c}{12}\log{\biggl[\frac{4}{\varepsilon_1\varepsilon_2}\sin^2{\biggl(\frac{F_t(\phi_1)-F_t(\phi_2)}{2}\biggr)}\biggr]}+(F_t\leftrightarrow \overbar{F}_t) -b'(1)\,,
    \label{eq:EE_flat_2}
\end{equation}
Let us also mention that the entropy \eqref{eq:EE_1} is obtained from a two-point function of twist operators $\mathcal{T}_n^{\text{H}}(\phi,t)$ and $\widetilde{\mathcal{T}}_n^{\text{H}}(\phi,t)$ which in the Heisenberg picture are scalar primary operators \eqref{eq:2D_scalar_operator} with scaling dimension $\Delta = \Delta_n = \frac{c}{12}(n-\frac{1}{n})$. Indeed, we can see that up to factors, the trace \eqref{eq:trace_rho_n_diff_inv} takes the form
\begin{equation}
    \tr{\rho(t)^n}\vert_{\text{D}}  = \bra{0} \mathcal{T}_n^{\text{H}}(\phi_1,t)\,\widetilde{\mathcal{T}}_n^{\text{H}}(\phi_2,t)\ket{0},
    \label{eq:twist_2_point}
\end{equation}
where we have used that the primary two-point function is given by \eqref{eq:scalar_2_point}. From this viewpoint, the interpretation of the factors $F_t'(\phi),\overbar{F}'_t(\phi)$ in the entropy \eqref{eq:EE_1} is that they arise from the adjoint action of $U(t) $ on the twists which transform as primaries \eqref{eq:primary_transformation_law}.

The formula \eqref{eq:EE_1} is problematic from the operator point of view. The appearance of the Weyl factor $\omega$ evaluated at the end-points of the interval in \eqref{eq:EE_1} agrees with previous results in the literature such as \cite{Shaposhnik:2022jzc}. However, this $\omega$-dependence is problematic, because the Hamiltonian operator \eqref{eq:HS_split} of the CFT is $\omega$-independent due to Weyl invariance. Therefore the time-evolved state $\ket{\Psi(t)} = U(t)\ket{0}$ and also the entropy $S(t)$, which is a property of the state alone, are independent of $\omega$. As alluded above, the solution to this puzzle is that we have computed the replica path integral in an incorrect scheme (the diffeomorphism invariant scheme), while it is the chirally split scheme which reproduces results consistent with a state interpretation. We will now show that the replica path integral in the chirally split scheme produces an entropy independent of $\omega$ as required.

\paragraph{Chirally split scheme.} We want to write the replica path integrals \eqref{eq:normal_Liouville} and \eqref{eq:regularized_Z_transformation} in the chirally split renormalization scheme. First, let us define the functions $\omega_{\text{e}}(w,\overbar{w})$ and $\varphi_{\text{e}}(w,\overbar{w})$ which analytically continue to the Weyl factors $\omega(x^-,x^+)$ and $\varphi(x^-,x^+)$ of the Lorentzian metric defined in \eqref{eq:conformally_flat_g} respectively. It follows that $\sigma_{\text{e}} = \omega_{\text{e}}+\varphi_{\text{e}}$. To move to the chirally split scheme, we must subtract the Liouville action as prescribed in \eqref{eq:chirally_split_scheme}. Therefore the Euclidean path integral in the chirally split scheme is given by\footnote{To go completely to the chirally split scheme, one has to also subtract the counterterm $K[\nu,\overbar{\nu}]$. However, this term must cancel in the ratio \eqref{eq:ratio}, because the result \eqref{eq:cylinder_ratio_final} is already chirally split in $(F_t,\overbar{F_t})$. Therefore this counterterm is not involved in the entanglement entropy and we will not include it here.}
\begin{equation}
    Z_{\text{F}}^{\text{reg}}[g_{\text{e}};S^1\times\mathbb{R}] \equiv Z_{\text{D}}^{\text{reg}}[g_{\text{e}};S^1\times\mathbb{R}]\,e^{-I_{\text{Lio}}^{\text{reg}}[\omega_{\text{e}},\hat{g}_{\text{e}};S^1\times\mathbb{R}]}\,.
    \label{eq:chirally_split_path_integral}
\end{equation}
The counterterm here appears in the ratio
\begin{equation}
    \frac{Z_{\text{D}}^{\text{reg}}[g_{\text{e}};S^1\times\mathbb{R}]}{Z_{\text{D}}^{\text{reg}}[\hat{g}_{\text{e}}; S^1\times\mathbb{R}]} = e^{ I_{\text{Lio}}^{\text{reg}}[\omega_{\text{e}},\hat{g}_{\text{e}};S^1\times\mathbb{R}]}\,,
    \label{eq:normal_Liouville_2}
\end{equation}
where we have defined $\hat{g}_{\text{e}} \equiv g_{\text{e}}\vert_{\omega_{\text{e}} = 0}$. Substituting to \eqref{eq:chirally_split_path_integral}, we see that the chirally split path integral becomes
\begin{equation}
    Z_{\text{F}}^{\text{reg}}[g_{\text{e}};S^1\times\mathbb{R}] = Z_{\text{D}}^{\text{reg}}[\hat{g}_{\text{e}};S^1\times\mathbb{R}] \,.
    \label{eq:Z_F}
\end{equation}
Therefore path integrals in the chirally split scheme are obtained from path integrals in the diffeomorphism invariant scheme by setting $\omega_{\text{e}} = 0$.\footnote{Up to a factor involving the counterterm $K[\nu,\overbar{\nu}]$ which is not relevant for the entanglement entropy; see the previous footnote.} In Lorentzian signature, the entropy $S(t)$ in the chirally split scheme is therefore given by $S(t) = S_{\text{D}}(t)\vert_{\omega = 0}$ which using \eqref{eq:EE_1} gives
\begin{equation}
    S(t) = \frac{c}{12}\log{\Biggl(\frac{4\sin^2{\bigl(\frac{F_t(\phi_1)-F_t(\phi_2)}{2}\bigr)}}{\varepsilon_1\varepsilon_2\, F_{t}'(\phi_1)\,F_{t}'(\phi_2)}\Biggr)}+(F_t\leftrightarrow \overbar{F}_t)-b'(1)\,,
    \label{eq:chirally_split_entropy}
\end{equation}
where $\varepsilon_{1,2}$ are geodesic cut-offs in the flat metric $\eta$.\footnote{The same result can also be obtained from the Rényi entropy which in the chirally split scheme with flat space cut-offs is $\tr{\rho(t)^n}\vert_{\text{F}} \equiv \mathcal{R}_{\text{F}}(g;\tilde{\varepsilon}_i) = \mathcal{R}_{\text{D}}(g\vert_{\omega_{\text{e}}=0} ;\tilde{\varepsilon}_i) = \mathcal{R}_{\text{D}}(e^{2\varphi}\eta ;\tilde{\varepsilon}_i) = \mathcal{R}_{\text{D}}(\eta ;e^{-\varphi}\tilde{\varepsilon}_i) $. Replacing $\tilde{\varepsilon}_i\rightarrow e^{-\varphi}\tilde{\varepsilon}_i$ in \eqref{eq:trace_rho_n_diff_inv_2} and computing the derivative with respect to $n$ at $n = 1$ yields \eqref{eq:chirally_split_entropy}.} As explained above, this is the correct expression for the entanglement entropy in a CFT driven by the metric \eqref{eq:curvedg}. It is a non-local function of the metric components, because $F_t,\overbar{F}_t$ are obtained as solutions of the first-order differential equations \eqref{eq:varphinunubar} in terms of $\nu,\overbar{\nu}$. Therefore the entropy is sensitive to the full history of the background metric on the time-interval $(0,t)$. A Lorentzian operator-algebraic derivation of the formula \eqref{eq:chirally_split_entropy} is left for future work.

Let us consider the static case $\partial_t \nu = \partial_t\overbar{\nu} = 0$ as defined in \eqref{eq:static_nu_nubar}. In this case, the functions $F_t,\overbar{F}_t$ are given by \eqref{eq:static_curve}. If we assume the initial conditions $F_0 = h \equiv p$ and $\overbar{F}_0 =\overbar{h} \equiv \overbar{p}$, the entropy becomes 
\begin{equation}
    S(t) = \frac{c}{12}\log{\Biggl(\frac{4\sin^2{\bigl(\frac{p(\phi_1)-p(\phi_2)}{2}\bigr)}}{\varepsilon_1\varepsilon_2\, p'(\phi_1)\,p'(\phi_2)}\Biggr)}+(p\leftrightarrow \overbar{p})-b'(1)\,.
    \label{eq:chirally_split_entropy_static}
\end{equation}
The result is time-independent, because as explained in Section \ref{subsec:unitary_evolution}, the global state \eqref{eq:time_ind_state} is time-independent up to a phase in this case. The formula \eqref{eq:chirally_split_entropy_static} computes the entanglement entropy in a global Virasoro pure state.

The formula \eqref{eq:chirally_split_entropy} in the chirally split scheme is consistent with previous formulae derived using twist operators in the inhomogeneous CFT literature \cite{Bernard:2019mqm,Lapierre:2020ftq,Fan:2020orx}. The static result \eqref{eq:chirally_split_entropy_static} is also consistent with \cite{Tonni:2017jom}. The formulae in \cite{Bernard:2019mqm,Lapierre:2020ftq,Fan:2020orx} are based on defining the Schrödinger picture twist operator to be independent of the Weyl factor $\omega$ as
\begin{equation}
    \mathcal{T}_n^{\text{S}}(\phi) = \mathcal{O}_{\Delta_n\slash 2}(\phi)\otimes \mathcal{O}_{\Delta_n\slash 2}(\phi)\,.
\end{equation}
Then it follows that \cite{Bernard:2019mqm,Lapierre:2020ftq,Fan:2020orx}
\begin{equation}
    \tr{\rho(t)^n} = \bra{\Psi(t)}\mathcal{T}_n^{\text{S}}(\phi)\,\widetilde{\mathcal{T}}_n^{\text{S}}(\phi)\ket{\Psi(t)} 
\end{equation}
which gives the same answer as \eqref{eq:trace_rho_n_diff_inv} up to constant factors, but with $\omega = 0$. Alternatively, moving to the chirally split scheme amounts to a Weyl rescaling of the curved space Heisenberg picture twist operators as $\mathcal{T}_n^{\text{H}}(\phi,t)\rightarrow e^{\Delta_n \omega}\,\mathcal{T}_n^{\text{H}}(\phi,t)$.

\paragraph{Comparison with previous literature.} We can see that for a metric of the form \eqref{eq:g_conformally_flat} with $\nu = -1$ and $\overbar{\nu} = 1$, corresponding to $F_t(\phi) = \phi - t$ and $\overbar{F}_t(\phi) = \phi + t$ , there is no entropy production in the chirally split scheme $S(t) - S(0) = 0$. However, in the diffeomorphism invariant scheme, the entropy production is controlled by the Weyl factor
\begin{equation}
    S_{\text{D}}(t) - S(0) = \frac{c}{6}\,(\omega(\phi_1,t) + \omega(\phi_2,t))\,.
\end{equation}
This result was used to calculate the entropy production in an expanding universe in \cite{Cotler:2022weg}. All of this generated entropy comes from the dependence of the cut-off on the Weyl factor and does not have a state interpretation.

We may also consider the case $\omega = 0$ and $F_t(\phi) = h(\phi-t)$, $\overbar{F}_t(\phi) = \overbar{h}(\phi+t)$ for which the metric is $g_{ab}\,dx^adx^b = e^{2\rho}\,dx^-dx^+$ with Weyl factor $e^{2\rho} \equiv h'(\phi-t)\,\overbar{h}'(\phi+t)$. In this case, the entropy in both diffeomorphism and chirally split schemes is given by
\begin{equation}
    S(t) = \frac{c}{6}\,(\rho(\phi_1,t)+\rho(\phi_2,t))+\frac{c}{12}\log{\biggl[\frac{4}{\varepsilon_1\varepsilon_2}\sin^2{\biggl(\frac{h(\phi_1-t)-h(\phi_2-t)}{2}\biggr)}\biggr]}+(h\leftrightarrow \overbar{h})\,,
    \label{eq:chirally_split_entropy_Fiola}
\end{equation}
which matches with the formula given in \cite{Fiola:1994ir} after mapping to Minkowski space.

\section{The holographic dual of a driven CFT}\label{sec:holography}

In this section, we revisit the holographic dual of a CFT driven by a time-dependent background metric in light of the results of the previous sections. This extends the holographic duals of a static inhomogeneous CFT \cite{Li:2025rzl} and of a slowly-driven CFT \cite{deBoer:2023lrd} to arbitrary driving. The holographic dual geometry allows us to reproduce field theory calculations of Section \ref{sec:CCrevisited} from the bulk.

\subsection{The bulk metric}
\label{subsec:hologrpahy_setup}

We are interested in solutions of 3D gravity that asymptote to the background metric \eqref{eq:curvedg} driving the CFT on the conformal boundary. In light-ray coordinates \eqref{eq:lightraycoords}, the boundary metric takes the conformally flat form \eqref{eq:conformally_flat_g}. A method yielding a bulk solution that asymptotes to a general conformally flat metric on the boundary was presented in \cite{Li:2025rzl}, however, only the bulk solution that gives a static boundary metric \eqref{eq:staticg} was used to describe (time-independent) inhomogeneous CFTs. Here we realize that this approach can also be used to derive the solution dual to an arbitrarily driven inhomogeneous CFT.

To see how the dual solution is found, we start with AdS$_3$ solution $(\mathbb{R}^2\times \mathbb{R}_+,G)$ where the AdS metric in Poincaré coordinates $ (z,x^-,x^+)$ is
\begin{equation}
    G_{\mu\nu}(X)\,dX^\mu dX^\nu = \frac{\ell^2}{z^2}\,(dz^2+dx^-dx^+)\,,
    \label{eq:poincare_metric}
\end{equation}
where $X^\mu = (z,x^a)$ denote bulk coordinates. The conformal boundary $\partial (\mathbb{R}^2\times \mathbb{R}_+) = \mathbb{R}^2$ at $z = 0$ is endowed with the flat Lorentzian metric
\begin{equation}
    \lim_{z\rightarrow 0}\frac{z^2}{\ell^2}\,G_{ab}(X)\,dX^a dX^b = \eta_{ab}\,dx^adx^b= dx^-dx^+
\end{equation}
so that the dual CFT lives on Minkowski space $(\mathbb{R}^2,\eta)$. The metric \eqref{eq:poincare_metric} is a solution of 3D gravity since solutions of Einstein's equations in three dimensions are locally AdS in an open neighborhood around any point. 

We want to find a bulk solution $(\mathcal{M},\widetilde{G})$ of 3D Einstein's equations with a cylinder boundary $\partial \mathcal{M} = S^1\times \mathbb{R}$ equipped with the metric $g$ \eqref{eq:curvedg} that drives the dual CFT. For the bulk metric, this amounts to the boundary condition
\begin{equation}
    \lim_{z\rightarrow 0}\frac{z^2}{\ell^2}\,\widetilde{G}_{ab}(X)\,dX^a dX^b =g_{ab}(x)\,dx^adx^b = e^{2\sigma}\,dx^-dx^+\,,
    \label{eq:3D_BC}
\end{equation}
where the Weyl factor $\sigma = \omega + \varphi$ with $\varphi = -\frac{1}{2}\log{F_t'(\phi)}-\frac{1}{2}\log{\overbar{F}_t'(\phi)}$ so that the boundary metric coincides with \eqref{eq:curvedg}. Since Einstein's equations are diffeomorphism covariant, a new solution is found by pulling back the Poincaré AdS metric \eqref{eq:poincare_metric} with a diffeomorphism $D\colon \mathcal{M}\rightarrow  \mathbb{R}^2\times \mathbb{R}_+$ as
\begin{equation}
\widetilde{G}_{\mu\nu}(X) =  \frac{\partial D^{\rho}}{\partial x^{\mu}}\frac{\partial D^{\sigma}}{\partial x^{\nu}}\,G_{\rho\sigma}(D(X))\,.
\label{eq:Gtilde}
\end{equation}
The condition $\partial \mathcal{M} = S^1\times \mathbb{R}$ requires that on the conformal boundary $z = 0$, the diffeomorphism $D$ acts as the two-dimensional conformal diffeomorphism $\mathcal{C}\colon S^1\times \mathbb{R} \rightarrow \mathbb{R}^2$ mapping a subset $x^{\pm}\in (-\pi,\pi)$ of the Lorentzian cylinder $(S^1\times \mathbb{R},\eta)$ to Minkowski space $(\mathbb{R}^2,\eta)$ parametrized by $x^\pm \in \mathbb{R}$. Explicitly, this conformal diffeomorphism is given by $\mathcal{C}(x^-,x^+) = (C_-(x^-),C_+(x^+))$ where \cite{Besken:2020snx}
\begin{equation}
    C_{\pm}(x^\pm) = \tan{\frac{x^\pm}{2}}\,,
    \label{eq:Cpm_tan}
\end{equation}
which is the Lorentzian version of the map \eqref{eq:tan_map_euclidean}. Together the boundary conditions \eqref{eq:3D_BC} and \eqref{eq:Cpm_tan} imply that
\begin{align}
    D^z(z,x^-,x^+) &= z e^{-2\sigma(x^-,x^+)}\,C_-'(x^-)\,C_+'(x^+) + \mathcal{O}(z^2)\,,\nonumber\\
    D^\pm(z,x^-,x^+) &= C_{\pm}(x^{\pm}) + \mathcal{O}(z^2)\,.
    \label{eq:ISTY_asymptotics}
\end{align}
In general, a diffeomorphism with asymptotics \eqref{eq:ISTY_asymptotics} takes the metric out of the Fefferman--Graham (FG) form at subleading orders in $z\rightarrow 0$ by generating a non-zero off-diagonal component $\widetilde{G}_{z \pm}\neq 0$ proportional to derivatives $\partial_{\pm}\sigma$ and $\partial_{\pm}C_{\pm}$. However, it is well known that due to the metric being asymptotically locally AdS, the metric may be brought back to the FG form with an additional diffeomorphism which does not spoil the boundary condition \eqref{eq:3D_BC} \cite{Fefferman:2007rka}. The result is a special diffeomorphism $P$ which we call the Imbimbo--Schwimmer--Theisen--Yankielowicz (ISTY) diffeomorphism \cite{Imbimbo:1999bj}: it preserves the FG form $\widetilde{G}_{z \pm} = 0$ and generates an arbitrary Weyl factor $\sigma$ on the conformal boundary.\footnote{The authors of \cite{Imbimbo:1999bj} called \eqref{eq:PBH_diffeomorphism} (in the special case of $C_{\pm}(x^\pm) = x^\pm$) a Penrose--Brown--Henneaux (PBH) diffeomorphism after \cite{Brown:1986nw,Penrose:1985bww} which has since become standard nomenclature in the literature. However, this naming is slightly misleading, because the diffeomorphism considered in \cite{Brown:1986nw} actually preserves the form of the (flat) boundary metric, in other words, it performs a conformal isometry instead of a Weyl transformation. We prefer to differentiate these two types of diffeomorphisms.} The general case involving both non-trivial $\sigma$ and $C_{\pm}$ in $d+1 = 3$ has been written down only recently \cite{Li:2025rzl} and it is explicitly
\begin{align}
	P^{z}(z,x^-,x^+) &= \frac{z\,e^{-\sigma}\,(\partial_-C_-\, \partial_+C_+)^{3\slash 2}}{\partial_-C_-\, \partial_+C_+ + \frac{1}{4} z^{2}\,e^{-2\sigma}\,(\mathcal{D}_-\partial_-C_-)(\mathcal{D}_+\partial_+C_+)}\,,\nonumber\\
	P^{\pm}(z,x^-,x^+)&=C_\pm(x^\pm) + \frac{\frac{1}{2}\,z^{2}\,e^{-2\sigma}\,(\partial_\pm C_\pm)^2\,(\mathcal{D}_\mp\partial_\mp C_\mp)}{\partial_-C_-\,\partial_+C_++\frac{1}{4}\,z^{2}\,e^{-2\sigma}\,(\mathcal{D}_-\partial_-C_-)(\mathcal{D}_+\partial_+C_+)}\,,
    \label{eq:PBH_diffeomorphism}
\end{align}
where we have introduced the ``covariant'' derivative $\mathcal{D}_{\pm} \equiv \partial_{\pm} -2\, (\partial_{\pm}\sigma)$ and $\partial_{\pm} = \frac{\partial}{\partial x^\pm}$. For $\sigma = 0$, this reduces to the diffeomorphism in \cite{Roberts:2012aq} which implements a conformal isometry preserving the boundary metric (conformal diffeomorphism with $C_{\pm}$ and a compensating Weyl transformation). Setting $C_{\pm}(x^\pm) = x^\pm$ and expanding in powers of $z\rightarrow 0$, \eqref{eq:PBH_diffeomorphism} reduces to the diffemorphism written down in \cite{Skenderis:2000in}.

Using \eqref{eq:PBH_diffeomorphism}, we find that the bulk metric \eqref{eq:Gtilde} with $D = P$ takes the form
\begin{equation}
	\widetilde{G}_{\mu\nu}(X)\,dX^{\mu}dX^{\nu} = 	\ell^{2}\,\biggl[\frac{1}{z^{2}}\,(dz^{2} + e^{2\sigma}\,dx^-dx^+ ) + g_{(2)ab}\,dx^adx^b + z^2g_{(4)ab}\,dx^adx^b\biggr]\,,
    \label{eq:Gtilde_solution}
\end{equation}
where explicitly
\begin{gather}
	g_{(2)\pm\pm} = \partial_\pm^{2}\sigma -(\partial_\pm\sigma)^{2}-\frac{1}{2}\,\{C_\pm(x^\pm),x^\pm\}\,,\quad g_{(2)-+} = \partial_-\partial_{+}\sigma\,, \label{eq:g2_sol}\\
    g_{(4)\pm\pm} = e^{-2\sigma}\,g_{(2)-+}\,g_{(2)\pm\pm}\,,\quad g_{(4)-+} = \frac{1}{2}\,e^{-2\sigma}\,\bigl(g_{(2)--}\,g_{(2)++}+g_{(2)-+}^2\bigr)\,.
    \label{eq:g2_g4_sol}
\end{gather}
Clearly the boundary condition \eqref{eq:3D_BC} is satisfied so that \eqref{eq:Gtilde_solution} is the correct solution dual to a CFT driven by the time-dependent background metric \eqref{eq:curvedg} on the Lorentzian cylinder. This extends the solution presented in \cite{deBoer:2023lrd} beyond the slow-driving regime.

The metric \eqref{eq:Gtilde_solution} is in FG form as advertised, and notably has a truncated FG expansion with terms only up to order $\mathcal{O}(z^2)$ appearing. This is non-trivial since the diffeomorphism \eqref{eq:PBH_diffeomorphism} itself has an infinite expansion. The reason that the expansion truncates is that we started from the Poincaré metric \eqref{eq:poincare_metric}: an ISTY diffeomorphisn does not generate terms $g_{(n\geq 6)}$ if they vanish in the initial metric $G$ \cite{Skenderis:2000in}. The FG form preserving ISTY diffeomorphism $P$ is enough for our purposes to find a bulk solution satisfying the boundary condition \eqref{eq:3D_BC}. This is without any loss of generality, and in the next section, this allows for a simple identification of the one-point function of the dual stress tensor from the subleading terms in the usual way. One could consider more general diffeomorphisms in which $\widetilde{G}_{\mu\nu}$ is not in the FG form, but this complicates the identification of the stress energy tensor one-point function from the metric; see \cite{Ciambelli:2019bzz,Ciambelli:2023ott,Arenas-Henriquez:2024ypo}.\footnote{We thank Dominik Neuenfeld for helping to clarify this point.}

\subsection{Holographic observables}
\label{subsec:hologrpahy_observables}

In Section \ref{subsec:corr_fn}, we computed in field theory the one-point function of the stress tensor, the two-point functions of two scalar operators of scaling dimension $\Delta$ and in Section \ref{subsec:EE} the entanglement entropy of an interval in the curved background. We will now reproduce all these results using the holographic dual geometry derived in the previous section.

\paragraph{Stress tensor.} We  consider 3D Einstein gravity with Lorentzian action
\begin{equation}
    I_{\text{EH}}[G] = \frac{1}{2\kappa}\int_{z>\epsilon} d^3X\sqrt{-G}\,\biggl(\mathcal{R} + \frac{2}{\ell^2}\biggr) + \frac{1}{\kappa}\int_{z = \epsilon}d^{2}x\sqrt{-\gamma}\,\biggl(K-\frac{1}{\ell}\biggr)\,,
    \label{eq:Einstein_action}
\end{equation}
where $\mathcal{R}$ is the Ricci scalar of $G$, $\gamma_{ab}$ is the induced metric of the $z = \epsilon$ boundary in the metric $G_{\mu\nu}$, $K_{ab} = \gamma^c_a\gamma^d_b\,\nabla_{c}n_d$ is the extrinsic curvature of the boundary with $K = \gamma^{ab}K_{ab}$ is its trace, $n_a$ is the outward-pointing unit normal vector of the boundary and $\kappa = 8\pi G_{\text{N}}$ in terms of the Newton's constant.

In the near-boundary limit, the Brown--York (BY) boundary stress tensor \cite{Brown:1992br} of the solution $\widetilde{G}$ is given by
\begin{equation}
    T^{\text{BY}}_{ab} \equiv \lim_{\epsilon\rightarrow 0}\frac{-2}{\sqrt{-\gamma}}\frac{\delta I_{\text{EH}}[\widetilde{G}]}{\delta \gamma^{ab}} = -\frac{1}{\kappa}\,\biggl(K_{ab}-K\gamma_{ab}+\frac{1}{\ell}\,\gamma_{ab} \biggr)\, .
     \label{eq:T_BY_definition}
\end{equation}
By substituting the metric \eqref{eq:Gtilde_solution}, we obtain (see also \cite{Henningson:1998ey,deHaro:2000vlm})
\begin{equation}
    T^{\text{BY}}_{ab} =  \frac{\ell}{\kappa}\,\bigl(g_{(2)ab}-g_{ab}\,g^{cd}\,g_{(2)cd}\bigr)\,.
    \label{eq:Einstein_variation}
\end{equation}
Substituting to \eqref{eq:Einstein_variation} and using the Brown--Henneaux formula $\ell\slash \kappa = c\slash (12 \pi)$ \cite{Brown:1986nw}, we obtain
\begin{equation}
     T^{\text{BY}}_{\pm\pm} = C_{\pm\pm}^{\text{D}} -\frac{c}{24\pi}\,\{C_\pm(x^\pm),x^\pm\}\,,\quad T^{\text{BY}}_{-+}= -C_{-+}^{\text{D}}-\frac{c}{12\pi}\,g_{-+}\nabla^2\sigma\,,
     \label{eq:T_BY_mid_step}
\end{equation}
where we have used that \eqref{eq:g2_sol} can be written in terms of the Liouville stress tensor \eqref{eq:CD_stress_tensor} of the Weyl factor $\sigma$ as
\begin{equation}
    g_{(2)\pm\pm} = \frac{12\pi}{c}\,C_{\pm\pm}^{\text{D}} -\frac{1}{2}\,\{C_\pm(x^\pm),x^\pm\}\,,\quad g_{(2)-+} =\frac{12\pi}{c}\,C_{-+}^{\text{D}}+g_{-+}\nabla^2\sigma\,.  
\end{equation}
Therefore, we can identify
\begin{equation}
    T^{\text{BY}}_{\pm\pm}(x^-,x^+) = \bra{0} T_{\pm\pm}(x^\pm)\ket{0}+C_{\pm\pm}^{\text{D}}(x^-,x^+)\,,
\end{equation}
where $T_{\pm\pm}(\phi)$ are the stress tensor operators defined in \eqref{eq:quantizedTs} and we have used \eqref{eq:L_n_vacuum} so that $ \{C_\pm(x^\pm),x^\pm\} = 1\slash 2 =  -\frac{24\pi}{c} \bra{0} T_{\pm\pm}(x^\pm)\ket{0}$ for the conformal diffeomorphism \eqref{eq:Cpm_tan}. Thus the BY stress tensor \eqref{eq:T_BY_definition} of the solution \eqref{eq:Gtilde_solution} computes the vacuum expectation value of the renormalized stress tensor operator \eqref{eq:TDW} in the diffeomorphism invariant scheme\footnote{In minimal subtraction with only the area counterterm added to the Einstein action.}
\begin{equation}
    T^{\text{BY}}_{\pm\pm}(x^-,x^+) = \bra{0}T_{\pm\pm}^{\text{D}}(x^-,x^+)\ket{0}.
    \label{eq:BY_stress_tensor}
\end{equation}
This same result was derived in \cite{deBoer:2023lrd} under the assumption of a slow-driving limit. Our calculation shows that the result holds also for arbitrary driving beyond the slow-driving regime.

The fact that the BY stress tensor \eqref{eq:T_BY_definition} computes the expectation value in the diffeomorphism invariant scheme is expected, because the Einstein action reproduces the $\text{Diff}\ltimes \text{Weyl}$ anomaly of the dual CFT as a Weyl anomaly. This follows since Einstein gravity in AdS clearly preserves diffeomorphism invariance in the boundary directions, but violates diffeomorphisms acting in the holographic radial direction due to the presence of the boundary, and these are the diffeomorphisms responsible for the Weyl transformations of the dual CFT. As discussed in \cite{deBoer:2023lrd}, one can move to the chirally split scheme by including an additional boundary term to the Einstein action \eqref{eq:Einstein_action} whose limit as $\epsilon\rightarrow 0$ gives the counterterm in equation \eqref{eq:chirally_split_scheme}. Such a boundary term does not modify the Dirichlet problem for the metric, because it depends only on the components of the boundary metric.\footnote{This problem is related to the holomorphic factorization of 3D gravity in Euclidean signature studied in \cite{Krasnov:2001cu}.}

\paragraph{Scalar fields.}

The two-point function of scalar fields of scaling dimension $\Delta $ is simplest to compute holographically in the limit $\Delta\rightarrow \infty $ where we can make use of the geodesic approximation. Precisely, the two-point function \eqref{eq:scalar_2_point} in this limit is \cite{Balasubramanian:1999zv}
\begin{equation}
    \langle \mathcal{O}(\phi_1,t_1)\,\mathcal{O}(\phi_2,t_2)\rangle = e^{-\Delta \widetilde{L}_{\text{reg}}(\phi_1,t_1;\phi_2,t_2)}\,,\quad \Delta\rightarrow \infty\,,
    \label{eq:geodesic_approximation}
\end{equation}
where $\widetilde{L}_{\text{reg}}(\phi_1,t_1;\phi_2,t_2)$ is the regularized length of the minimal\footnote{There can be multiple geodesics connect the boundary points and at leading order in $\Delta\rightarrow \infty$ only the one with the smallest regularized length contributes.} bulk geodesic anchored at $(\phi_i,t_i)$ on the conformal boundary. The length is computed in the bulk geometry \eqref{eq:Gtilde_solution} dual to the instantaneous state of the driven CFT and regulated by integrating the length only up to a small value of $z $ near the conformal boundary. We leverage the knowledge of how the driven metric \eqref{eq:Gtilde_solution} arises from the Poincaré metric via the ISTY diffeomorphism $P$ given in \eqref{eq:PBH_diffeomorphism} to compute observables. Concretely, we first compute the observables of interest in Poincaré coordinates and then map using $P$ to the driven space-time to obtain the final result.

In the Poincaré AdS metric $G$ \eqref{eq:poincare_metric}, the geodesic length between any two bulk points $X_i = (z_i,x^-_i,x^+_i)$ is given by 
\begin{equation}
L(X_1,X_2) = \log{\left(\frac{1 + \sqrt{1-\xi_{12}^{2}}}{\xi_{12}} \right)}, \quad \xi_{12} = \frac{2z_1z_2}{z_1^{2}+z_2^{2}+(x_1^- - x_2^-)(x_1^+ - x_2^+)}\,.
\label{eq:geodesic_length}
\end{equation}
Because pure AdS has trivial topology, this is the length of the unique geodesic connecting the two points.

Now the unique minimal geodesic between two points $X_1$ and $X_2$ in the metric $\widetilde{G}$ is the preimage of the geodesic in the metric $G$ under the diffeomorphism \eqref{eq:PBH_diffeomorphism}. Since geodesic length transforms as a biscalar under diffeomorphisms, the length of this geodesic in the metric $\widetilde{G}$ is simply
\begin{equation}
    \widetilde{L}(X_1;X_2)= L(P(X_1);P(X_2))\,,
    \label{eq:Ltilde}
\end{equation}
where $P$ is the ISTY diffeomorphism \eqref{eq:PBH_diffeomorphism}. In terms of the transformed chordal distance variable
\begin{equation}
    \widetilde{\xi}_{12} \equiv  \frac{2P^z(X_1)\,P^z(X_2)}{P^z(X_1)^{2}+P^z(X_2)^{2}+(P^-(X_1)-P^-(X_2))(P^+(X_1)-P^+(X_2))}\,,
    \label{eq:transformed_chordal}
\end{equation}
the length \eqref{eq:Ltilde} takes the same form as \eqref{eq:geodesic_length}.

To connect to the geodesic approximation of the scalar two-point function in the driven boundary CFT, let us first consider a specific geodesic in FG coordinates $(z,x^-,x^+)$, namely the one anchored at the conformal boundary at $(\phi_i,t_i)$. This requires taking $z_i\rightarrow 0$ so we consider the geodesic between the bulk points $X_i = (\varepsilon_i,x^-_i,x^+_i)$ where $\varepsilon_i \rightarrow 0^+$ are regulators and $x_i^- = F_{t_i}(\phi_i)$ and $x_i^+ = F_{t_i}(\phi_i)$. The length of such a geodesic defines the regularized geodesic length, which in the case of the metric $\widetilde{G}$ is explicitly
\begin{equation}
    \widetilde{L}_{\text{reg}}(\phi_1,t_1;\phi_2,t_2)\equiv \widetilde{L}(\varepsilon_1,F_{t_1}(\phi_1),\overbar{F}_{t_1}(\phi_1);\varepsilon_2,F_{t_2}(\phi_2),\overbar{F}_{t_2}(\phi_2))\,.
    \label{eq:L_tilde_reg}
\end{equation}
The important point here is that the cut-offs $z_i = \varepsilon_i$ are the same irrespectively of the metric used to compute the length. Using the formula \eqref{eq:Ltilde}, we obtain in the $\varepsilon_i\rightarrow 0^+$ limit
\begin{equation}
     \widetilde{L}_{\text{reg}} = \frac{1}{2}\log\left[\frac{16}{\varepsilon_1^2\varepsilon_2^2}\,e^{2\sigma (\phi_1,t_1)+2\sigma (\phi_2,t_2)} \sin^2\left(\frac{x^-_1-x^-_2}{2}\right)\sin^2\left(\frac{x^+_1-x^+_2}{2}\right)\right] + \ldots\,,
    \label{eq:LinDrivenSpacetimeNonReg}
\end{equation}
where ellipsis denote terms that vanish in the limit and $\sigma = \omega + \varphi$ with $\varphi = -\frac{1}{2}\log{F_t'(\phi)}-\frac{1}{2}\log{\overbar{F}_t'(\phi)}$. Therefore we obtain
\begin{equation}
     e^{-\Delta \widetilde{L}_{\text{reg}}}= \Biggl[\frac{\varepsilon_1^2\varepsilon_2^2\,e^{-2\,(\omega(\phi_1,t_1)+\omega(\phi_2,t_2))}\,F_{t_1}'(\phi_1)\,\overbar{F}_{t_1}'(\phi_1)\,F_{t_2}'(\phi_2)\,\overbar{F}_{t_2}'(\phi_2)}{16\sin^2{\bigl(\frac{F_{t_1}(\phi_1)-F_{t_2}(\phi_2)}{2}\bigr)}\sin^2{\bigl(\frac{\overbar{F}_{t_1}(\phi_1)-\overbar{F}_{t_2}(\phi_2)}{2}\bigr)}}\Biggr]^{\Delta\slash 2}\,,
\end{equation}
and the geodesic approximation \eqref{eq:geodesic_approximation} agrees with the field theory result \eqref{eq:scalar_2_point} upon fixing the normalization constant $b_{\mathcal{O}} = (-1)^{\Delta\slash 2}\,(2\pi)^4\,\varepsilon_1^{\Delta}\varepsilon_2^{\Delta}$. This amounts to a cut-off dependent normalization of the operators which can be interpreted as wave function renormalization. Equivalently, one could also consider renormalized length with the divergence subtracted in the geodesic approximation, which leads to an order one value for $b_{\mathcal{O}}$.

\paragraph{Entanglement entropy from the RT formula.} For the driven CFT, let us now holographically compute the entanglement entropy \eqref{eq:EE} of the boundary interval $(\phi_1,\phi_2)\in S^1$ on a constant time slice at time $t$. The Ryu-Takayanagi (RT) prescription \cite{Ryu:2006bv,Ryu:2006ef} instructs us to compute the area of the minimal area codimension-two extremal surface $\mathcal{S}\subset \mathcal{M}$ anchored at the endpoints $(\phi_1,t)$ and $(\phi_2,t)$ to find
\begin{align}
    S_{\text{RT}}(t)=\frac{\text{Area}\,(\mathcal{S})}{4 G_N}\,.
    \label{eq:RT}
\end{align}
Because the bulk space-time is three-dimensional in our setup, the surface $\mathcal{S}$ is a geodesic in the driven space-time geometry \eqref{eq:Gtilde_solution} and its area is given by the regularized geodesic length $\text{Area}\,(\mathcal{S}) = \widetilde{L}_{\text{reg}}(X_1;X_2)$ between the bulk points $X_1=(\varepsilon_1,F_t(\phi_1),\overbar{F}_t(\phi_1))$ and $X_2=(\varepsilon_2,F_t(\phi_2),\overbar{F}_t(\phi_2))$ as defined in \eqref{eq:L_tilde_reg}. The calculation of $\widetilde{L}_{\text{reg}}$ has already been carried out in the previous paragraph. By substituting \eqref{eq:LinDrivenSpacetimeNonReg} into \eqref{eq:RT}, using again the Brown--Henneaux formula $\ell\slash (8\pi G_{\text{N}}) = c\slash (12 \pi)$ \cite{Brown:1986nw} and $\sigma = \omega + \varphi$ with $\varphi = -\frac{1}{2}\log{F_t'(\phi)}-\frac{1}{2}\log{\overbar{F}_t'(\phi)}$, we find
\begin{align}
    S_{\text{RT}}(t) = \frac{c}{6}\,(\omega(\phi_1,t) + \omega(\phi_2,t))+\frac{c}{12}\Biggl[\log{\Biggl(\frac{4\sin^2{\bigl(\frac{F_t(\phi_1)-F_t(\phi_2)}{2}\bigr)}}{\varepsilon_1\varepsilon_2\, F_{t}'(\phi_1)\,F_{t}'(\phi_2)}\Biggr)}+(F_t\leftrightarrow \overbar{F}_t)\Biggr]\,.
    \label{eq:EE_holo}
\end{align}
This matches with the result \eqref{eq:EE_1} of the field theory calculation in the diffeomorphism invariant scheme $S_{\text{RT}}(t) = S_{\text{D}}(t)$ upon setting the additive constant $b'(1)$, not fixed by the field theory calculation, to zero.

Note, however, that the interpretation of the cut-offs $\varepsilon_i$ appearing in the entropy are a priori different: in the field theory result, they are radii of geodesic disks in the flat Euclidean metric that are cut-out to regularize the entropy, while here, they are cut-offs in the holographic radial coordinate $z_i = \varepsilon_i$. The two are related, because a boundary disk of radius $\varepsilon$ can be extended into the bulk as a hemisphere $z^2 + \vert w\vert^2 = \varepsilon^2 $ which is minimal area surface in the Euclidean continuation $x^- = w$, $x^+ = \overbar{w}$ of the Poincaré AdS metric \eqref{eq:poincare_metric}. The tip of the hemisphere is located at $z = \varepsilon$ so that cutting of the geodesic at $z_i = \varepsilon_i$ amounts to cutting of two hemispheres of radii $\varepsilon_i$ in the bulk.

The fact that the RT formula \eqref{eq:RT} reproduces the entropy in the diffeomorphism invariant scheme is expected, because it is derived from the Einstein action using the Lewkowycz--Maldacena (LM) method \cite{Lewkowycz:2013nqa} and the Einstein action reproduces the $\text{Diff}\ltimes \text{Weyl}$ anomaly in the diffeomorphism invariant scheme as discussed above. Including the boundary term, responsible for shifting the scheme to the chirally split scheme, to the Einstein action and redoing the LM derivation produces an additional boundary term to the RT formula. This term subtracts the $\omega$-dependent pieces in \eqref{eq:EE_holo} reproducing the chirally split field theory result \eqref{eq:chirally_split_entropy}.

To conclude, we want to draw attention to this scheme dependence: when observables are calculated in the operator formalism of driven inhomogeneous CFTs with deformed Virasoro Hamiltonians—as shown, the chirally split renormalization scheme underlies this approach—then one should not expect that gravity calculations starting from the Einstein action reproduce these results, since standard gravity calculations are tied to the diffeomorphism invariant scheme. In order to find agreement, one needs to include an additional boundary term in the Einstein action \eqref{eq:Einstein_action} for the gravity calculations. A core finding of this article is the proper identification of which renormalization scheme implicitly underlies which of the standard calculations performed in the literature.

\section{Conclusion and future directions}
\label{sec:conclusion}

In this work, we have revisited and extended the relation between driven inhomogeneous CFTs and CFTs in evolving curved space-times. The purpose was to give a precise Lorentzian dictionary between diffeomorphism invariant curved space-time physics of two-dimensional CFTs and the operator (Hamiltonian) formulation of the theory. 

We treated various renormalization schemes carefully. As a key lesson, the chirally split scheme, which preserves Weyl symmetry but breaks diffeomorphism invariance, emerged as the one that enables to connect to the standard operator point of view, namely a driven inhomogeneous CFT with a deformed Virasoro Hamiltonian. In this scheme, the c-number terms in the Heisenberg picture Hamiltonian precisely agree with Schwarzian terms coming from the adjoint action of a unitary representation of the Virasoro group on the stress tensor.

We furthermore carefully analyzed scheme dependent quantities such as expectation values of energy and entanglement entropy. The curved space entanglement entropy has a consistent interpretation as a property of the evolving quantum state in the operator formulation of the CFT only when calculated in the chirally split scheme. We also revisited the holographic dual description of the driven CFT, extending previous bulk geometries to arbitrary time-dependent driving. The holographic calculation starting from the Einstein action agrees with field theory observables calculated in the diffeomorphism invariant scheme. We drew attention to the fact that to obtain agreement with field theory observables in the chirally split renormalization scheme, an additional boundary term has to be included in the Einstein action. 

Our results provide a handbook on how to correctly treat Lorentzian two-dimensional CFTs on evolving curved space-times, from both the operator and holographic perspectives, and establish a model with a wide range of applicability, especially for questions about the time-dependence of observables. Let us conclude by outlining possible extensions and future directions of our work.

\paragraph{Driving in the presence of boundaries.} A natural extension of our work is to introduce one-dimensional timelike boundaries, with (conformal) boundary conditions imposed, into the driven CFT. The case of static, nonmoving boundaries amounts to considering a two-dimensional CFT on the Riemannian manifold $(\mathcal{I}\times \mathbb{R}, g)$, where $\mathcal{I}\subset S^1$ is a connected interval of the circle. A more involved setting is to let the positions of the two boundaries be functions of time $t$.

The holographic dual of such driven boundary CFT in three-dimensional gravity may be obtained by the inclusion of constant tension end-of-the-world branes along the lines of \cite{Takayanagi:2011zk}. The embeddings of the branes are required to satisfy the boundary Einstein equation in the dual geometry \eqref{eq:Gtilde_solution} of the driven CFT derived in Section \ref{sec:holography}. Such embeddings are easily obtained using the diffeomorphism \eqref{eq:PBH_diffeomorphism} from known solutions \cite{Takayanagi:2011zk} in the Poincaré AdS geometry \eqref{eq:poincare_metric}.

Related setups have been studied previously in the literature: see \cite{BarberoG:2016nkv} for static boundaries and, in the context of the dynamical Casimir effect, \cite{Moore:1970tmc} for moving boundaries. In the setup of \cite{BarberoG:2016nkv}, a free theory is quantized along non-trivial foliations of $\mathcal{I}\times \mathbb{R}$ in a flat background metric. This is related to the curved space-time setting above, since, similarly to the case without boundaries, non-trivial foliations are equivalent to coupling the CFT to a non-trivial background metric. The main feature of these setups is the absence of a unitary operator implementing the evolution \cite{BarberoG:2016nkv,Moore:1970tmc}. Therefore, the problem arising in higher-dimensional theories on curved backgrounds reappears already in two dimensions in the presence of boundaries. We plan to consider driven boundary CFTs and the revisit the existence of a unitary operator in future work.

\paragraph{Driving with additional background fields.} The CFT considered in this work is driven only by a time-dependent background metric. It would be interesting to include driving due to other evolving classical background fields, such as a gauge field coupled to a current or a source coupled to a scalar field. The former case has been studied \cite{Isler:1987ax}. From the operator point of view, the latter is related to the primary-deformed Virasoro circuit studied in \cite{Erdmenger:2024xmj}.

In the case of driving with a source for a scalar field, the scaling dimension $\Delta$ of the field affects the physics. When the field is classically relevant, $\Delta<2$, the driving triggers a non-trivial renormalization group (RG) flow, leading to the running of the space-time-dependent source. The renormalization of a space-time-dependent coupling can be dealt with using methods of the local renormalization group developed in \cite{Osborn:1991gm}. As a result, the driving becomes a function of the energy scale of the theory. Space-time-dependent sources also lead to additional Weyl anomalies in the stress tensor \cite{Osborn:1991gm,Jack:2013sha}, which could be matched with the Heisenberg evolution.

From the holographic perspective, sources for additional CFT operators correspond to non-normalizable modes for matter fields in the bulk. Driving with sources triggers an evolution of these fields whose backreaction can produce black holes along the lines of \cite{Anous:2016kss}. It would be interesting to relate the real-time Lorentzian evolution studied here with the Euclidean methods used to study black hole formation in \cite{Anous:2016kss}.\footnote{We thank Julian Sonner for a discussion on this point.}

\paragraph{Complexity.} It is natural to ask whether our model is useful for investigating field theory quantities beyond the ones studied in this article. A field theory measure of recent interest in the literature is complexity whose holographic dual description is still under investigation (see \cite{Chapman:2021jbh, Nandy:2024evd,Baiguera:2025dkc,Rabinovici:2025otw} for reviews). In the context of driven inhomogeneous CFTs, or equivalently, Virasoro quantum circuits, a lot of focus has been put on quantum circuit complexity \cite{Caputa:2018kdj,Flory:2020eot,Flory:2020dja,Erdmenger:2020sup,Erdmenger:2021wzc,Erdmenger:2022lov}. Another class of complexity measures of interest is the Krylov spread complexity \cite{Balasubramanian:2022tpr}, for which a precise bulk manifestation as extremal volume has recently been uncovered in quantum mechanical models in \cite{Rabinovici:2023yex,Heller:2024ldz} and, for the complexity rate in two-dimensional CFTs, as proper momentum of an infalling particle in the dual spacetime \cite{Caputa:2024sux,He:2024pox,Fan:2024iop}. It would be interesting to compare holographic proposals with Krylov spread complexity (and its rate) in cases beyond Möbius or $SL_k(2,\mathbb{R})$ driving of inhomogeneous CFTs. The holographic bulk geometry constructed in this paper could be used to test such proposals.

\paragraph{Schrödinger evolution in black hole space-times.} An interesting application of our results would be to study unitary CFT time-evolution in two-dimensional black-hole backgrounds, including both static cases and time-evolving backgrounds with horizon formation. It is of interest to keep track of the intermediate evolution of the Schrödinger picture state along different slices of the collapsing black-hole space-time \cite{Giddings:2020dpb,Giddings:2021ipt,Giddings:2022sss}. Our results could allow a more detailed analysis of the evolution of stress-energy and entanglement entropy in these situations.

\paragraph{Random background metrics.} The background metric in our situation is classical and non-dynamical. There are essentially two avenues to go beyond a fixed choice of a metric: first is to consider an ensemble average over metrics while the second is to make it dynamical and quantize. Effects of ensemble averaging with Gaussian noise were studied in \cite{Bernard:2019mqm,Christopoulos:2021imm}. A dynamical quantized metric can be described by two-dimensional quantum gravity such as JT gravity or by first-order theories of the type discussed in \cite{Katanaev:1986wk} (see \cite{Strobl:1999wv} for a review).\footnote{We thank Dmitri Vassilevich for discussion on the quantization of these models.} The combined unitary evolution of the metric and the CFT could lead to interesting non-trivial effects, while a fixed background metric discussed here would arise from a coherent state for the gravity sector. The evolution of the backreacted system is interesting, because gravity is in general sensitive to the renormalization scheme.

\paragraph{Non-unitary driving.} The evolution is unitary due to the components of the background metric being real valued functions of the coordinates. Non-unitary evolution arises immediately if the metric picks up an imaginary part and becomes complex. In the context of driven inhomogeneous CFTs, non-unitary evolution has been considered in \cite{Wen:2024bzm,Lapierre:2025zsg} with emphasis on complex metrics in \cite{Wen:2024bzm}. Our framework allows for a straightforward generalization of the simple complex metric considered in \cite{Wen:2024bzm}.

\paragraph{Entanglement entropy of a Virasoro state.} In Section \ref{subsec:EE}, we derived the entanglement entropy of an interval in the instantaneous state generated by the driving. As a special case, we obtained a formula \eqref{eq:chirally_split_entropy_static} for the entropy of a static Virasoro state. The derivation relies on the properties of the Euclidean path integral in the diffeomorphism invariant scheme and on the inclusion of finite counterterms to move to the chirally split scheme. We argued that the result is compatible with a state interpretation, but we did not prove it. Therefore, we would like to have a direct derivation of \eqref{eq:chirally_split_entropy_static} in the operator formulation, which does not rely on the Euclidean path integral. This would clarify the role played by the cut-off from the operator perspective and its absence in the relative entropy between Virasoro states \cite{Hollands:2019czd,Panebianco:2019plp}. We hope to report progress on this problem in the near future.

\acknowledgments

We thank Maria Stella Adamo, Jan de Boer, Horacio Casini, Markus B. Fröb, Michal P. Heller, Jonathan Karl, Arnab Kundu, Ren\'e Meyer, Dominik Neuenfeld, Leo Shaposhnik, Julian Sonner and Dmitri Vassilevich for useful discussions. J.~E.~and J.~K. are supported by the Deutsche Forschungsgemeinschaft (DFG, German Research Foundation) through the German-Israeli Project Cooperation (DIP) grant ‘Holography and the Swampland’, as well as under Germany’s Excellence Strategy through the W\"{u}rzburg-Dresden Cluster of Excellence on Complexity and Topology in Quantum Matter - ct.qmat (EXC 2147, project-id 390858490). The project was also supported by  DFG individual grant ER 301/8-1. J.~E.~and J.~K. would like to thank the IIP in Natal and the organizers of the workshop ``Quantum Gravity, Holography and Quantum Information'' at IIP Natal for hospitality during the final stages of this project. T.~S. is supported by the Research Foundation - Flanders (FWO) doctoral fellowship 11I5425N.

\begin{appendix}

\section{Classical Ward identities and Hamiltonian}\label{app:ward_identities}

In this appendix, we derive the classical $\text{Diff}\ltimes \text{Weyl}$ Ward identities for a $d$-dimensional CFT on a (semi-)Riemannian manifold $(M,g)$ with a curved Lorentzian metric $g$. The derivation produces conditions for the classical stress tensor of the CFT both inside the manifold $ M$ and on its boundary $\partial M$. We use the boundary identities to derive a formula for the classical Hamiltonian of the theory in terms of the stress tensor.

\subsection{Diffeomorphisms and Weyl transformations}\label{app:ward_identities_derivation}

Let us consider an action of the type
\begin{equation}
	I[\Phi,g;M] = \int_M d^{d}x\sqrt{-g}\,\mathcal{L}\,,
    \label{eq:CFT_action}
\end{equation}
where $M$ is a manifold with a spacelike boundary $\partial M$. We assume that the variation of the action with respect to $\Phi$ and $g$ is given by
\begin{equation}
	\delta I[\Phi,g;M] = \int_{M} d^{d}x\,\sqrt{-g}\,\biggl(-\frac{1}{2}\,T_{ab}^{\text{cl}}\,\delta g^{ab}+E_{\alpha}\,\delta \Phi^{\alpha}\biggr)+\int_{\partial M} d^{d-1}\hat{x}\sqrt{\gamma}\,n_a\,\Pi^{a}_{\alpha}\,\delta \Phi^{\alpha}\,,
	\label{eq:generalvariation}
\end{equation}
where $n^a = g^{ab}\,n_a$ is the inward-pointing, in this case past-pointing, unit normal vector of $\partial M$ that satisfies $n_an^a = -1$ and $\gamma_{ij}$ is the induced metric of $\partial M$.\footnote{The sign of the boundary term arising from the divergence theorem depends on whether the boundary is spacelike or timelike. The usual convention is that the boundary term comes with an overall plus sign, but the direction of the normal vector is chosen differently for space- and timelike boundaries. In this convention, which we follow here, the normal vector must be taken to be inward-pointing (outward-pointing) when the boundary is spacelike (timelike).\label{eq:boundary_sign_footnote}} We have also introduced an index $\alpha$ (such as a space-time or a spinor index) for the fundamental field. If $\mathcal{L}$ depends only on first derivatives of $ \Phi $, we have explicitly
\begin{equation}
	E_{\alpha} = \frac{\partial \mathcal{L}}{\partial \Phi^{\alpha}} - \nabla_a\biggl(\frac{\partial \mathcal{L}}{\partial(\nabla_a\Phi^{\alpha})}\biggr)\,,\quad \Pi^{a}_{\alpha} = \frac{\partial \mathcal{L}}{\partial(\nabla_a\Phi^{\alpha})}\,.
    \label{eq:E_Pi}
\end{equation}
Let the embedding of $\partial M$ be given by $x^{a} = B^{a}(\hat{x})$ where $\hat{x}^i$ with $i = 1,\ldots,d-1$ denote coordinates on $\partial M$. The variation of the action with respect to the embedding is given by \cite{flanders1973differentiation,Reddiger:2019lqh}
\begin{equation}
	\delta_BI[\Phi,g;M] = \int_{\partial M} d^{d-1}\hat{x}\sqrt{\gamma}\,\mathcal{L}\,n_a \delta B^{a}\,,
    \label{eq:embedding_var}
\end{equation}
where the overall sign is a plus since $n^a$ is past-pointing (see footnote \ref{eq:boundary_sign_footnote}). Clearly the action changes only if the variation of the embedding is transverse to the boundary $n_a\delta B^a \neq 0$.

\paragraph{Diffeomorphisms.} We will assume that under a diffeomorphism $D$, the action satisfies
\begin{equation}
	I[D^{*}\Phi,D^{*}g;D^{-1}(M)] = I[\Phi,g;M]\,,
	\label{eq:diffweylinvariance}
\end{equation}
where $D^{*}$ denotes pull-back with $D$ under which 
\begin{equation}
		(D^*g)_{ab}(x) = \frac{\partial D^{c}}{\partial x^{a}}\frac{\partial D^{d}}{\partial x^{b}}\,g_{cd}(D(x))\,.
\end{equation}
The explicit formula for $(D^*\Phi)^\alpha $ depends on the type of the index $\alpha$. Equation \eqref{eq:diffweylinvariance} is a generalization of diffeomorphism invariance in the presence of boundaries: when the diffeomorphism does not move the boundary $D(M) = M$, the condition reduces to $I[D^{*}\Phi,D^{*}g;M] = I[\Phi,g;M]$.

At the infinitesimal level $D^{a}(x) = x^{a} + \xi^{a}(x)$, the pull-backs are given by
\begin{equation}
	\delta_D\Phi^{\alpha} = \pounds_\xi\Phi^{\alpha}\,,\quad \delta_Dg^{ab} = \pounds_\xi g^{ab} = -\nabla^{a}\xi^{b}-\nabla^{b}\xi^{a}\,,\quad \delta_D B^{a}(\hat{x}) = -\xi^{a}(B(\hat{x}))\,,
\end{equation}
where $\pounds_\xi$ denotes the Lie derivative in the direction of the vector field $\xi$. The change in the embedding of the boundary follows from the fact that the embedding of $ \partial(D^{-1}(M)) $ is given by $D^{-1}(B(\hat{x}))$ and the appearance of the inverse function explains the minus sign in $\delta_D B^{a}$. Expanding \eqref{eq:diffweylinvariance} to linear order in $\xi$ by using \eqref{eq:generalvariation} and \eqref{eq:embedding_var}, we obtain after integration by parts
\begin{align}
	0=\delta_DI[\Phi,g;M] &=\int_{M} d^{d}x\sqrt{-g}\,(-\xi^{a}\,\nabla^{b}T_{ab}^{\text{cl}} + E_{\alpha}\,\pounds_\xi \Phi^{\alpha})\nonumber\\
	&+\int_{\partial M}d^{d-1}\hat{x}\sqrt{\gamma}\,(\xi^an^{b}\,T_{ab}^{\text{cl}}+n_a\,\Pi^{a}_{\alpha}\,\pounds_\xi \Phi^{\alpha}-n_a\xi^{a}\mathcal{L}) \,.\label{eq:infinitesimaldiffeoinvariance}
\end{align}
The vanishing bulk term gives requires
\begin{equation}
		\nabla^{b}T_{ab}^{\text{cl}} = E_{\alpha}\pounds_a\Phi^{\alpha}\,.
\end{equation}
where $\pounds_a$ denotes the Lie derivative in the direction of the vector field that generates constant-$x^a$ coordinate lines. On the other hand, the vanishing of the boundary term gives the identity
\begin{align}
	(n_a\,\Pi^{a}_{\alpha}\,\pounds_\xi\Phi^{\alpha}-n_a\xi^a\mathcal{L})\lvert_{\partial M}\,= -\xi^{a}n^{b}\,T_{ab}^{\text{cl}}\lvert_{\partial M}\,.\label{eq:boundarydiffward}
\end{align}
Since the choice of $\partial M$ was arbitrary, this formula is valid on any spacelike slice. We will use it below in Appendix \ref{app:Hamiltonian_derivation}.

\paragraph{Weyl transformations.} We assume that the action is also Weyl invariant
\begin{equation}
    I[\widehat{\Phi},\widehat{g};M] = I_M[\Phi,g;M]\,,
    \label{eq:Weyl_invariance}
\end{equation}
where we have defined
\begin{equation}
    \widehat{g}_{ab}(x) = e^{2\chi(x)}\,g_{ab}(x)\,,\quad \widehat{\Phi}^\alpha(x) = e^{-\Delta_\Phi\chi(x)}\,\Phi^\alpha(x)\,.
\end{equation}
For an infinitesimal Weyl transformation $\chi(x) = \delta\chi(x)\ll 1$, we obtain
\begin{equation}
    \delta_\chi g^{ab} = -2\,\delta\chi\, g^{ab},\quad \delta_\chi \Phi^\alpha = -\Delta_{\Phi}\,\delta\chi\, \Phi^\alpha\,.
\label{eq:gJweyl}
\end{equation}
Using \eqref{eq:generalvariation}, Weyl invariance amounts to
\begin{equation}
0 = \delta_\chi I_M = \int_{M} d^{d}x\sqrt{-g}\,\delta \chi\,(g^{ab}\,T_{ab}^{\text{cl}}  -\Delta_{\Phi}\, E_\alpha\,\Phi^\alpha) - \int_{\partial M}d^{d-1}\hat{x}\sqrt{\gamma}\,\delta\chi\,\Delta_{\Phi}\,n_{a}\Pi^a_\alpha\,\Phi^\alpha\,.
\label{eq:deltaI}
\end{equation}
This is valid for all $\delta\chi$ so that vanishing of the bulk term implies
\begin{equation}
g^{ab}\,T_{ab}^{\text{cl}} = \Delta_{\Phi}\, E_\alpha\,\Phi^\alpha\,.
\label{eq:WeylWard}
\end{equation}
In addition, when $\Delta_\Phi \neq 0$, vanishing of the boundary term requires
\begin{equation}
    n_{a}\,\Pi^a_\alpha\,\Phi^\alpha\lvert_{\partial M}\, = 0\,.
    \label{eq:boundaryweyl}
\end{equation}
This is to be understood as an additional boundary condition on $\Phi$ to ensure Weyl invariance at the classical level. In theories with $\Delta_\Phi = 0$, such as in $d = 2$ for free scalar fields, \eqref{eq:boundaryweyl} is not necessary.

\subsection{Derivation of the classical Hamiltonian}\label{app:Hamiltonian_derivation}

Let us now consider the action \eqref{eq:CFT_action} on a manifold $M_s $ whose boundary $\partial M_s = m_s\cup m_0$ consists of two disjoint spacelike surfaces such that $m_0$ is the past and $m_s$ the future boundary respectively. In coordinates $x = (\phi,t)$,\footnote{Here we assume that $\phi$ parametrizes $d-1$ spatial directions instead of being just a single periodic coordinate on the one-dimensional circle as in the main text.} we assume that $m_s$ is located at $t = s$ which implies that its embedding is given by $B^t_s(\phi) = s$ where we have assumed that $\phi = B^{\phi}_s(\hat{x}) \equiv \hat{x}$. Let $\Phi_s(\phi,t)$ be a solution of the equations of motion $E(\Phi_s) = 0$ with boundary conditions at the two boundaries that fix the solution uniquely. Assuming for simplicity a scalar field, the boundary conditions are Dirichlet conditions
\begin{equation}
    \Phi_s(\phi,s)= \Psi(\phi)\,,\quad \Phi_s(\phi,0) = \Psi_0(\phi)\,.
    \label{eq:scalar_dirichlet}
\end{equation}
The dependence on $s$ in $\Phi_s$ arises through the first boundary condition. The generalization to non-scalar fields is discussed in the end.

The classical Hamiltonian is defined as the Hamilton--Jacobi variation of the action as
\begin{equation}
	H_{\text{cl}}(t) \equiv -\tfrac{d}{ds}I[\Phi_s,g;M_s]\Big\vert_{s = t}\,,
	\label{eq:HamiltonianHJ}
\end{equation}
where the metric $g$ is understood to be a fixed background field so that the total derivative $\tfrac{d}{ds}$ acts only on $M_s$ and $\Phi_s$. In particular, only the embedding of the future boundary $m_s$ is varied. Explicitly,
\begin{equation}
    \tfrac{d}{ds} B^a_s(\phi) = \delta^a_t\,,\quad \tfrac{d}{ds} \Phi_s(\phi,t) = \dot{\Phi}_s(\phi,t)\,,\quad \tfrac{d}{ds} g^{ab}(\phi,t) = 0\,,
    \label{eq:d_ds_action}
\end{equation}
where in $\dot{\Phi}_s$ the dot denotes derivative with respect to the subscript. Computing the Hamiltonian \eqref{eq:HamiltonianHJ} by applying \eqref{eq:generalvariation} and \eqref{eq:embedding_var} to the variations \eqref{eq:d_ds_action}, we obtain
\begin{equation}
	H_{\text{cl}}(t) = -\int_{m_t} d^{d-1}\phi\sqrt{\gamma}\,(n_a\,\Pi^{a}_{\alpha}\,\dot{\Phi}_t^\alpha+n_t\,\mathcal{L})+ \int_{m_0} d^{d-1}\phi\sqrt{\gamma}\,n_a\,\Pi^{a}_{\alpha}\,\dot{\Phi}_t^\alpha\,,
    \label{eq:Hamiltonian_midstep}
\end{equation}
where $n_t = n_a\,\delta^a_t$ and there are only boundary terms since $\Phi_s$ is on-shell. Notice that the variation of the scalar field vanishes on the past boundary $\dot{\Phi}_s(\phi,0) = 0$ by the Dirichlet boundary condition \eqref{eq:scalar_dirichlet}. Therefore the boundary term at $m_0$ vanishes and the Hamiltonian is completely supported on the future boundary $m_t$.

Now notice that the variation \eqref{eq:d_ds_action} can be interpreted as an infinitesimal diffeomorphism $D^a = x^a + \xi^a$ in the direction of the vector field $\xi^a = -\delta^a_t$. The action of this diffeomorphism on the embedding of $m_s$ clearly matches with \eqref{eq:d_ds_action}: $\delta_D B^a_s(\phi) = -\xi^a(\phi) = \tfrac{d}{ds}B^a_s(\phi)$. The action on a scalar field also matches if we use the Dirichlet boundary condition \eqref{eq:scalar_dirichlet} to yield
\begin{equation}
    0 = \tfrac{d}{ds} \Psi(\phi) = \tfrac{d}{ds} \Phi_s(\phi,s) = \dot{\Phi}_s(\phi,s) + \partial_s \Phi_s(\phi,s)\,,
\end{equation}
where $\partial_t$ means derivative with respect to the second argument inside the brackets in $\Phi_s(\phi,t)$. Thus we obtain for a scalar field
\begin{equation}
    \tfrac{d}{ds} \Phi_s(\phi,t)\vert_{s=t} = \dot{\Phi}_t(\phi,t) = -\partial_t \Phi_t(\phi,t) = \xi^a \partial_a\Phi_t(\phi,t) = \pounds_\xi \Phi_t(\phi,t) \,.
    \label{eq:using_dirichlet}
\end{equation}
In other words, the on-shell scalar solution changes upon varying the location of the future boundary $m_s$ as a Lie derivative in the direction of the vector field $\xi^a = -\delta^a_t$. Thus we can write the Hamiltonian \eqref{eq:Hamiltonian_midstep} as
\begin{equation}
    H_{\text{cl}}(t) = -\int_{m_t} d^{d-1}\phi\sqrt{\gamma}\,(n_a\,\Pi^{a}_{\alpha}\,\pounds_\xi \Phi-n_a\xi^a \mathcal{L})\,.
\end{equation}
Using the boundary diffeomorphism Ward identity \eqref{eq:boundarydiffward}, we thus get
\begin{equation}
    H_{\text{cl}}(t) = \int_{m_t} d^{d-1}\phi\sqrt{\gamma}\,\xi^{a}n^{b}\,T_{ab}^{\text{cl}}\,.
    \label{eq:final_Hamiltonian_app}
\end{equation}
Formula \eqref{eq:Noetherchargeformula} of the main text is recovered upon using $\xi^a = -\zeta^a$.

We have assumed that $\Phi$ is a scalar field, but the derivation here works for any non-scalar field as long as the response of the solution $\Phi_s$ to a variation of the location of the future $m_s$ under a change in $s$ can be written as a Lie derivative. We only proved this explicitly for a scalar field which required that the boundary condition imposed at $m_s$ is a Dirichlet condition. For non-scalar fields, imposing a simple Dirichlet condition is not sufficient: we see that equation \eqref{eq:using_dirichlet} is violated, because the Lie derivative includes additional terms on top of $\xi^a \partial_a\Phi_t$. Therefore, the Dirichlet boundary condition has to be modified such that $\tfrac{d}{ds} \Phi_s\vert_{s=t} = \pounds_\xi \Phi_t$ holds. We expect the modified boundary conditions to involve pull-backs of space-time indices to the boundary, in other words, only tensorial objects intrinsic to the boundary are kept fixed.\footnote{We have not proven this explicitly.} Regardless, it is natural to assume that a boundary condition implying $\tfrac{d}{ds} \Phi_s\vert_{s=t} = \pounds_\xi \Phi_t$ exists for all fields so that the expression \eqref{eq:final_Hamiltonian_app} for the Hamiltonian is general.

\section{$\text{Diff}\ltimes \text{Weyl}$ anomaly from the heat kernel}\label{app:heatkernel}

In this appendix, we compute the renormalized $\text{Diff}\ltimes \text{Weyl}$ Ward identities using point-splitting and the heat kernel expansion. This produces the $\text{Diff}\ltimes \text{Weyl}$ anomaly as a mixed diffeomorphism-Weyl anomaly which can be translated to a pure Weyl anomaly by addition of counterterms. For a conformal scalar field in $d = 2,4$ dimensions, our calculations reproduce previous results.

As shown in Appendix \ref{app:ward_identities}, the classical Ward identities are given by
\begin{equation}
    \nabla^{b}T_{ab}^{\text{cl}}(x) = E_\alpha(\Phi(x))\,\pounds_a\Phi^\alpha(x)\,,\quad g^{ab}(x)\,T_{ab}^{\text{cl}}(x) = \Delta_{\Phi}\, E_\alpha(\Phi(x))\,\Phi^\alpha(x)\,.
    \label{eq:diffweylward_app}
\end{equation}
At the classical level, these can be equivalently written as the $\epsilon\rightarrow 0$ limit of
\begin{align}
    \nabla^{b}T_{ab}^{\text{reg}}(x) &= \frac{1}{2}\,[E_\alpha(\Phi(x)),\pounds_a\Phi^\alpha(x+\epsilon)]_+\,,\nonumber\\
    g^{ab}(x)\,T_{ab}^{\text{reg}}(x) &= \frac{\Delta_{\Phi}}{2}\,[E_\alpha(\Phi(x)),\Phi^\alpha(x+\epsilon)]_+\,,
    \label{eq:diffweylward_app_reg}
\end{align}
where the anti-commutator between two local operators is defined as
\begin{equation}
	[A(x_1),B(x_2)]_+ \equiv A(x_1)\,B(x_2)+A(x_2)\,B(x_1)
\end{equation}
and we have defined the point-split regularized stress tensor that satisfies $\lim_{\epsilon\rightarrow 0}T_{ab}^{\text{reg}}(x) = T_{ab}^{\text{cl}}(x)$ (see also \cite{Christensen:1978yd}). At the quantum level, both the left- and the right-hand sides of \eqref{eq:diffweylward_app_reg} are divergent in the limit $\epsilon\rightarrow 0$, but we take \eqref{eq:diffweylward_app_reg} to be the regularized definition of the Ward identities. They can be further renormalized by subtracting divergences on both sides leaving Ward identities for the renormalized stress tensor operator $T_{ab}^{\text{ren}}(x)$.

To isolate finite terms that remain after renormalization on the right-hand side, we use the fact that the divergence structure is controlled by the anti-commutator of $\Phi$. These divergences can be found by first considering divergences of the time-ordered product
\begin{equation}
    \overleftarrow{\mathcal{T}}\{\Phi(x)\,\Phi(x')\} = G_{\text{F}}(x,x') + \Phi^2(x) + \ldots \,\quad x'\rightarrow x\,,
\end{equation}
where $\Phi^2(x)$ is a suitably normal-ordered composite product operator, ellipsis denote terms that vanish in the limit $x'\rightarrow x$, the time-ordering is defined as
\begin{equation}
	\overleftarrow{\mathcal{T}}\{\Phi(\phi_1,t_1)\,\Phi(\phi_2,t_2)\} \equiv \Phi(\phi_1,t_1)\,\Phi(\phi_2,t_2)\;\Theta(t_1 - t_2)+\Phi(\phi_2,t_2)\,\Phi(\phi_1,t_1)\; \Theta(t_2-t_1)
	\label{eq:lor_time_ordering}
\end{equation}
and the Feynman propagator $G_{\text{F}}(x,x') $ in the background metric $g$ is the solution of the equation
\begin{equation}
    E_\alpha\,G_{\text{F}}(x,x') =\frac{1}{\sqrt{-g}}\,\delta^{(d)}(x-x')\,.
    \label{eq:Feynman_propagator}
\end{equation}
Here the equation of motion operator acts only on the first argument $x$. As long as the field $\Phi$ is real valued, the singularities of the anti-commutator are controlled by the real part of the Feynman propagator $G_{\text{F}}(x_1,x_2)$ as
\begin{equation}
    [\Phi(x),\Phi(x')]_+ = 2\,\text{Re}\,G_{\text{F}}(x,x') +  \Phi^2(x) + \ldots\,,\quad x'\rightarrow x\,.
    \label{eq:anti_comm_div}
\end{equation}
The singularity structure of the real part of the Feynman propagator can be obtained by using the heat kernel expansion \cite{Adler:1976jx,Christensen:1976vb}. The heat kernel $\mathcal{G}(x,x';s)$ is defined as the solution of the equations
\begin{equation}
	\partial_{s}\mathcal{G}(x,x';s)= -E_\alpha\,\mathcal{G}(x,x';s)\,,\quad \mathcal{G}(x,x';0) = \frac{1}{\sqrt{-g}}\,\delta^{(d)}(x-x')\,.
    \label{eq:heat_kernel_def}
\end{equation}
It follows that the Feynman propagator \eqref{eq:Feynman_propagator} can be written as
\begin{equation}
    G_{\text{F}}(x,x') = \lim_{\lambda\rightarrow 0^+}\int_{\lambda}^\infty ds\,\mathcal{G}(x,x';s)\,,
    \label{eq:GF_heat_kernel}
\end{equation}
where $\lambda$ is another regulator independent from the point-splitting regulator. Equation \eqref{eq:GF_heat_kernel} together with \eqref{eq:heat_kernel_def} imply
\begin{equation}
    E_\alpha G_{\text{F}}(x,x')  = \lim_{\lambda\rightarrow 0^+}\mathcal{G}(x,x';\lambda)\,.
    \label{eq:E_acting_GF}
\end{equation}
By substituting the short distance expansion \eqref{eq:anti_comm_div} to \eqref{eq:diffweylward_app_reg} and using \eqref{eq:E_acting_GF}, we obtain
\begin{equation}
    \nabla^{b}T_{ab}^{\text{reg}}(x) = \lim_{\lambda\rightarrow 0^+}\text{Re}\,\partial_a\mathcal{G}(x,x';\lambda)\,,\quad g^{ab}(x)\,T_{ab}^{\text{reg}}(x) = \Delta_\Phi\lim_{\lambda\rightarrow 0^+}\text{Re}\,\mathcal{G}(x,x';\lambda)\,,
\end{equation}
where the derivative on the right-hand side of the first equation acts on $x'\equiv x + \epsilon$ and we have also assumed that the normal-ordered operator is on-shell $E_\alpha(\Phi^2) = 0$ as usual for Heisenberg operators. 

Now we can take the coincidence limit $x' = x+\epsilon\rightarrow x$ and define the renormalized stress tensor
\begin{equation}
    T^{\text{ren}}_{ab}(x) \equiv \lim_{\epsilon\rightarrow 0}\bigl(T_{ab}^{\text{reg}}(x) + T_{ab}^{\text{ct}}(x)\bigr)\,,
\end{equation}
where $T_{ab}^{\text{ct}}(x)$ are divergent counterterms. Assuming the coincidence limit $x'\rightarrow x$ commutes with the $\lambda\rightarrow 0^+$ limit, we obtain
\begin{align}
    \nabla^{b}T_{ab}^{\text{ren}}(x) = \frac{1}{2}\lim_{\lambda\rightarrow 0^+}\text{Re}\,\partial_a\mathcal{G}(x,x;\lambda)+\nabla^{b}T_{ab}^{\text{ct}}(x)\,,\nonumber\\
    g^{ab}(x)\,T_{ab}^{\text{ren}}(x) = \Delta_\Phi\lim_{\lambda\rightarrow 0^+}\text{Re}\,\mathcal{G}(x,x;\lambda) + g^{ab}(x)\,T_{ab}^{\text{ct}}(x)\,,\label{eq:ren_ward_identities_heat_kernel}
\end{align}
where now the derivative in the first equation acts on $x$ and we have used the fact that
\begin{equation}
    \frac{d}{d x}B(x,x) = 2\lim_{x'\rightarrow x}\frac{\partial}{\partial x'}\,B(x,x') 
    \label{eq:factor_of_two}
\end{equation}
for a symmetric function $B(x,x') = B(x',x)$.

The heat kernel in Lorentzian signature can be written as \cite{Christensen:1976vb,Adler:1976jx,Christensen:1978yd}
\begin{equation}
	\mathcal{G}(x,x';s) = \frac{1}{(4\pi i s)^{d\slash 2}}\,e^{-\frac{\sigma(x,x')}{2is}}\,\Omega(x,x'; s)\,,
    \label{eq:heat_kernel_Omega}
\end{equation}
where we have the expansion
\begin{equation}
	\Omega(x,x';s) = \sum_{n=0}^{\infty}a_n(x,x')\, (is)^{n}\,,
    \label{eq:Omega_expansion}
\end{equation}
where $a_n$ are known as Seeley--DeWitt coefficients. Together \eqref{eq:heat_kernel_Omega} and \eqref{eq:Omega_expansion} imply a short-time expansion of the heat kernel
\begin{equation}
	\mathcal{G}(x,x;\epsilon) = \mathcal{D}(x;\epsilon)+\frac{1}{(4\pi)^{d\slash 2}}\,a_{d\slash 2}(x,x) + \mathcal{O}( \epsilon)\,,\quad \epsilon\rightarrow 0^+\,,
    \label{eq:G_divergences}
\end{equation}
where $\mathcal{D}(x;\epsilon)$ is a sum of negative powers of $\epsilon$ and is a purely divergent term. The divergent term can be subtracted with a counterterm that satisfies
\begin{equation}
    \nabla^{b}T_{ab}^{\text{ct}}(x) = -\frac{1}{2}\,\partial_a\mathcal{D}(x;\epsilon)\,,\quad g^{ab}(x)\,T_{ab}^{\text{ct}}(x) =  -\Delta_{\Phi}\,\mathcal{D}(x;\epsilon)\,.
    \label{eq:T_ct}
\end{equation}
Substituting \eqref{eq:G_divergences} and \eqref{eq:T_ct} to \eqref{eq:ren_ward_identities_heat_kernel}, we obtain the renormalized Ward identities
\begin{equation}
    \nabla^{b}T_{ab}^{\text{ren}}(x) = \frac{1}{2\,(4\pi)^{d\slash 2}}\,\partial_a a_{d\slash 2}(x,x)\,,\quad g^{ab}(x)\,T_{ab}^{\text{ren}}(x) = \frac{\Delta_\Phi}{(4\pi)^{d\slash 2}}\, a_{d\slash 2}(x,x)\,.
    \label{eq:ren_ward_identities_app}
\end{equation}
We see that the $\text{Diff}\ltimes \text{Weyl}$ anomaly appears as a mixed diffeomorphism-Weyl anomaly. One can reduce it to a pure Weyl anomaly by the inclusion of finite counterterms. This amounts to a redefinition of the renormalized stress tensor as
\begin{equation}
	T_{ab}^{\text{D}} \equiv T_{ab}^{\text{ren}} - \frac{1}{2}\frac{1}{(4\pi)^{d\slash 2}}\, a_{d\slash 2}(x,x)\,g_{ab}\,.
\end{equation}
The Weyl anomaly in the diffeomorphism invariant scheme becomes
\begin{equation}
	\nabla^{b}T_{ab}^{\text{D}} = 0\,,\quad  g^{ab}\,T_{ab}^{\text{D}} = \frac{1}{(4\pi)^{d\slash 2}}\left(\Delta_{\Phi}-\frac{d}{2}\right)\, a_{d\slash 2}(x,x)\,.
 \label{eq:weylanomaly}
\end{equation}

\paragraph{Conformal scalar field.}  Let us consider a scalar field in $d $ space-time dimensions with a non-minimal coupling to the metric and with action
\begin{equation}
	I[\Phi,g] = -\frac{1}{2}\int d^d x\sqrt{-g}\,\bigl( g^{ab}\,\partial_a\Phi\,\partial_b \Phi + \xi\,\Phi^{2} R\bigr)\,.
	\label{eq:classical_scalar_field}
\end{equation}
With the definitions \eqref{eq:generalvariation}, the equation of motion is given by
\begin{equation}
    E(\Phi) = (\nabla^2-\xi R)\,\Phi \,,
    \label{eq:scalar_E}
\end{equation}
and the classical stress tensor is
\begin{align}
    T^{\text{cl}}_{ab} &= (1-2\xi)\,\partial_a\Phi\,\partial_b \Phi-\frac{1}{2}\,(1-4\xi)\,g_{ab}\,g^{cd}\,\partial_c\Phi\,\partial_d \Phi\nonumber\\
    &\quad + \xi \,\Phi\,\biggl(R_{ab} - \frac{1}{2}\,R\,g_{ab}-2\,(\nabla_a \nabla_b - g_{ab}\nabla^{2})\biggr)\,\Phi\,.
    \label{eq:scalar_T}
\end{align}
The action \eqref{eq:classical_scalar_field} is Weyl invariant if the non-minimal coupling $\xi$ and the Weyl dimension $\Delta_\Phi$ of the scalar field are chosen as
\begin{equation}
    \xi = \frac{d-2}{4\,(d-1)}\,,\quad \Delta_\Phi = \frac{d-2}{2}\,.
    \label{eq:xi_DeltaPhi}
\end{equation}
One can confirm that the stress tensor \eqref{eq:scalar_T} satisfies the classical Ward identities \eqref{eq:diffweylward} with the values \eqref{eq:xi_DeltaPhi}.

Now substituting \eqref{eq:xi_DeltaPhi} to the Ward identities \eqref{eq:weylanomaly} in the diffeomorphism invariant scheme, we obtain
\begin{equation}
	\nabla^{b}T_{ab}^{\text{D}} = 0\,,\quad  g^{ab}\,T_{ab}^{\text{D}} = -\frac{1}{(4\pi)^{d\slash 2}}\,a_{d\slash 2}(x,x)\,.
 \label{eq:eq:ren_ward_identities_app_weyl_2}
\end{equation}
For a generic $\xi$, the first three coefficients are given by\footnote{See for example Appendix A of \cite{McAvity:1990we}.}
\begin{gather}
	a_0(x,x) = 1\,,\quad a_1(x,x) = \biggl(\frac{1}{6}-\xi\biggr)\, R\,,\\
a_2(x,x) = \frac{1}{2}\biggl(\frac{1}{6}-\xi\biggr)^{2}\,R^{2}+\frac{1}{180}(R_{abcd}R^{abcd}-R_{ab}R^{ab})+\biggl(\frac{1}{30} - \frac{1}{6}\,\xi\biggr)\,\nabla^{2}R\,.
\end{gather}
For a conformal scalar field \eqref{eq:xi_DeltaPhi}, we obtain
\begin{gather}
	a_1(x,x) = -\frac{d-4}{12\,(d-1)}\, R\,,\\
a_2(x,x) = \frac{(d-4)^2}{288\,(d-1)^2}\,R^{2}+\frac{1}{180}(R_{abcd}R^{abcd}-R_{ab}R^{ab})-\frac{d-6}{120\,(d-1)}\,\nabla^{2}R\,.
\end{gather}
In $d = 2$, we obtain
\begin{equation}
	 a_1(x,x) = \frac{1}{6}\, R\,,\quad a_2(x,x) = \frac{1}{72}\,R^{2}+\frac{1}{180}(R_{abcd}R^{abcd}-R_{ab}R^{ab})+\frac{1}{30} \,\nabla^{2}R\,,
    \label{eq:coeffiecients_d_2}
\end{equation}
while in $d = 4$, we have
\begin{equation}
	a_1(x,x) = 0\,,\quad a_2(x,x) = \frac{1}{180}\,(R_{abcd}R^{abcd}-R_{ab}R^{ab}+\nabla^{2}R)\,.
    \label{eq:coeffiecients_d_4}
\end{equation}
Therefore the heat kernel has the expansion
\begin{equation}
	\mathcal{G}(x,x'; s) =\begin{dcases}
	    \frac{1}{4\pi s} + \frac{1}{24\pi}\,R+ \mathcal{O}(s)\,,\quad &d = 2\,,\\
        \frac{1}{ 16\pi^{2}s^{2}} + \frac{1}{180}\,(R_{abcd}R^{abcd}-R_{ab}R^{ab}+\nabla^{2}R) + \mathcal{O}( s)\,,\quad &d = 4\,.
	\end{dcases} 
\end{equation}
By substituting the coefficients to \eqref{eq:eq:ren_ward_identities_app_weyl_2} and using $\Delta_\Phi = 0$ and $\Delta_\Phi = 1$ in two and four dimensions respectively, we obtain
\begin{equation}
	g^{ab}\,T_{ab}^{\text{D}} =
\begin{dcases}
  -\frac{1}{24\pi}\,R\,,\quad &d = 2\,,\\
  -\frac{1}{2880\pi^{2}}\,(R_{abcd}R^{abcd}-R_{ab}R^{ab}+\nabla^{2}R)\,,\quad &d = 4\,.
\end{dcases}
\end{equation}
For $d = 2$ is the usual Weyl anomaly for free scalar field with central charge $c = 1$. For $d = 4$ it matches with \cite{Christensen:1976vb,Christensen:1978yd} (including the overall sign) and the appearance of $\Delta_\Phi = 1$ in \eqref{eq:weylanomaly} is necessary to obtain the correct result.

\section{$\text{Diff}\ltimes \text{Weyl}$ anomaly in different renormalization schemes}\label{app:actions_anomalies}

In this Appendix, we consider the $\text{Diff}\ltimes \text{Weyl}$ anomaly in detail in two renormalization schemes: diffeomorphism invariant and chirally split schemes. In particular, we write down the Ward identities explicitly in coordinates $(\phi,t)$ used to parametrize the background metric \eqref{eq:curvedg} of the CFT. In addition, we derive the anomaly contributions $C_{ab}^{\text{D}}$ and $C_{ab}^{\text{F}}$ to the renormalized stress tensor in the two schemes respectively by varying the effective actions.

\subsection{Diffeomorphism invariant scheme}\label{subapp:diff_invariant_scheme}

We will start with the diffeomorphism invariant scheme.

\paragraph{Variation of the effective action.} The solution of the Ward identities \eqref{eq:C_D_Wards} is given by
\begin{equation}
	C_{ab}^{\text{D}} = -\frac{2}{\sqrt{-g}}\frac{\delta A_{\text{D}}[g]}{\delta g^{ab}}\,,
    \label{eq:Cab_D_app}
\end{equation}
where $A_{\text{D}}[g]$ is the Polyakov action \eqref{eq:Polyakov_text}. Given the metric $g$ \eqref{eq:curvedg}, the flat metric is $\eta = e^{-2\sigma}g$. Thus it follows that
\begin{equation}
    A_{\text{D}}[\eta] = A_{\text{D}}[e^{-2\sigma}g]  = I_{\text{Lio}}[-\sigma,g]+A_{\text{D}}[g]\,,
\end{equation}
where the Liouville action is defined in \eqref{eq:Liouville_action_text}. We obtain the result
\begin{equation}
    A_{\text{D}}[g] = -I_{\text{Lio}}[-\sigma,g]+A_{\text{D}}[\eta]\,.
\end{equation}
Substituting to \eqref{eq:Cab_D_app} gives
\begin{equation}
    C_{ab}^{\text{D}} = \biggl(-C_{ab}^{\text{Lio}} +\frac{\delta \sigma}{\delta g^{ab}}\,E_{\text{Lio}}\biggr)\bigg\vert_{\chi = -\sigma}\,,
    \label{eq:Cab_D_app_2}
\end{equation}
where we have defined
\begin{equation}
    C_{ab}^{\text{Lio}}\equiv -\frac{2}{\sqrt{-g}}\frac{\delta I_{\text{Lio}}[\chi,g]}{\delta g^{ab}}\,,\quad E_{\text{Lio}} \equiv -\frac{2}{\sqrt{-g}}\frac{\delta I_{\text{Lio}}[\chi,g]}{\delta \chi}\,.
\end{equation}
The Liouville stress tensor becomes explicitly
\begin{equation}
    C_{ab}^{\text{Lio}} = -\frac{c}{12\pi}\,\biggl[\nabla_{a}\chi\,\nabla_b\,\chi\,-\nabla_a\nabla_b\,\chi+g_{ab}\,\biggl(\nabla^2 \chi-\frac{1}{2}\, \nabla^{c}\chi\,\nabla_c\chi\biggr)\biggr]\,,
    \label{eq:Liouville_mid_step}
\end{equation}
where $\nabla_a$ is the covariant derivative compatible with $g$. On the other hand, we have
\begin{equation}
    E_{\text{Lio}} = \frac{c}{12\pi}\,( -2\,\nabla^2\chi +  R )\,.
\end{equation}
Since $\eta = e^{-2\sigma}g$, we have $0 = e^{2\sigma}\,(R + 2\,\nabla^2\sigma)$ so that
\begin{equation}
    E_{\text{Lio}}\big\vert_{\chi = -\sigma} = -\frac{c}{6\pi}\,\nabla^2(\chi + \sigma)\big\vert_{\chi = -\sigma} = 0\,.
\end{equation}
Therefore substituting this and \eqref{eq:Liouville_mid_step} to \eqref{eq:Cab_D_app_2}, we obtain
\begin{equation}
    C_{ab}^{\text{D}} = \frac{c}{12\pi}\,\biggl[\nabla_{a}\sigma\,\nabla_b\,\sigma+\nabla_a\nabla_b\,\sigma-g_{ab}\,\biggl(\nabla^2 \sigma+\frac{1}{2}\, \nabla^{c}\sigma\,\nabla_c\sigma\biggr)\biggr]\,.
\end{equation}

\paragraph{Ward identities.} The covariant derivatives of a symmetric rank-2 tensor field $C_{ab}^{\text{D}}$ in the metric \eqref{eq:curvedg} take the form
\begin{align}
	\nabla^{b}C_{-b}^{\text{D}} &= 2e^{-2\sigma}\,[\partial_+ C_{--}^{\text{D}}+(\partial_--2\partial_-\sigma)\,C_{-+}^{\text{D}}]\,,\\
	\nabla^{b}C_{+b}^{\text{D}} &= 2e^{-2\sigma}\,[\partial_- C_{++}^{\text{D}}+(\partial_+-2\partial_+\sigma)\,C_{-+}^{\text{D}}]\,,
    \label{eq:general_covariant_derivatives}
\end{align}
where $\partial_{\pm}\equiv \frac{\partial}{\partial x^{\pm}}$ are derivatives with respect to the light-ray coordinates \eqref{eq:lightraycoords} and $e^{-2\sigma} = e^{-2\omega}\,F_t'\,\overbar{F}_t'$. Using these, the Ward identities \eqref{eq:C_D_Wards} become
\begin{equation}
	\partial_+ C_{--}^{\text{D}}+(\partial_--2\partial_-\varphi)\,C_{-+}^{\text{D}} = 0\,,\;\;\, \partial_- C_{++}^{\text{D}}+(\partial_+-2\partial_+\sigma)\,C_{-+}^{\text{D}} = 0\,,\;\;\, C_{-+}^{\text{D}} = -\frac{c}{96\pi}\,e^{2\sigma}R\,.
\end{equation}
Substituting the third equation to the first two, we obtain
\begin{equation}
	\partial_+ C_{--}^{\text{D}}-\frac{c}{96\pi}\,e^{2\sigma}\partial_- R = 0\,,\quad \partial_- C_{++}^{\text{D}}-\frac{c}{96\pi}\,e^{2\sigma}\partial_+ R = 0\,,\quad C_{-+}^{\text{D}} = -\frac{c}{96\pi}\,e^{2\sigma}R\,.
	\label{eq:diffeo_wards_midstep}
\end{equation}
From \eqref{eq:lightraycoords} it follows that
\begin{equation}
	\partial_- =-\frac{1}{F_t'(\phi)}\frac{\partial_t - \overbar{\nu}\, \partial_\phi}{\overbar{\nu}-\nu}, \quad \partial_+ = \frac{1}{\overbar{F}_t'(\phi)}\frac{\partial_t - \nu\, \partial_\phi}{\overbar{\nu}-\nu}\,,
	\label{eq:tildepartial}
\end{equation}
which imply
\begin{align}
	F_t'(\phi)^{2}\,\overbar{F}_t'(\phi)\,(\overbar{\nu}-\nu)\,\partial_+ C_{--}^{\text{D}} &= (\partial_t - \nu\,\partial_\phi-2\partial_\phi \nu)\,(F_t'^{2}C_{--}^{\text{D}})\,, \nonumber\\
	\overbar{F}_t'(\phi)^{2}\,F_t'(\phi)\,(\overbar{\nu}-\nu)\,\partial_- C_{++}^{\text{D}} &=(\partial_t - \overbar{\nu}\,\partial_\phi-2\partial_\phi \overbar{\nu})\,(\overbar{F}_t'^{2}C_{++}^{\text{D}})\,.\label{eq:partial_pm_identity}
\end{align}
where we have used the identities
\begin{equation}
	(\partial_t - \nu\,\partial_\phi-h\,\partial_\phi \nu)\,F_t'(\phi)^{h} = 0\,,\quad (\partial_t - \overbar{\nu}\,\partial_\phi-h\,\partial_\phi \overbar{\nu})\,\overbar{F}_t'(\phi)^{h} = 0
\end{equation}
for $h = 2$. Substituting \eqref{eq:partial_pm_identity} to \eqref{eq:diffeo_wards_midstep}, we obtain finally
\begin{align}
	(\partial_t - \nu\, \partial_\phi-2\partial_\phi \nu)\, (F_t'^{2}C_{--}^{\text{D}})-\frac{c}{96\pi}\,\,e^{2\omega}\,(\partial_t - \overbar{\nu}\, \partial_\phi)\, R &= 0\,,\\
	 (\partial_t - \overbar{\nu}\, \partial_\phi-2\partial_\phi \overbar{\nu})(\overbar{F}_t'^{2} C_{++}^{\text{D}})-\frac{c}{96\pi}\,e^{2\omega}\,(\partial_t - \nu\, \partial_\phi)\, R &= 0\,,\\
     F_t'\overbar{F}_t'\,C_{-+}^{\text{D}} + \frac{c}{96\pi}\,e^{2\omega}R&=0\,.
\end{align}

\subsection{Chirally split scheme}\label{subapp:chirally_split_scheme}

Now we move on to the chirally split scheme.

\paragraph{Derivation of the effective action.} The derivation of the decomposition \eqref{eq:Polyakov_decomposition_text} of the Polyakov action of the metric \eqref{eq:curvedg} was done in \cite{deBoer:2023lrd}. Here we will briefly review the derivation and correct a factor of two error in the final result.

To prove the decomposition, we need to show that the Liouville action $I_{\text{Lio}}[\varphi,\eta]$ of $\varphi =-\frac{1}{2}\log{F_t'}  -\frac{1}{2}\log{\overbar{F}_t'} $ decomposes as in \eqref{eq:Liouville_varphi}. First, in a general coordinate system we have\footnote{Notice that in \cite{deBoer:2023lrd} the coefficient of the Polyakov action, and therefore also of the Liouville action, contains an additional factor of $1\slash 2$.}
\begin{equation}
	I_{\text{Lio}}[\varphi,\eta] = \frac{c}{24\pi}\int d^2x\sqrt{-\eta}\,\eta^{ab}\,\partial_a\varphi\,\partial_b\varphi\,.
\end{equation}
In light-ray coordinates $x^{\pm}$, we have\footnote{This factor of $1\slash 2$ coming from the determinant in light-ray coordinates was missed in the calculation of \cite{deBoer:2023lrd}.} $\sqrt{-\eta} = \frac{1}{2}$ and $\eta^{ab}\,\partial_a\varphi\,\partial_b\varphi = 4\,\partial_-\varphi\,\partial_+\varphi $ so that
\begin{equation}
	I_{\text{Lio}}[\varphi,\eta] = \frac{c}{12\pi}\int dx^{-}dx^{+}\,\partial_-\varphi\,\partial_+\varphi\,.
\end{equation}
Substituting the definition of $\varphi$ and integrating by parts twice one of the terms, we obtain
\begin{align}
	&I_{\text{Lio}}[\varphi,\eta]\label{eq:ILio_varphi_midstep}\\
	&= -\frac{c}{48\pi}\int dx^{-}dx^{+}\,(\partial_-\log{F_t'}\,\partial_+ \log{F_t'} +\partial_-\log{\overbar{F}_t'}\,\partial_+ \log{\overbar{F}_t'} +2\,\partial_-\log{\overbar{F}_t'}\,\partial_+ \log{F_t'})\,.\nonumber
\end{align}
Now using \eqref{eq:tildepartial}, we obtain the identities \cite{deBoer:2023lrd}
\begin{align}
	F_t'\overbar{F}_t'\,(\overbar{\nu}-\nu)\,\partial_-\log{F_t'}\,\partial_{+}\log{F_t'}&= -(\partial_{\phi} \nu)\biggl(\frac{\partial_{\phi} \nu}{\overbar{\nu}-\nu}-\partial_{\phi}\log{F_t'}\biggr)\,,\nonumber\\
F_t'\overbar{F}_t'\,(\overbar{\nu}-\nu)\,\partial_-\log{\overbar{F}_t'}\,\partial_{+}\log{\overbar{F}_t'} &= -(\partial_{\phi} \overbar{\nu})\biggl(\frac{\partial_{\phi} \overbar{\nu}}{\overbar{\nu}-\nu}+\partial_{\phi}\log{\overbar{F}_t'}\biggr)\,,\nonumber\\
	F_t'\overbar{F}_t'\,(\overbar{\nu}-\nu)\,\partial_-\log{\overbar{F}_t'}\,\partial_{+}\log{F_t'}&= -\frac{(\partial_{\phi} \nu)(\partial_{\phi} \overbar{\nu})}{\overbar{\nu}-\nu}\,,
	\label{derivativeproducts}
\end{align}
from which it follows that
\begin{align}
	&F_t'\overbar{F}_t'\,(\overbar{\nu}-\nu)(\partial_-\log{F_t'}\,\partial_+ \log{F_t'} +\partial_-\log{\overbar{F}_t'}\,\partial_+ \log{\overbar{F}_t'} +2\,\partial_-\log{\overbar{F}_t'}\,\partial_+ \log{F_t'})\\
	&=(\partial_\phi \nu)\,\partial_\phi\log{F_t'}-(\partial_\phi \overbar{\nu})\,\partial_\phi\log{\overbar{F}_t'}-\frac{(\partial_\phi \nu+\partial_\phi \overbar{\nu})^2}{\overbar{\nu}-\nu}\,.
\end{align}
Substituting to \eqref{eq:ILio_varphi_midstep} after performing the change of integration variables\footnote{The sign of the Jacobian determinant here is fixed by the metric $\eta_{ab}(x)\,dx^adx^b = dx^-dx^+ = F_t'\,\overbar{F}_t'\,(d\phi+\nu\,dt)(d\phi+\overbar{\nu}\,dt)$ via $ d^2x\sqrt{-\eta} = dx^-dx^+\,\frac{1}{2} = d\phi dt\,\frac{1}{2}\,F_t'\overbar{F}_t'\,(\overbar{\nu}-\nu) $.} $dx^{-}dx^{+} = d\phi dt\,F_t'\overbar{F}_t'\,(\overbar{\nu}-\nu) $, we obtain\footnote{This differs from the result of \cite{deBoer:2023lrd} by a factor of $-1\slash 2$ once the factor of $1\slash 2$ difference in the definition of the Polyakov action is accounted for. The factors here are crucial in obtaining the correct $-\frac{c}{24\pi}$ prefactor for the Schwarzian terms appearing in the chirally split stress tensor.}
\begin{equation}
	I_{\text{Lio}}[\varphi,\eta]= -\frac{c}{48\pi}\int d\phi dt\,\biggl[(\partial_\phi \nu)\,\partial_\phi\log{F_t'}-(\partial_\phi \overbar{\nu})\,\partial_\phi\log{\overbar{F}_t'}-\frac{(\partial_\phi \nu+\partial_\phi \overbar{\nu})^2}{\overbar{\nu}-\nu}\biggr]\,.
\end{equation}
Integrating the first two terms by parts in $\phi$ gives equation \eqref{eq:Liouville_varphi} of the main text.

\paragraph{Variation of the effective action.} In the chirally split scheme, we have
\begin{equation}
	C_{ab}^{\text{F}} = -\frac{2}{\sqrt{-g}}\frac{\delta A_{\text{F}}[g]}{\delta g^{ab}}\,,
	\label{eq:Chatdef_app}
\end{equation}
where $A_{\text{F}}[g] = \Gamma[\nu] + \overbar{\Gamma}[\overbar{\nu}]$ and $\Gamma,\overbar{\Gamma}$ are defined in \eqref{appdecomposition}. The functional derivatives are explicitly
\begin{align}\label{Ttildepp}
	-\frac{2}{\sqrt{-g}}\frac{\delta}{\delta g^{--}}   &=-\frac{1}{F_t'^{2}}\left( \frac{\delta}{\delta \nu} + \frac{1}{\overbar{\nu}-\nu}\frac{\delta}{\delta \omega}\right)\,,\\
	\label{Ttildemm}
	-\frac{2}{\sqrt{-g}}\frac{\delta}{\delta g^{++}}    &=\frac{1}{\overbar{F}_t'^{2}}\left( \frac{\delta}{\delta \overbar{\nu}} - \frac{1}{\overbar{\nu}-\nu}\frac{\delta}{\delta \omega}\right)\,,\\
	-\frac{2}{\sqrt{-g}}\frac{\delta}{\delta g^{-+}}   &= \frac{1}{F_t'\,\overbar{F}_t'}\frac{1}{\overbar{\nu}-\nu} \frac{\delta}{\delta \omega}\,.
	\label{pathintegralvevs}
\end{align}
Thus it follows that
\begin{equation}
	C_{--}^{\text{F}} = -\frac{1}{F_t'^{2}}\frac{\delta \Gamma[\nu]}{\delta \nu}\,,\quad C_{++}^{\text{F}} = \frac{1}{\overbar{F}_t'^{2}}\frac{\delta \overbar{\Gamma}[\overbar{\nu}]}{\delta \overbar{\nu}}\,,\quad C_{-+}^{\text{F}} = 0\,.
	\label{eq:Chatcomponents}
\end{equation}
The variations here may be computed following \cite{Lazzarini:1990xid}. First we obtain after integration by parts
\begin{equation}
	\delta \Gamma =\frac{c}{48\pi}\int d\phi dt\, (\delta\nu\,\partial_\phi^{2}\log{F_t'}+\partial_\phi^{2}\nu\,\delta\log{F_t'})\,.
	\label{eq:deltaGamma_1}
\end{equation}
From the identity
\begin{equation}
	(\partial_t - \nu\, \partial_\phi)\,\log{F_t'} = \partial_\phi \nu\,,
	\label{eq:identity_1}
\end{equation}
we obtain
\begin{equation}
	\partial_\phi^2 \nu=  (\partial_t - \nu\, \partial_\phi)\,\partial_\phi\log{F_t'}- \partial_\phi\nu\,\partial_\phi\log{F_t'}\,.
	\label{eq:identity_3}
\end{equation}
Substituting to the second term in \eqref{eq:deltaGamma_1}, it becomes
\begin{equation}
	\frac{c}{48\pi}\int d\phi dt\, \partial_\phi^2\nu\,\delta\log{F_t'}=\frac{c}{48\pi}\int d\phi dt\, ((\partial_t - \nu\, \partial_\phi)\,\partial_\phi\log{F_t'}\,\delta\log{F_t'}- \partial_\phi\nu\,\partial_\phi\log{F_t'}\,\delta\log{F_t'})\,.
	\label{eq:second_term_0}
\end{equation}
In the last term, integrating $\partial_\phi\nu$ by parts, we obtain
\begin{align}
    \frac{c}{48\pi}\int d\phi dt\, \partial_\phi^2\nu\,\delta\log{F_t'}&=\frac{c}{48\pi}\int d\phi dt\, (\partial_t \partial_\phi\log{F_t'}\,\delta\log{F_t'}+ \nu\,\partial_\phi\log{F_t'}\,\partial_\phi\delta\log{F_t'})\nonumber\\
    &=-\frac{c}{48\pi}\int d\phi dt\,  \partial_\phi\log{F_t'}\,(\partial_t-\nu\,\partial_\phi)\,\delta\log{F_t'}\,.
    \label{eq:second_term}
\end{align}
Now taking the variation of \eqref{eq:identity_1} gives
\begin{equation}
	(\partial_t - \nu\, \partial_\phi)\,\delta\log{F_t'}=\delta\nu\, \partial_\phi\log{F_t'} + \partial_\phi \delta \nu\,.
	\label{eq:identity_2}
\end{equation}
Substituting to \eqref{eq:second_term} gives
\begin{align}
	\frac{c}{48\pi}\int d\phi dt\, \partial_\phi^2\nu\,\delta\log{F_t'} &= -\frac{c}{48\pi}\int d\phi dt\,  \partial_\phi\log{F_t'}\,(\delta\nu\, \partial_\phi\log{F_t'} + \partial_\phi \delta \nu)\\
	&= \frac{c}{48\pi}\int d\phi dt\, \delta\nu\,(- (\partial_\phi\log{F_t'})^2 + \partial_\phi^2\log{F_t'})\,.
\end{align}
Substituting to \eqref{eq:deltaGamma_1} gives finally
\begin{equation}
	\delta \Gamma
	=\frac{c}{48\pi}\int d\phi dt\, \delta\nu\,(2\,\partial_\phi^{2}\log{F_t'}- (\partial_\phi\log{F_t'})^2) = \frac{c}{24\pi}\int d\phi dt\, \delta\nu\,\{F_t(\phi),\phi\}\,.
	\label{eq:deltaGamma_final}
\end{equation}
The calculation of $\delta \overbar{\Gamma}$ is analogous with an overall minus sign. Therefore from \eqref{eq:Chatcomponents} we obtain
\begin{equation}
	C_{--}^{\text{F}} = -\frac{c}{24\pi}\frac{1}{F_t'(\phi)^{2}}\{F_t(\phi),\phi\}\,,\quad C_{++}^{\text{F}} = -\frac{c}{24\pi}\frac{1}{\overbar{F}_t'(\phi)^{2}}\{\overbar{F}_t(\phi),\phi\}\,,\quad C_{-+}^{\text{F}} = 0\,.
    \label{eq:CF_schwarzians}
\end{equation}

\paragraph{Ward identities.} Let us now derive the Ward identities satisfied by \eqref{eq:CF_schwarzians}. First, the covariant derivatives of $C_{ab}^{\text{F}}$ in the metric \eqref{eq:curvedg} take the form \eqref{eq:general_covariant_derivatives}. Since $C_{-+}^{\text{F}} = 0$ by $g^{ab}\,C_{ab}^{\text{F}} = 0$, we obtain
\begin{equation}
	\nabla^{a}C_{\pm a}^{\text{F}} = 2e^{-2\sigma}\,\partial_{\mp}C_{\pm\pm}^{\text{F}}\,.
	\label{eq:covariantderivativeChat}
\end{equation}
Using the identities \eqref{eq:partial_pm_identity} in \eqref{eq:covariantderivativeChat}, we obtain
\begin{align}
	\nabla^{a}C_{- a}^{\text{F}} &= \frac{2e^{-2\omega}}{\overbar{\nu}-\nu}\frac{1}{F_t'}\,(\partial_t -\nu\,\partial_\phi - 2\partial_\phi\nu)\,(F_t'^{2}C_{--}^{\text{F}})\,,\nonumber\\
	\nabla^{a}C_{+ a}^{\text{F}} &= -\frac{2e^{-2\omega}}{\overbar{\nu}-\nu}\frac{1}{\overbar{F}_t'}\,(\partial_t -\overbar{\nu}\,\partial_\phi - 2\partial_\phi\overbar{\nu})\,(\overbar{F}_t'^{2}C_{++}^{\text{F}})\,.
	\label{eq:covariants}
\end{align}
Now the Schwarzian derivative satisfies
\begin{equation}
	(\partial_t -\nu\,\partial_\phi - 2\partial_\phi\nu)\,\{F_t(\phi),\phi\} = \partial_{\phi}^{3}\nu\,.
\end{equation}
Thus substituting \eqref{eq:CF_schwarzians} to \eqref{eq:covariants}, we obtain
\begin{equation}
    \nabla^{a}C_{- a}^{\text{F}} = -\frac{c}{24\pi}\frac{2e^{-2\omega}}{\overbar{\nu}-\nu}\frac{1}{F_t'}\,\partial_\phi^3\nu\,,\quad \nabla^{a}C_{+ a}^{\text{F}} = \frac{c}{24\pi}\frac{2e^{-2\omega}}{\overbar{\nu}-\nu}\frac{1}{\overbar{F}_t'}\,\partial_\phi^3\overbar{\nu}\,,
    \label{eq:diffeo_Ward_chirally_split}
\end{equation}
which together with $g^{ab}\,C_{ab}^{\text{F}} = 0$ are the Ward identities in the chirally split scheme. Equivalently, writing the covariant derivatives explicitly, \eqref{eq:diffeo_Ward_chirally_split} can be written as
\begin{align}
	(\partial_t -\nu\,\partial_\phi - 2\partial_\phi\nu)\,(F_t'^{2}C_{--}^{\text{F}}) =  -\frac{c}{24\pi}\,\partial_\phi^{3} \nu\,,\quad (\partial_t -\overbar{\nu}\,\partial_\phi - 2\partial_\phi\overbar{\nu})\,(\overbar{F}_t'^{2}C_{++}^{\text{F}}) =  -\frac{c}{24\pi}\,\partial_\phi^{3} \overbar{\nu}\,,
\end{align}
together with $C_{-+}^{\text{F}} = 0$. The same equations have also appeared in a holographic context in \cite{Grumiller:2016pqb}.

\section{Chirally split scheme from point-splitting in space}\label{app:point_splitting_space}

The action of a free scalar field is given by
\begin{equation}
	I[\Phi,g]= \int d^{2}x\sqrt{-g}\,\mathcal{L} = -\frac{1}{2}\int d^{2}x\sqrt{-g}\,g^{ab}\,\partial_{a}\Phi\,\partial_b \Phi\,.
\end{equation}
For the metric \eqref{eq:curvedg}, it becomes
\begin{equation}
	I[\Phi,g] = \int d\phi dt\,\frac{1}{\overbar{\nu}-\nu}\,[(\partial_{t} - \nu\,\partial_\phi)\,\Phi\,][(\partial_{t} - \overbar{\nu}\,\partial_\phi)\,\Phi\,]
\end{equation}
The classical stress tensor is
\begin{equation}
	 T_{\pm\pm}^{\text{cl}} = (\partial_{\pm}\Phi)^{2}\,,\quad T_{-+}^{\text{cl}} =0\,.
\end{equation}
The equations of motion are $\partial_-\partial_+ \Phi = 0$. The conjugate momentum at time $t$ is defined as the variation of the on-shell action $I[\Phi_t,g;M_t]$ with respect to the boundary value $\Psi_t(\phi)$ of the scalar field on the future boundary $m_t$ of $M_t$ using definitions of Appendix \ref{app:Hamiltonian_derivation}. Explicitly
\begin{equation}
    \Pi(\phi,t) \equiv \frac{I[\Phi_t,g;M_t]}{\delta\Psi_t(\phi)} = \sqrt{\gamma}\,n_a\,\frac{\partial \mathcal{L}}{\partial(\partial_a \Phi)}\bigg\vert_{m_t}\,,
\end{equation}
where we have used \eqref{eq:generalvariation} and \eqref{eq:E_Pi}. Using $\sqrt{\gamma} = e^{\omega} $ and $n_a = \frac{1}{2}\,e^{\omega}\,(\overbar{\nu}-\nu)\,\delta^t_a $ \eqref{eq:n_down_def}, we obtain $\sqrt{\gamma}\,n_a = \sqrt{-g}\,\delta_a^t$. It follows that
\begin{equation}
    \Pi(\phi,t) = \frac{\partial (\sqrt{-g}\,\mathcal{L})}{\partial \dot{\Phi}(\phi,t)} = \frac{(\partial_{t} - \nu\,\partial_\phi)\,\Phi}{\overbar{\nu}-\nu}+\frac{(\partial_{t} - \overbar{\nu}\,\partial_\phi)\,\Phi}{\overbar{\nu}-\nu} = -F_t'(\phi)\,\partial_-\Phi +\overbar{F}_t'(\phi)\,\partial_+\Phi\,,
\end{equation}
where we used \eqref{eq:tildepartial}. We quantize the theory by promoting $\Phi$ and $\Pi$ into Heisenberg picture operators $\Phi_{\text{H}}$ and $\Pi_{\text{H}}$ which satisfy canonical equal-time commutation relations
\begin{equation}
    [\Phi_{\text{H}}(\phi_1,t),\Pi_{\text{H}}(\phi_2,t)] = i\,\delta_{2\pi}(\phi_1-\phi_2)\,,
    \label{eq:canonical_commutation_relation}
\end{equation}
where the $2\pi$-periodic delta function $\delta_{2\pi}(\phi) = \frac{1}{2\pi}\sum_{n=-\infty}^\infty e^{in\phi}$. The Heisenberg picture field operator satisfies the equation of motion and therefore takes the form
\begin{equation}
    \Phi_{\text{H}}(\phi,t) = \Phi_-(x^-) + \Phi_+(x^+) = \Phi_-(F_t(\phi)) + \Phi_+(\overbar{F}_t(\phi))\,.
\end{equation}
The conjugate momentum becomes
\begin{gather}
    \Pi_{\text{H}}(\phi,t) = -F_t'(\phi)\,\Pi_-(x^-)+\overbar{F}_t'(\phi)\,\Pi_+(x^+)\,,\label{eq:conjugate_momentum_operator}\\
    \Pi_-(x^-) \equiv \partial_-\Phi_-(x^-) =  \Phi_-'(F_t(\phi))\,,\quad \Pi_+(x^+) \equiv \partial_+\Phi_+(x^+) = \Phi_+'(\overbar{F}_t(\phi))\,.
\end{gather}
The canonical equal-time commutation relation \eqref{eq:canonical_commutation_relation} is equivalent to the chiral commutation relations
\begin{equation}
    [\Phi_{\pm}(\phi_1),\Pi_{\pm}(\phi_2)] = \pm\frac{i}{2}\,\delta_{2\pi}(\phi_1-\phi_2)\mp\frac{1}{2}\frac{i}{2\pi}\,,\quad [\Phi_{\pm}(\phi_1),\Pi_{\mp}(\phi_2)] = \mp \frac{1}{4}\frac{i}{2\pi}\,.
    \label{eq:chiral_commutation_relations}
\end{equation}
To see this, substitute \eqref{eq:conjugate_momentum_operator} into \eqref{eq:canonical_commutation_relation}, the canonical commutator becomes
\begin{align}
    &[\Phi_{\text{H}}(\phi_1,t),\Pi_{\text{H}}(\phi_2,t)]\nonumber\\
    &= -F_t'(\phi_2)\,[\Phi_{-}(F_t(\phi_1)),\Pi_{-}(F_t(\phi_2))]+\overbar{F}_t'(\phi_2)\,[\Phi_{+}(\overbar{F}_t(\phi_1)),\Pi_{+}(\overbar{F}_t(\phi_2))]\nonumber\\
    &-F_t'(\phi_2)\,[\Phi_{+}(F_t(\phi_1)),\Pi_{-}(F_t(\phi_2))]+\overbar{F}_t'(\phi_2)\,[\Phi_{-}(\overbar{F}_t(\phi_1)),\Pi_{+}(\overbar{F}_t(\phi_2))]\,.
\end{align}
Using \eqref{eq:chiral_commutation_relations}, we obtain
\begin{equation}
    [\Phi_{\text{H}}(\phi_1,t),\Pi_{\text{H}}(\phi_2,t)] = \frac{i}{2}\,F_t'(\phi_2)\,\delta_{2\pi}(F_t(\phi_1)-F_t(\phi_2))+\frac{i}{2}\,\overbar{F}_t'(\phi_2)\,\delta_{2\pi}(\overbar{F}_t(\phi_1)-\overbar{F}_t(\phi_2))\,.
\end{equation}
Due to the identity $f'(\phi_2)\,\delta_{2\pi}(f(\phi_1)-f(\phi_2)) = \delta_{2\pi}(\phi_1-\phi_2)$ for a function $f'(\phi)>0$, the canonical commutation relation \eqref{eq:canonical_commutation_relation} is recovered.

From \eqref{eq:chiral_commutation_relations}, we obtain
\begin{equation}
    [\Pi_{\pm}(\phi_1),\Pi_{\pm}(\phi_2)] = \pm \frac{i}{2}\,\delta_{2\pi}'(\phi_1-\phi_2)\,,\quad [\Pi_-(\phi_1),\Pi_+(\phi_2)] = 0\,.
    \label{eq:momentum_self_commutator}
\end{equation}
Therefore we can write
\begin{equation}
    \Pi_-(\phi) = -\frac{1}{\sqrt{2}}\,(J(\phi)\otimes \mathbf{1})\,,\quad \Pi_+(\phi) = \frac{1}{\sqrt{2}}\,(\mathbf{1}\otimes \overbar{J}(\phi))\,,
    \label{eq:Pi_J_relation}
\end{equation}
where $\overbar{J}(\phi) = J(-\phi)$ and $J(\phi)$ satisfies the $U(1)$ current algebra
\begin{equation}
    [J(\phi_1),J(\phi_2)] = -i\,\delta_{2\pi}'(\phi_1-\phi_2)\,.
    \label{eq:U1_current_algebra}
\end{equation}
The relative minus sign between the two terms in the conjugate momentum \eqref{eq:conjugate_momentum_operator} leads to the minus sign in the $\Pi_+$ commutator \eqref{eq:momentum_self_commutator} which requires the introduction of $\overbar{J}$ in \eqref{eq:Pi_J_relation}. The minus sign in the first equation \eqref{eq:Pi_J_relation} is needed for the mode expansions below.

\paragraph{Mode expansions.} We expand the chiral scalar field operators as\footnote{We use the same conventions as in \cite{Torre:1997zs}, but note their light-ray coordinates are $\pm x^{\pm}$.}
\begin{equation}
	\Phi_{\pm}(x^{\pm}) = \frac{1}{2}\frac{1}{\sqrt{2\pi}}\,(Q\pm P x^{\pm})+\frac{1}{\sqrt{2}}\sum_{n=1}^{\infty}\frac{1}{\sqrt{2\pi n}}\,[a_n^{\pm}\,e^{\mp inx^{\pm}} + (a_n^{\pm})^{\dagger}\,e^{\pm in x^{\pm}}]
\end{equation}
so that the chiral momenta become
\begin{equation}
	\Pi_{\pm}(x^{\pm}) = \pm\frac{1}{2}\frac{1}{\sqrt{2\pi}}\,P+\frac{i}{\sqrt{2}}\sum_{n=1}^{\infty}\sqrt{\frac{n}{2\pi}}\,[\mp a_n^{\pm}\,e^{\mp in x^{\pm}} \pm  (a_n^{\pm})^{\dagger}\,e^{\pm in x^{\pm}}]\,.
\end{equation}
Defining
\begin{equation}
    a_n^- \equiv a_n\otimes \mathbf{1}\,,\quad a_n^+ \equiv \mathbf{1}\otimes a_n\,,
\end{equation}
the chiral commutation relations \eqref{eq:chiral_commutation_relations} (and hence also the canonical commutation relation \eqref{eq:canonical_commutation_relation}) turn out to be equivalent to
\begin{equation}
    [a_n,a_m^\dagger] = \delta_{nm}\,,\quad [Q,P] = i\,.
\end{equation}
Expanding the $U(1)$ current as
\begin{equation}
    J(\phi) = \frac{1}{2\pi}\sum_{n=-\infty}^{\infty} J_n\,e^{in\phi}
    \label{eq:J_mode_expansion}
\end{equation}
and using \eqref{eq:Pi_J_relation}, we obtain\footnote{The minus sign in the first definition of \eqref{eq:Pi_J_relation} is needed for this equation to hold.}
\begin{equation}
    J_0 = \sqrt{\pi}\,P\,,\quad J_n = \sqrt{2\pi}\times\begin{dcases}
    -i\sqrt{n}\,a_{n}\,,\quad &n\geq 1\\
    i\sqrt{-n}\,a_{-n}^{\dagger}\,,\quad &n\leq -1
    \end{dcases}\,.
\end{equation}
We see that $J_n^{\dagger} = J_{-n}$ and that $J_n$ with negative $n$ are creation operators as usual.

\paragraph{Stress tensor from the $U(1)$ current algebra.} Substituting the mode expansion \eqref{eq:J_mode_expansion} to the commutator \eqref{eq:U1_current_algebra}, we obtain
\begin{equation}
    [J_n,J_m] = 2\pi n\,\delta_{n,-m}\,.
    \label{eq:current_mode_expansion}
\end{equation}
Defining the normal-ordered product
\begin{equation}
    \mathcal{N}\{J_nJ_m\} \equiv J_nJ_m\,\Theta(-n) + J_mJ_n\,\Theta(n)\,,
\end{equation}
we can define Virasoro generators
\begin{equation}
    L_n \equiv \frac{1}{2\pi}\frac{1}{2}\sum_{m=-\infty}^{\infty}\mathcal{N}\{J_{n-m}J_m\}-\frac{1}{24}\,\delta_{n,0} \,,
\end{equation}
which satisfy the Virasoro algebra \eqref{eq:stresstensorexp} with central charge $c = 1$ \cite{Goddard:1986ee}. Therefore the stress tensor operator \eqref{eq:stresstensorexp} becomes
\begin{equation}
    T(\phi)  = \frac{1}{(2\pi)^2}\frac{1}{2}\sum_{n,m=-\infty}^{\infty}\mathcal{N}\{J_{n}J_m\}\,e^{i(n+m)\phi}-\frac{1}{2\pi}\frac{1}{24}\,,
    \label{eq:T_normal_ordering}
\end{equation}
where we have renamed the summation variable $n\rightarrow n+m$.

We will now consider the anti-commutator of the currents. In terms of the normal-ordered product, we obtain
\begin{equation}
    [J_n,J_m]_+  = 2\pi\,\vert n \vert\,\delta_{n,-m}+2\,\mathcal{N}\{J_nJ_m\}\,,
    \label{eq:mode_anti_commutator}
\end{equation}
where we have used \eqref{eq:current_mode_expansion}. Thus using the mode expansion \eqref{eq:J_mode_expansion} and \eqref{eq:mode_anti_commutator}, we obtain
\begin{equation}
    \frac{1}{4}\,[J(\phi_1),J(\phi_2)]_+ = \frac{1}{4\pi}\frac{1}{2}\sum_{n=-\infty}^\infty \vert n \vert\,e^{i n(\phi_1-\phi_2)}  + \frac{1}{(2\pi)^2}\frac{1}{2}\sum_{n,m=-\infty}^{\infty}\mathcal{N}\{J_{n}J_m\}\,e^{i n\phi_1+im\phi_2}\,.
    \label{eq:anti_comm_midstep}
\end{equation}
The Fourier series here converges
\begin{equation}
   \frac{1}{2}\sum_{n=-\infty}^\infty \vert n \vert\,e^{i n\phi} = -\frac{1}{4\sin^{2}(\frac{\phi}{2})} \equiv \partial_{\phi_1}\mathcal{P}\left(\frac{1}{2\tan\bigl(\frac{\phi}{2}\bigr)}\right) \,,
    \label{eq:Fourier_expansion_absolute_n}
\end{equation}
where $\mathcal{P}$ denotes the Cauchy principal value. Expanding the last term of \eqref{eq:anti_comm_midstep} in powers of $\phi_1-\phi_2$, we obtain
\begin{equation}
    \frac{1}{4}\,[J(\phi_1),J(\phi_2)]_+ = -\frac{1}{4\pi}\frac{1}{4\sin^{2}(\frac{\phi_1-\phi_2}{2})} + T(\phi_1)+\frac{1}{2\pi}\frac{1}{24} + \mathcal{O}(\phi_1-\phi_2)\,,\quad \phi_2\rightarrow \phi_1\,,
    \label{eq:anti_commutator_expansion}
\end{equation}
where we have used \eqref{eq:T_normal_ordering}. This allows us to extract the stress tensor via point-splitting as
\begin{equation}
	T(\phi) = \lim_{\epsilon\rightarrow 0}\biggl(\frac{1}{4}\,[J(\phi+\epsilon),J(\phi)]_++\frac{1}{4\pi\epsilon^{2}}\biggr)\,,
    \label{eq:T_from_point_splitting}
\end{equation}
where we have used that
\begin{equation}
	\lim_{\epsilon\rightarrow 0}\biggl(-\frac{1}{4 \pi}\frac{1}{4\sin^{2}(\frac{\epsilon}{2})}+\frac{1}{4\pi\epsilon^{2}}\biggr) =-\frac{1}{2\pi}\frac{1}{24}\,.
\end{equation}

\paragraph{Point-splitting in space.} On-shell, the classical stress tensor is $T_{\pm\pm}^{\text{cl}} = (\partial_\pm\Phi_\pm)^2$ which in terms of the conjugate momentum gives
\begin{equation}
		T_{--}^{\text{cl}}(\phi,t) = \Pi_-(F_t(\phi))^2\,,\quad T_{++}^{\text{cl}}(\phi,t) = \Pi_+(\overbar{F}_t(\phi))^2\,.
\end{equation}
Replacing $\Pi_-$ with the Heisenberg picture operator is not well defined due to the singularities in the operator product. Thus we will perform point-splitting by replacing the product with one-half times the anti-commutator (the one-half ensures it reduces to the classical product if the operators were c-numbers) and by displacing one of the factors in space $\phi$. Namely, we define the renormalized operator
\begin{equation}
    F_t'(\phi)^2\,T_{--}^{\text{F}}(\phi,t) \equiv \lim_{\epsilon\rightarrow 0}\biggl[\frac{1}{2}\,F_t'(\phi+\epsilon)\,F_t'(\phi)\,[\Pi_-(F_t(\phi+\epsilon)),\Pi_-(F_t(\phi))]_++\frac{1}{4\pi\epsilon^{2}}\biggr]\,,
\label{eq:point_splitting_app}
\end{equation}
and similarly definition for $\overbar{F}_t'^2\,T_{++}^{\text{F}}$ in terms of $\Pi_+$ and $\overbar{F}_t$. The usage of the anti-commutator ensures that the definition is Hermitian. Using \eqref{eq:Pi_J_relation}, we obtain
\begin{equation}
    F_t'(\phi)^2\,T_{--}^{\text{F}}(\phi,t) = \lim_{\epsilon\rightarrow 0}\biggl[\frac{1}{4}\,F_t'(\phi+\epsilon)\,F_t'(\phi)\,[J(F_t(\phi+\epsilon)),J(F_t(\phi))]_+\otimes \mathbf{1}+\frac{1}{4\pi\epsilon^{2}}\biggr]\,.
\end{equation}
Substituting the anti-commutator \eqref{eq:anti_commutator_expansion} and using
\begin{equation}
    \lim_{\epsilon\rightarrow 0}\biggl[-\frac{1}{4\pi}\frac{F_t'(\phi+\epsilon)\,F_t'(\phi)}{4\sin^{2}{\bigl(\frac{F_t(\phi+\epsilon)-F_t(\phi)}{2}\bigr)}}+\frac{1}{4\pi\epsilon^2}\biggr] = -\frac{1}{2\pi}\frac{1}{24}\,F_t'(\phi)^2 - \frac{1}{24\pi}\{F_t(\phi),\phi\}\,,
\end{equation}
we obtain
\begin{equation}
    F_t'(\phi)^2\,T_{--}^{\text{F}}(\phi,t) = F_t'(\phi)^2\,T(F_t(\phi))\otimes \mathbf{1} - \frac{1}{24\pi}\{F_t(\phi),\phi\}\,.
\end{equation}
This matches with \eqref{eq:chirally_split_stress_tensor} for $c = 1$.

\section{Heisenberg picture tensor fields}\label{app:Heisenberg_tensors}

In this appendix, we give a general description of the quantization of classical tensor fields. Let us consider a rank-$n$ tensor field $\mathcal{O}_{a_1\ldots a_n}$ which is composite field constructed from $\Phi$ and possibly from the background metric $g$.\footnote{Or more generally, a tensor field of an interacting theory which does not have an underlying free field description with a fundamental field.}. Under $\psi = (D,\chi) \in \text{Diff}\ltimes \text{Weyl}$, it transforms as
\begin{equation}
    (\psi \mathcal{O})_{a_1\ldots a_n}(x) =  e^{-\Delta_{\text{w}}\chi(D(x))}\,\frac{\partial D^{b_1}}{\partial x^{a_1}}\cdots\frac{\partial D^{b_n}}{\partial x^{a_n}}\,\mathcal{O}_{b_1\ldots b_n}(D(x))
    \label{eq:O_transformation}
\end{equation}
where $\Delta_{\text{w}}$ is the Weyl weight of the field. In light-ray coordinates, the metric $g$ is conformally flat with Weyl factor $e^{2\sigma}$. It follows that the components $\mathcal{O}_{-\ldots- + \ldots +}$ of the tensor field in light-ray coordinates are given by
\begin{equation}
	\mathcal{O}_{-\ldots- + \ldots +}(x^-,x^+) = e^{-\Delta_{\text{w}}\sigma(x^-,x^+)}\,\mathcal{O}^{\eta}_{-\ldots- + \ldots +}(x^-,x^+)\,,
	\label{eq:differentweylframes}
\end{equation}
where $\mathcal{O}^{\eta}_{-\ldots- + \ldots +}$ are the components in the flat background metric $\eta$. Here we have assumed that the tensor is completely symmetric to simplify the notation, but the analysis is valid for arbitrary tensor fields.

Upon quantization $\mathcal{O}^{\eta}$ is promoted to an operator that transforms appropriately under conformal isometries of $\eta$. Let $\psi_{\mathcal{C}}= (\mathcal{C},\omega_\mathcal{C})\in\text{Diff}\ltimes \text{Weyl}$ be a conformal isometry of the flat metric $\eta$ defined by the equation $\psi_{\mathcal{C}} \eta = \mathcal{C}^*(e^{2\omega_{\mathcal{C}}}\eta) = \eta $. Then explicitly $\mathcal{C}(x^-,x^+) = (C(x^-),\overbar{C}(x^+))$ and $e^{-2\omega_\mathcal{C}(C(x^-),\overbar{C}(x^+))} = C'(x^-)\,\overbar{C}'(x^+) $ so that using $\frac{\partial \mathcal{C}^{a}}{\partial x^{b}} = \partial_{\pm}C_{\pm}(x^{\pm})\, \delta^{a}_{\pm}\,\delta^{\pm}_{b}$ in equation \eqref{eq:O_transformation}, we obtain that the components transform as
\begin{equation}
    (\psi_{\mathcal{C}}\,\mathcal{O})_{-\ldots -+\ldots+}^{\eta}(x^-,x^+) =C_{-}'(x^{-})^{h}\,C_{+}'(x^{+})^{\overbar{h}}\,\mathcal{O}_{-\ldots -+\ldots+}^{\eta}(C_-(x^-),C_+(x^+))\,,
    \label{eq:conformal_isometry_flat_space}
\end{equation}
where the weights
\begin{equation}
    h = \frac{\Delta_{\text{w}}}{2} + n_-\,,\quad \overbar{h} = \frac{\Delta_{\text{w}}}{2} + n_+
\end{equation}
and $n_\pm$ denote the number of $\pm$ components such that $n_- + n_+ = n$. Often, one introduces the scaling dimension $\Delta \equiv h + \overbar{h} = \Delta_{\text{w}} + n$ and the spin $s \equiv h - \overbar{h} = n_--n_+$ of the field component. Only for a scalar field $n = 0$ the scaling dimension coincides with the Weyl weight $\Delta = \Delta_{\text{w}}$.

The transformation \eqref{eq:conformal_isometry_flat_space} implies that $\mathcal{O}_{-\ldots -+\ldots+}^{\eta}$ is promoted to an operator as
\begin{equation}
    \mathcal{O}_{-\ldots -+\ldots+}^{\eta}(x^-,x^+) = \mathcal{O}_{h}(x^-)\otimes \mathcal{O}_{\overbar{h}}(x^+)\,,
\end{equation}
where $\mathcal{O}_{h}(\phi)$ is a chiral primary operator defined by its transformation under the adjoint action of the Virasoro group
\begin{equation}
	V_F\,\mathcal{O}_{h}(\phi)\,V_F^\dagger = F'(\phi)^{h}\,\mathcal{O}_{h}(F(\phi))\,,\quad \overbar{\mathcal{O}}_{h}(\phi) \equiv \mathcal{O}_{h}(-\phi)\,,
	\label{eq:primary_transformation_law}
\end{equation}
with the normalization $\mathcal{O}_{0}(\phi) = \mathbf{1}$.

It follows that conformal isometries $(C,\overbar{C})$ \eqref{eq:conformal_isometry_flat_space} of the flat metric are implemented correctly by the adjoint action of the Virasoro group element $V_C\otimes \overbar{V}_{\overbar{C}}$ as
\begin{equation}
	(\psi_{\mathcal{C}}\mathcal{O}^{\eta}_{-\ldots -+\ldots+})(x^-,x^+) = (V_C\otimes \overbar{V}_{\overbar{C}})\,\mathcal{O}^{\eta}_{-\ldots -+\ldots+}(x^-,x^+)\,(V_C\otimes \overbar{V}_{\overbar{C}})^\dagger\,.
\end{equation}
From \eqref{eq:differentweylframes}, the curved space field operator becomes
\begin{equation}
    \mathcal{O}_{-\ldots -+\ldots+}(x^-,x^+) = e^{-\Delta_{\text{w}}\omega(x^-,x^+)}\,F_t'(\phi)^{\frac{\Delta_{\text{w}}}{2}}\,\overbar{F}_t'(\phi)^{\frac{\Delta_{\text{w}}}{2}}\,\mathcal{O}_{h}(x^-)\otimes \mathcal{O}_{\overbar{h}}(x^+)\,.
    \label{eq:curved_space_field}
\end{equation}
The Heisenberg picture operator is defined as
\begin{equation}
    \mathcal{O}^{\text{H}}_{-\ldots -+\ldots+}(\phi,t) \equiv F_t'(\phi)^{n_-}\,\overbar{F}_t'(\phi)^{n_+}\,\mathcal{O}_{-\ldots -+\ldots+}(x^-,x^+)\,,
\end{equation}
which is a generalization of \eqref{eq:Heisenberg_stress_tensor} to tensor fields. Substituting \eqref{eq:curved_space_field}, we obtain
\begin{equation}    
\mathcal{O}^{\text{H}}_{-\ldots -+\ldots+}(\phi,t) = e^{-\Delta_{\text{w}}\omega(\phi,t)}\,F_t'(\phi)^{h}\,\overbar{F}_t'(\phi)^{\overbar{h}}\,\mathcal{O}_{h}(F_t(\phi))\otimes \mathcal{O}_{\overbar{h}}(\overbar{F}_t(\phi))
\label{eq:Heisenberg_tensor_field}
\end{equation}
so that the Schrödinger operator is
\begin{equation}
    \mathcal{O}^{\text{S}}_{-\ldots -+\ldots+}(\phi,t) = e^{-\Delta_{\text{w}}\omega(\phi,t)}\,\mathcal{O}_{h}(\phi)\otimes \mathcal{O}_{\overbar{h}}(\phi)\,,
\end{equation}
which contains explicit time-dependence via the Weyl factor $\omega(\phi,t)$ due to the non-zero Weyl weight.

The above analysis contains the conjugate momentum $\Pi_{\text{H}}$ \eqref{eq:conjugate_momentum_operator_text} of the free scalar field and the operator (non-Schwarzian part) of the stress tensor $T_{\pm\pm}^{\text{H}}$ as special cases of vanishing Weyl weight $\Delta_{\text{w}} = 0$. The conjugate momentum $\Pi_{\text{H}}$ is a difference of two $n = 1$ fields with $(n_-,n_+) = (1,0)$ and $(n_-,n_+) = (0,1)$ while the stress tensor components $T_{\pm\pm}^{\text{H}}$ \eqref{eq:Heisenberg_stress_tensor} have $(n_-,n_+) = (2,0)$ and $(n_-,n_+) = (0,2) $ respectively.

\section{Computation of regularized replica path integrals}\label{app:regularized_liouville}

In this appendix, we compute ratios of regularized Euclidean path integrals over the replica manifold and study its transformation under diffeomorphisms and Weyl transformations of the metric. The necessary preliminaries and notation for the calculation of the Liouville action in Appendix \ref{subapp:reg_liouville_calculation} are first given in Appendix \ref{eq:simple_example}. The calculation of Appendix \ref{subapp:reg_liouville_calculation} is essentially a review of the results presented in \cite{Estienne:2025uhh} applied to our setup. The $\text{Diff}\ltimes \text{Weyl}$ transformation properties are considered in Appendix \ref{subapp:Liouville_wards}.

\subsection{Simple illustrative example: area of a replicated disk}\label{eq:simple_example}

To illustrate subtleties in the replica calculation and the associated notation, we will first consider a simple illustrative example: the calculation of the area of the $n$-fold cover of flat the unit disk branched at the origin. In the next section, we will consider the more complicated case of a Liouville action.

Let us consider the Riemannian manifold $(\mathbb{D},\eta)$ which is the open unit disk $\mathbb{D} = \{ w = r e^{i\theta}\in \mathbb{C}\,\vert\, 0 \leq r < 1\,, \theta \in [0,2\pi)\}$ in the complex plane equipped with the flat (Euclidean) metric\footnote{In this appendix, we drop the subscript in $g_{\text{e}}$ and $\eta_{\text{e}}$ compared to the main text so that $g$ and $\eta$ denote Euclidean metrics.\label{note:dropping_e}}
\begin{equation}
    \eta_{ab}(x)\,dx^adx^b = dwd\overbar{w} = dr^2 + r^2d\theta^2\,.
\end{equation}
The area of the disk is given by
\begin{equation}
    \vol{(\mathbb{D},\eta)} = \int_{\mathbb{D}}d^2x\,\sqrt{\eta} = \int_{0}^{2\pi} d\theta \int_0^1 dr\,r = \pi\,.
\end{equation}
Let us then consider the replicated flat disk modeled as the Riemannian manifold $(\mathbb{D},\mathcal{P}^*\eta)$ where the ramified covering map $\mathcal{P}\colon \mathbb{D}\rightarrow \mathbb{D}$ is given by
\begin{equation}
    \mathcal{P}^w(w,\overbar{w}) = P(w) = w^n\,,\quad \mathcal{P}^{\overbar{w}}(w,\overbar{w}) = \overbar{P}(\overbar{w}) = \overbar{w}^n\,.
    \label{eq:Ptilde_app}
\end{equation}
As a result, $(\mathbb{D},\mathcal{P}^*\eta)$ is an $n$-fold cover of $(\mathbb{D},\eta)$ branched at the origin $w = 0$. Notice that the ramified map $\mathcal{P}$ is $n$-to-$1$, because, for example, all of the $n$ points $ re^{\frac{2\pi i k}{n}}$ with $k = 1,2,\ldots, n$ are mapped to the same point $r^n e^{2\pi ik} = r^n$.

The volume of the replicated flat disk $(\mathbb{D},\mathcal{P}^*\eta)$ is given by
\begin{equation}
    \vol{(\mathbb{D},\mathcal{P}^*\eta)} =\int_{\mathbb{D}}d^2x\,\sqrt{\mathcal{P}^*\eta}\,.
    \label{eq:volume_replicated_disk}
\end{equation}
The pull-back of the flat metric with \eqref{eq:simple_example} is given by
\begin{equation}
    (\mathcal{P}^*\eta)_{ab}\,dx^adx^b = P'(w)\,\overbar{P}'(\overbar{w})\,dwd\overbar{w} = n^2\,\vert w\vert^{n-1}\,dwd\overbar{w} =  n^2r^{2(n-1)}\,(dr^2 + r^2d\theta^2)\,.
\end{equation}
Therefore we obtain
\begin{equation}
    \vol{(\mathbb{D},\mathcal{P}^*\eta)} = \int_{0}^{2\pi} d\theta \int_0^1 dr\,r \,n^2r^{2(n-1)} = 2\pi n^2\,\int_0^1 dr\,r^{2n-1} = n \pi\,,
\end{equation}
which gives the expected answer
\begin{equation}
    \vol{(\mathbb{D},\mathcal{P}^*\eta)} = n\vol{(\mathbb{D},\eta)}\,.
    \label{eq:replica_volume}
\end{equation}
We want to emphasize that in the integral \eqref{eq:volume_replicated_disk} the integration domain is the unit disk on the complex plane. In other words, the integral over the replicated disk is \textit{not} performed over the domain $\theta \in (0,2\pi n)$, but only over $\theta \in (0,2\pi)$. The factor of $n$ in the final result \eqref{eq:replica_volume} comes from the Jacobian determinant of the ramified map. The geometric interpretation is that the disk $\mathbb{D}$ equipped with the metric $\mathcal{P}^*\eta$ when embedded in Euclidean space is an $n$-sheeted surface.

There is a second way to obtain the relation \eqref{eq:replica_volume} involving the factor of $n$ as follows. First we write
\begin{equation}
    \vol{(\mathbb{D},\mathcal{P}^*\eta)} =\int_{\mathbb{D}}d^2x\,\sqrt{\mathcal{P}^*\eta} = \int_{\mathbb{D}}d^2x\sqrt{\eta}\det{\partial_a \mathcal{P}^b}\,.
\end{equation}
Performing a change of integration variables $x \rightarrow \mathcal{P}(x)$, we obtain
\begin{equation}
    \vol{(\mathbb{D},\mathcal{P}^*\eta)}  = \int_{\mathcal{P}^{-1}(\mathbb{D})}d^2x\,\sqrt{\eta} = \int_{\mathcal{P}^{-1}(\mathbb{D})}d\theta dr\,r\,.
\end{equation}
Now since $\mathbb{D}$ is covered by $\theta \in (0,2\pi) $, we see that the preimage $\mathcal{P}^{-1}(\mathbb{D}) = \mathbb{D}\sqcup \ldots \sqcup \mathbb{D}$, which is a disjoint union of $n$ disks, is covered by $\theta \in (0,2\pi n)$. Therefore we recover \eqref{eq:replica_volume}:
\begin{equation}
    \vol{(\mathbb{D},\mathcal{P}^*\eta)}  = \int_{0}^{2\pi n} d\theta \int_0^1 dr\,r = n\pi\,.
\end{equation}

\subsection{Regularized Liouville action of the replica manifold}\label{subapp:reg_liouville_calculation}

In this appendix, we compute ratios of Euclidean path integrals of the replica manifold modeled as $(\mathbb{R}^2,\mathcal{P}^* g)$ and it is a branched cover of the plane $(\mathbb{R}^2,g)$ equipped with the Euclidean (see footnote \ref{note:dropping_e}) metric $g$. The replica manifold is defined by the ramified map $\mathcal{P}\colon \mathbb{R}^2\rightarrow  \mathbb{R}^2$, and assuming it is branched at the origin, $\mathcal{P}$ is given by \eqref{eq:Ptilde_app}.

We consider both a curved $g$ and a flat $\eta$ metric on $\mathbb{R}^2$. Explicitly
\begin{equation}
    g_{ab}\,dx^{a}dx^{b} = e^{2\sigma(w,\overbar{w})}\,\eta_{ab}\,dx^{a}dx^{b}\,,\quad   \eta_{ab}\,dx^{a}dx^{b} = dwd\overbar{w}\,.
\end{equation}
Their pull-backs $\widetilde{g} \equiv \mathcal{P}^*g$ and $\widetilde{\eta} \equiv \mathcal{P}^*\eta$ are given by
\begin{equation}
	\widetilde{g}_{ab}\,dx^{a}dx^{b} = e^{2\widetilde{\sigma}(w,\overbar{w})}\,\eta_{ab}\,dx^{a}dx^{b}\,,\quad   \widetilde{\eta}_{ab}\,dx^{a}dx^{b} = e^{2\kappa(w,\overbar{w})}\,dwd\overbar{w}\,,
    \label{eq:pull_back_metrics}
\end{equation}
where the Weyl factors are
\begin{gather}
\widetilde{\sigma}(w,\overbar{w}) = \kappa(w,\overbar{w})+ \Sigma(w,\overbar{w})\,,\\
    \kappa(w,\overbar{w}) \equiv \frac{1}{2}\log{(P'(w)\,\overbar{P}'(\overbar{w}))}\,,\quad \Sigma(w,\overbar{w}) \equiv \sigma(P(w),\overbar{P}(\overbar{w}))\,. 
    \label{eq:sigmatilde_split}
\end{gather}

\paragraph{Conical singularity.} The replica manifold, modeled as $(\mathbb{R}^2,\mathcal{P}^*g)$, is an $n$-fold branched cover of $(\mathbb{R}^2,g)$. The Weyl factor $\widetilde{\sigma}$ \eqref{eq:sigmatilde_split} has a logarithmic singularity at $w = 0$ of the form
\begin{equation}
	\widetilde{\sigma}(w,\overbar{w}) = \frac{\gamma}{2} \log{\vert w\vert^2} + \sigma(0)+\log{(\gamma+1)}+\mathcal{O}(w^{\gamma+1})+\mathcal{O}(\overbar{w}^{\gamma+1})\,,\quad w,\overbar{w}\rightarrow 0\,,
    \label{eq:sigmatilde}
\end{equation}
where we have defined $\gamma \equiv n-1$. In the metric $\widetilde{g}$, the logarithmic singularity corresponds to a conical singularity of excess angle $2\pi (n-1)$ at $w = 0$ which reflects the fact that $M_n$ is an $n$-fold cover of $\mathbb{R}^2$. Let us show this in detail.

Near $w = 0$, the metric has the expansion
\begin{equation}
    \widetilde{g}_{ab}\,dx^{a}dx^{b}  = (\gamma+1)^{2}\,\vert w\vert^{2\gamma}\,e^{2\sigma(0)}\,dwd\overbar{w}+\ldots\,,\quad w,\overbar{w}\rightarrow 0\,.
\end{equation}
Introducing polar coordinates $(r,\theta) \in (0,\infty)\times (0,2\pi)$ via $w = r e^{i\theta}$ and $\overbar{w} =r e^{-i\theta} $, we obtain
\begin{equation}
	\widetilde{g}_{ab}\,dx^{a}dx^{b} = (\gamma+1)^{2}\,r^{2\gamma}\,e^{2\sigma(0)}\,(dr^2 + r^{2}d\theta^{2}) + \ldots\,,\quad r \rightarrow 0\,.
\end{equation}
Defining a new set of polar coordinates
\begin{equation}
	\tilde{r} = e^{\sigma(0)}\,r^{\gamma+1}\,,\quad \tilde{\theta} = (\gamma+1)\,\theta \,,
    \label{eq:conical_flat_coordinates}
\end{equation}
the metric takes the flat form near the origin
\begin{equation}	
	\widetilde{g}_{ab}\,dx^{a}dx^{b} =d\tilde{r}^2 + \tilde{r}^{2}d\tilde{\theta}^{2} + \ldots\,.
\end{equation}
Since $\theta \in (0,2\pi)$, we have $\tilde{\theta} \in (0,2\pi\,(\gamma+1))$. Thus there is a conical singularity of excess angle $2\pi\,(\gamma+1)-2\pi  = 2\pi \gamma = 2\pi\,(n-1)\geq 0$ for $n\geq 1$.

\paragraph{Computation of regularized Liouville actions.} As explained in the main text, ratios of Euclidean path integrals in metrics $\widetilde{g} = e^{2\widetilde{\sigma}}\eta$,  $\widetilde{\eta} = e^{2\kappa}\eta$ and $g = e^{2\sigma}\eta$ are given by
\begin{equation}
    \frac{Z_{\text{D}}^{\text{reg}}[\widetilde{g};\mathbb{R}^2]}{Z_{\text{D}}^{\text{reg}}[\eta;\mathbb{R}^2]} = e^{I_{\text{Lio}}^{\text{reg}}[\widetilde{\sigma},\eta;\mathbb{R}^2]}\,,\quad \frac{Z_{\text{D}}^{\text{reg}}[\widetilde{\eta};\mathbb{R}^2]}{Z_{\text{D}}^{\text{reg}}[\eta;\mathbb{R}^2]} =e^{I_{\text{Lio}}^{\text{reg}}[\kappa,\eta;\mathbb{R}^2]}\,,\quad \frac{Z_{\text{D}}^{\text{reg}}[g;\mathbb{R}^2]}{Z_{\text{D}}^{\text{reg}}[\eta;\mathbb{R}^2]} = e^{I_{\text{Lio}}^{\text{reg}}[\sigma,\eta;\mathbb{R}^2]}\,.
\end{equation}
It follows that
\begin{equation}
    \frac{Z_{\text{D}}^{\text{reg}}[\widetilde{g};\mathbb{R}^2]}{Z_{\text{D}}^{\text{reg}}[g;\mathbb{R}^2]^n} = e^{I_{\text{Lio}}^{\text{reg}}[\widetilde{\sigma},\eta;\mathbb{R}^2]-I_{\text{Lio}}^{\text{reg}}[\kappa,\eta;\mathbb{R}^2]-nI_{\text{Lio}}^{\text{reg}}[\sigma,\eta;\mathbb{R}^2]}\,\frac{Z_{\text{D}}^{\text{reg}}[\widetilde{\eta};\mathbb{R}^2]}{Z_{\text{D}}^{\text{reg}}[\eta;\mathbb{R}^2]^n}\,.
    \label{eq:ratio_transformation}
\end{equation}
In all path integrals and Liouville actions in this expression, the cut-off radius is the same and equal to $\tilde{\delta}$. We will compute the combination of Liouville actions appearing here in the exponent when the ramified map is given by \eqref{eq:Ptilde_app} with branch points at $w = 0$ and at $w = \infty$. The generalization to any number of branch points in locations $w = w_i$ is eminent.

First, we compute the regularized Liouville action of $\widetilde{\sigma}$ in the flat metric which is
\begin{equation}
	I_{\text{Lio}}^{\text{reg}}[\widetilde{\sigma},\eta;\mathbb{R}^2] = \frac{c}{24\pi}\int_{\mathbb{R}^2\,\backslash  B_{\tilde{\delta}}(0;\widetilde{g})} d^2x\sqrt{\eta}\left( \eta^{ab}\,\partial_a \widetilde{\sigma}\,\partial_b\widetilde{\sigma} +\widetilde{\sigma} R_\eta \right),
    \label{eq:regularized_liouville_app}
\end{equation}
where the Ricci scalar vanishes $R_{\eta} = 0$. The geodesic disk is explicitly defined as
\begin{equation}
    B_{\delta_i}(w_i;g) \equiv \{(w,\overbar{w})\in \mathbb{R}^2\,\vert\,L_g(w_i,\overbar{w}_i;w,\overbar{w}) < \delta_i \}\,,
\end{equation}
where $L_g(w_1,\overbar{w}_1;w_2,\overbar{w}_2)$ is the geodesic distance between the points $(w_1,\overbar{w}_1)$ and $(w_2,\overbar{w}_2)$ in the metric $g$. The geodesic distance from the origin in the metric $\widetilde{g}$ is measured by the coordinate $\tilde{r}$ in \eqref{eq:conical_flat_coordinates} so that the boundary $\partial B_{\tilde{\delta}}(0;\widetilde{g})$ is located at $\tilde{r} = \tilde{\delta}$. This corresponds to
\begin{equation}
    r = \varepsilon \equiv e^{-\frac{\sigma(0)}{\gamma+1}}\,\tilde{\delta}^{\frac{1}{\gamma+1}}
    \label{eq:varepsilon}
\end{equation}
in $(r,\theta)$ coordinates. Notice that the integration domain is $\mathbb{R}^2\,\backslash  B_{\tilde{\delta}}(0;\widetilde{g})$, which in $(r,\theta)$ coordinates corresponds to $r>\varepsilon$ and $0 < \theta < 2\pi$ (not to $0<\theta<2\pi n$). This subtlety with the integration domain was illustrated in the simple example of Appendix \ref{eq:simple_example}.

Integrating the bulk integral by parts, the action \eqref{eq:regularized_liouville_app} can be written as
\begin{equation}
	I_{\text{Lio}}^{\text{reg}}[\widetilde{\sigma},\eta;\mathbb{R}^2] = -\frac{c}{24\pi}\int_{\mathbb{R}^2\,\backslash  B_{\tilde{\delta}}(0;\widetilde{g})} d^2x\sqrt{\eta}\, \widetilde{\sigma}\,\nabla^2\widetilde{\sigma} +\frac{c}{24\pi}\int_{\partial B_{\tilde{\delta}}(0;\widetilde{g})}dx\sqrt{h}\,n^a\,\widetilde{\sigma}\,\partial_a\widetilde{\sigma}\,,
    \label{eq:Liouville_partially_integrated_0}
\end{equation}
where $\nabla_a$ is the covariant derivative of $\eta$, $n^a$ is the outward-pointing unit normal vector and $h$ is the induced metric of the boundary. In coordinates $(r,\theta)$, the boundary is at $r = \varepsilon$ and we have
\begin{equation}
    \sqrt{h} = \varepsilon + \mathcal{O}(\varepsilon^2)\,,\quad n^a = -\delta^a_r+\mathcal{O}(\varepsilon^2)\,,\quad \varepsilon\rightarrow 0\,.
    \label{eq:sqrth_n_app}
\end{equation}
The minus sign in the normal vector is due to the fact that the integration region domain is $r>\varepsilon$ so that $n^a$ points towards the origin. Substituting \eqref{eq:sigmatilde}, the boundary term in \eqref{eq:Liouville_partially_integrated_0} is explicitly
\begin{align}
    \int_{\partial B_{\tilde{\delta}}(0;\widetilde{g})}dx\sqrt{h}\,n^a\,\widetilde{\sigma}\,\partial_a\widetilde{\sigma}
    &= -(\gamma\log{\varepsilon}+ \log{(\gamma+1})+\sigma(0)  + \mathcal{O}(\varepsilon^{\gamma+1}))\nonumber\\
    &\times\int_0^{2\pi} d\theta \,\varepsilon\,\partial_r(\gamma\log{r} + \log{(\gamma+1})+\sigma(0)+ \mathcal{O}(r^{\gamma+1}))\vert_{r = \varepsilon}\nonumber\\
    &= -2\pi\,(\gamma^2\log{\varepsilon} + \gamma\log{(\gamma+1)}+\gamma\, \sigma(0)+ \mathcal{O}(\varepsilon^{\gamma+1}))\,.
\end{align}
Substituting \eqref{eq:varepsilon}, we obtain
\begin{align}
	&I_{\text{Lio}}^{\text{reg}}[\widetilde{\sigma},\eta;\mathbb{R}^2]\label{eq:Liouville_partially_integrated}\\
    &= -\frac{c}{24\pi}\int_{\mathbb{R}^2\,\backslash  B_{\tilde{\delta}}(0;\widetilde{g})} d^2x\sqrt{\eta}\, \widetilde{\sigma}\,\nabla^2\widetilde{\sigma}-\frac{c}{12}\biggl[\frac{\gamma^2}{\gamma+1}\log{\tilde{\delta}}+\frac{\gamma}{\gamma+1}\,\sigma(0) + \gamma\log{(\gamma+1})+ \mathcal{O}(\tilde{\delta})\biggr]\,.\nonumber
\end{align}
This matches with \cite{Estienne:2025uhh} once the logarithmic divergence is subtracted with a counterterm.

Since $\kappa = \widetilde{\sigma}\vert_{\sigma = 0}$ and $\widetilde{\eta} = \widetilde{g}\vert_{\sigma = 0} $, we obtain from \eqref{eq:Liouville_partially_integrated} that
\begin{align}
	&I_{\text{Lio}}^{\text{reg}}[\kappa,\eta;\mathbb{R}^2]\label{eq:Liouville_partially_integrated_2}\\
    &= -\frac{c}{24\pi}\int_{\mathbb{R}^2\,\backslash  B_{\tilde{\delta}}(0;\widetilde{\eta})} d^2x\sqrt{\eta}\, \kappa\,\nabla^2\kappa-\frac{c}{12}\biggl[\frac{\gamma^2}{\gamma+1}\log{\tilde{\delta}} + \gamma\log{(\gamma+1})+ \mathcal{O}(\tilde{\delta})\biggr]\,.\nonumber
\end{align}
In the first term, we can replace $B_{\tilde{\delta}}(0;\widetilde{\eta})$ with $B_{\tilde{\delta}}(0;\widetilde{g})$ since their difference vanishes in the $\tilde{\delta}\rightarrow 0$ limit. What remains is to compute
\begin{align}
    I_{\text{Lio}}^{\text{reg}}[\Sigma,\widetilde{\eta};\mathbb{R}^2] &= \frac{c}{24\pi}\int_{\mathbb{R}^2\,\backslash  B_{\tilde{\delta}}(0;\widetilde{g})} d^2x\sqrt{\eta}\left( \eta^{ab}\,\partial_a \Sigma\,\partial_b\Sigma -2\,\Sigma\,\nabla^2\kappa \right)\\
    &= -\frac{c}{24\pi}\int_{\mathbb{R}^2\,\backslash  B_{\tilde{\delta}}(0;\widetilde{g})} d^2x\sqrt{\eta}\left( \Sigma\,\nabla^2\Sigma +2\,\Sigma\,\nabla^2\kappa \right),
    \label{eq:remaining}
\end{align}
where we have used $\sqrt{\widetilde{\eta}}\,\widetilde{\eta}^{ab} = \sqrt{\eta}\,\eta^{ab}$, $\sqrt{\widetilde{\eta}}\,R_{\widetilde{\eta}} = \sqrt{\eta}\,(R_{\eta}-2\,\nabla^2\kappa) =-\sqrt{\eta}\,2\,\nabla^2\kappa $, $\nabla_a$ is associated with $\eta$ and that the boundary term $n^a\,\Sigma\,\partial_a\Sigma\vert_{\partial B_{\tilde{\delta}}(0;\widetilde{g})}$ vanishes, because $\Sigma$ \eqref{eq:sigmatilde_split} does not contain logarithmic singularities. We also used that the geodesic disk is in the metric $e^{2\Sigma}\,\widetilde{\eta} = \widetilde{g}$.

Combining \eqref{eq:Liouville_partially_integrated}, \eqref{eq:Liouville_partially_integrated_2} and \eqref{eq:remaining}, we finally obtain
\begin{align}
    &I_{\text{Lio}}^{\text{reg}}[\widetilde{\sigma},\eta;\mathbb{R}^2] -I_{\text{Lio}}^{\text{reg}}[\kappa,\eta;\mathbb{R}^2]- I_{\text{Lio}}^{\text{reg}}[\Sigma,\widetilde{\eta};\mathbb{R}^2]\\
    &= -\frac{c}{12}\frac{\gamma}{\gamma+1}\,\sigma(0) +\frac{c}{24\pi}\int_{\mathbb{R}^2\,\backslash  B_{\tilde{\delta}}(0;\widetilde{g})} d^2x\sqrt{\eta}\,(\Sigma\,\nabla^2\kappa-\kappa\,\nabla^2\Sigma)\,,
    \label{eq:final_midstep}
\end{align}
where we have used that
\begin{equation}
    \widetilde{\sigma}\,\nabla^2\widetilde{\sigma}-(\Sigma\,\nabla^2\Sigma +2\,\Sigma\,\nabla^2\kappa)-\kappa\,\nabla^2\kappa = -(\Sigma\,\nabla^2\kappa-\kappa\,\nabla^2\Sigma)\,.
\end{equation}
Using Green's identity, we obtain
\begin{align}
    &\int_{\mathbb{R}^2\backslash  B_{\tilde{\delta}}(0;\widetilde{g})} d^2x\sqrt{\eta}\,(\Sigma\,\nabla^2\kappa-\kappa\,\nabla^2\Sigma)\\
    &= \int_{\partial B_{\tilde{\delta}}(0;\widetilde{g})}dx\sqrt{h}\,n^a\,(\Sigma\,\partial_a\kappa-\kappa\,\partial_a\Sigma)\nonumber\\
    &= -\int_{0}^{2\pi}d\theta\,\varepsilon\,(\Sigma\,\partial_r(\gamma\log{r}+\log{(\gamma+1)} + \ldots)-(\gamma\log{r}+\log{(\gamma+1)} + \ldots)\,\partial_r\Sigma)\vert_{r = \varepsilon}\nonumber\\
    &= -2\pi\gamma\,\sigma(0) + \mathcal{O}(\tilde{\delta})+ \mathcal{O}(\tilde{\delta}^{\frac{1}{\gamma+1}}\log{\tilde{\delta}})\,,
\end{align}
where we have used \eqref{eq:sqrth_n_app}, \eqref{eq:varepsilon} and the finite term comes only from the first term of the third line. Substituting to \eqref{eq:final_midstep} and taking the strict $\tilde{\delta}\rightarrow 0$ limit, we obtain \cite{Estienne:2025uhh}
\begin{equation}
    \lim_{\tilde{\delta}\rightarrow 0}\Bigl(I_{\text{Lio}}^{\text{reg}}[\widetilde{\sigma},\eta;\mathbb{R}^2] -I_{\text{Lio}}^{\text{reg}}[\kappa,\eta;\mathbb{R}^2]- I_{\text{Lio}}^{\text{reg}}[\Sigma,\widetilde{\eta};\mathbb{R}^2]\Bigr) = -\frac{c}{12}\frac{\gamma\,(\gamma+2)}{\gamma+1}\,\sigma(0)\,.
    \label{eq:final_result_single_point}
\end{equation}
Instead of \eqref{eq:remaining}, we can also write the third action in an alternative way starting from its defining expression
\begin{equation}
    I_{\text{Lio}}^{\text{reg}}[\Sigma,\widetilde{\eta};\mathbb{R}^2] = \frac{c}{24\pi}\int_{\mathbb{R}^2\,\backslash  B_{\tilde{\delta}}(0;\widetilde{g})} d^2x\sqrt{\widetilde{\eta}}\left( \widetilde{\eta}^{ab}\,\partial_a \Sigma\,\partial_b\Sigma +\Sigma\, R_{\widetilde{\eta}} \right).
\end{equation}
Since $\Sigma = \sigma \circ \mathcal{P}$, by a change of integration variables $y = \mathcal{P}(x)$, we obtain
\begin{equation}
    I_{\text{Lio}}^{\text{reg}}[\Sigma,\widetilde{\eta};\mathbb{R}^2] = \frac{c}{24\pi}\int_{\mathcal{P}^{-1}(\mathbb{R}^2\,\backslash  B_{\tilde{\delta}}(0;\widetilde{g}))} d^2y\sqrt{\eta}\left( \eta^{ab}\,\partial_a \sigma\,\partial_b\sigma+\sigma  R_{\eta} \right).
\end{equation}
As in the illustrative example of Appendix \ref{eq:simple_example}, $\mathcal{P}^{-1}(\mathbb{R}^2\,\backslash  B_{\tilde{\delta}}(0;\widetilde{g}))$ is a disjoint union of $n$ copies of $\mathbb{R}^2\,\backslash  B_{\tilde{\delta}}(0;\widetilde{g})$. Thus the integral gives a factor of $n$ and we obtain \cite{Estienne:2025uhh}
\begin{equation}
    I_{\text{Lio}}^{\text{reg}}[\Sigma,\widetilde{\eta};\mathbb{R}^2] = n\,I_{\text{Lio}}^{\text{reg}}[\sigma,\eta;\mathbb{R}^2]\,.
\end{equation}
As a result, \eqref{eq:final_result_single_point} gives a formula for the exponent in \eqref{eq:ratio_transformation}.

The result \eqref{eq:final_result_single_point} generalizes to any number of branch points in arbitrary locations, because the derivation above goes through in the same way with boundary terms localizes around all logarithmic singularities. The generalization is given by \cite{Estienne:2025uhh}
\begin{equation}
    \lim_{\tilde{\delta}\rightarrow 0}\Bigl(I_{\text{Lio}}^{\text{reg}}[\widetilde{\sigma},\eta;\mathbb{R}^2]-I_{\text{Lio}}^{\text{reg}}[\kappa,\eta;\mathbb{R}^2]-n\,I_{\text{Lio}}[\sigma,\eta;\mathbb{R}^2]\Bigr) = -\frac{c}{12}\sum_{i=1}^k\frac{\gamma_i\,(\gamma_i+2)}{\gamma_i+1}\,\sigma(w_i,\overbar{w}_i)\,.
    \label{eq:Liouville_combination_general_app}
\end{equation}

\subsection{$\text{Diff}\ltimes \text{Weyl}$ Ward identities}\label{subapp:Liouville_wards}

Let us consider the ratio
\begin{equation}
    \mathcal{R}_{(\mathbb{R}^2,\eta)}^{\tilde{\varepsilon}_1,\ldots,\tilde{\varepsilon}_k}(w_1,\overbar{w}_1;\ldots;w_k,\overbar{w}_k) = \frac{Z_{\text{D}}^{\text{reg}}[\mathcal{P}^*\eta;\mathbb{R}^2]}{Z_{\text{D}}^{\text{reg}}[\eta;\mathbb{R}^2]^n}\,,
\end{equation}
where $\mathcal{P}\colon \mathbb{R}^2\rightarrow \mathbb{R}^2$ has $k$ branch points at $\{w_i\}_{i=1}^k$ and the integration domains of both path integrals have holes $\mathcal{B}(w_1,\overbar{w}_1;\ldots;w_k,\overbar{w}_k) \equiv \cup_{i=1}^k B_{\tilde{\varepsilon}_i}(w_i,\mathcal{P}^*\eta)$ cut around the branch points. Thus $\tilde{\varepsilon}_i$ are the geodesic radii in the metric $\mathcal{P}^*\eta$. We will analyze how this behaves under diffeomorphisms and Weyl transformations of the flat metric $\eta$.

For Weyl transformations of the flat metric, the formula \eqref{eq:Liouville_combination_general_app} implies that \cite{Estienne:2025uhh}
\begin{equation}
    \mathcal{R}_{(\mathbb{R}^2,e^{2\chi}\eta)}^{\tilde{\varepsilon}_1,\ldots,\tilde{\varepsilon}_k}(w_1,\overbar{w}_1;\ldots;w_k,\overbar{w}_k) = e^{-\frac{c}{12}\sum_{i=1}^k \frac{\gamma_i\,(\gamma_i+2)}{\gamma_i+1}\,\chi(w_i,\overbar{w}_i)}\,\mathcal{R}_{(\mathbb{R}^2,\eta)}^{\tilde{\varepsilon}_1,\ldots,\tilde{\varepsilon}_k}(w_1,\overbar{w}_1;\ldots;w_k,\overbar{w}_k)\,.
    \label{eq:Weyl_ward_ratio_app}
\end{equation}
Because the regulated path integral is defined in a diffeomorphism invariant manner, we obtain
\begin{equation}
    \mathcal{R}_{(\mathbb{R}^2,D^*\eta)}^{\tilde{\varepsilon}_1,\ldots,\tilde{\varepsilon}_k}(w_1,\overbar{w}_1;\ldots;w_k,\overbar{w}_k) = \mathcal{R}_{(D(\mathbb{R}^2),\eta)}^{\tilde{\varepsilon}_1,\ldots,\tilde{\varepsilon}_k}(D(w_1,\overbar{w}_1);\ldots;D(w_k,\overbar{w}_k))\,.
    \label{eq:diff_ward_ratio_app}
\end{equation}
Combining the Weyl \eqref{eq:Weyl_ward_ratio_app} and diffeomorphism \eqref{eq:diff_ward_ratio_app} transformation properties we can derive conformal Ward identities. Let us consider for simplicity the case $k = 2$ relevant for the calculation of the entropy in the main text. The combined transformation in this case takes the form
\begin{align}
    &\mathcal{R}_{(\mathbb{R}^2,D^*(e^{2\chi}\eta))}^{\tilde{\varepsilon}_1,\tilde{\varepsilon}_2}(w_1,\overbar{w}_1;w_2,\overbar{w}_2)\nonumber\\
    &= e^{-\Delta_n((\chi\circ D)(w_1,\overbar{w}_1)+(\chi\circ D)(w_2,\overbar{w}_2))}\,\mathcal{R}_{(D(\mathbb{R}^2),\eta)}^{\tilde{\varepsilon}_1,\tilde{\varepsilon}_2}(D(w_1,\overbar{w}_1);D(w_2,\overbar{w}_2))\,,
    \label{eq:flat_space_transformation}
\end{align}
where $\Delta_n = \frac{c}{12}(n-\frac{1}{n})$. Consider a conformal isometry $D = \mathcal{C}$ of the flat metric
\begin{equation}
    \mathcal{C}(w,\overbar{w}) = (C(w),\overbar{C}(\overbar{w}))\,,\quad \chi(C(w),\overbar{C}(\overbar{w})) = -\frac{1}{2}\log{C'(w)}-\frac{1}{2}\log{C'(\overbar{w})}\,,
\end{equation}
which satisfy
\begin{equation}
    (\mathcal{C}^*e^{2\chi}\eta)_{ab}(x) = e^{2\chi(\mathcal{C}(x))}\,\frac{\partial \mathcal{C}^{c}}{\partial x^{a}}\frac{\partial \mathcal{C}^{d}}{\partial x^{b}}\,\eta_{cd}(\mathcal{C}(x))= \eta_{ab}(x)
\end{equation}
since $\eta_{w\overbar{w}}(w,\overbar{w}) = 1\slash 2$ $\eta_{ww} = \eta_{\overbar{w}\overbar{w}} = 0$ are constants. We obtain the conformal Ward identity
\begin{align}
    &\mathcal{R}_{(\mathbb{R}^2,\eta)}^{\tilde{\varepsilon}_1,\tilde{\varepsilon}_2}(w_1,\overbar{w}_1;w_2,\overbar{w}_2)\\
    &= [C'(w_1)\,\overbar{C}'(\overbar{w}_1)\,C'(w_2)\,\overbar{C}'(\overbar{w}_2)]^{\Delta_n\slash 2}\,\mathcal{R}_{(\mathcal{C}(\mathbb{R}^2),\eta)}^{\tilde{\varepsilon}_1,\tilde{\varepsilon}_2}(C(w_1),\overbar{C}(\overbar{w}_1);C(w_2),\overbar{C}(\overbar{w}_2))\,.
    \label{eq:Ward_identity}
\end{align}
We will restrict to the $SL(2,\mathbb{C})$ subgroup given by
\begin{equation}
    C(w) = \frac{a w+b}{c w+ d}\,,\quad \overbar{C}(\overbar{w}) = \frac{\overbar{a} \overbar{w}+\overbar{b}}{\overbar{c} \overbar{w}+ \overbar{d}}\,,
\end{equation}
where $\overbar{a},\overbar{b},\overbar{c},\overbar{d}\in\mathbb{C}$ are complex conjugates of $a,b,c,d\in\mathbb{C}$ that satisfy $ad-bc = 1$. This subgroup preserves the plane $\mathcal{C}(\mathbb{R}^2) = \mathbb{R}^2$ (assuming the point at infinity $\{\infty\}$ is included), so that in this case, \eqref{eq:Ward_identity} reduces to an equation for $\mathcal{R}_{(\mathbb{R}^2,\eta)}^{\tilde{\varepsilon}_1,\tilde{\varepsilon}_2}$ with the solution
\begin{equation}
    \mathcal{R}_{(\mathbb{R}^2,\eta)}^{\tilde{\varepsilon}_1,\tilde{\varepsilon}_2}(w_1,\overbar{w}_1;w_2,\overbar{w}_2) = \frac{B_n(\tilde{\varepsilon}_1,\tilde{\varepsilon}_2)}{[(w_1-w_2)(\overbar{w}_1-\overbar{w}_2)]^{\Delta_n}}\,.
\end{equation}
Here $ B_n( \tilde{\varepsilon}_1, \tilde{\varepsilon}_2)$ is a constant not fixed by \eqref{eq:Ward_identity}. The dependence on the cut-offs $\tilde{\varepsilon}_{1,2}$ can be fixed as follows. First, we write
\begin{equation}
    \mathcal{R}_{(\mathbb{R}^2,\eta)}^{\tilde{\varepsilon}_1,\tilde{\varepsilon}_2}(w_1,\overbar{w}_1;w_2,\overbar{w}_2) \equiv \mathcal{R}[\eta;\tilde{\varepsilon}_1,\tilde{\varepsilon}_2]
\end{equation}
making explicit the dependence on the chosen geodesic radii $ \tilde{\varepsilon}_i$. The geodesic distance $L_{\mathcal{P}^*\eta}(w_i,\overbar{w}_i;w,\overbar{w}) \equiv L_i[\eta]$ from the point $(w_i,\overbar{w}_i)$ in the metric $\mathcal{P}^*\eta$ is a functional of the metric satisfying $L_i[\lambda^2\eta] = \lambda L_i[\eta]$. Therefore the conditions $L_i[\eta]<  \tilde{\varepsilon}_i$, defining the geodesic disks, are invariant under the simultaneous scaling $\eta\rightarrow \lambda\eta$ and $ \tilde{\varepsilon}_i\rightarrow \lambda  \tilde{\varepsilon}_i$. Thus we obtain
\begin{equation}
    \mathcal{R}[\lambda^2\eta;\lambda \tilde{\varepsilon}_1,\lambda \tilde{\varepsilon}_2] = \mathcal{R}[\eta; \tilde{\varepsilon}_1, \tilde{\varepsilon}_2]\,.
    \label{eq:invariance_under_rescaling}
\end{equation}
On the other hand, \eqref{eq:flat_space_transformation} implies $\mathcal{R}[\lambda^2\eta; \tilde{\varepsilon}_1, \tilde{\varepsilon}_2] = \lambda^{-\Delta_n}\,\mathcal{R}[\eta; \tilde{\varepsilon}_1, \tilde{\varepsilon}_2]$ so that \eqref{eq:invariance_under_rescaling} implies
\begin{equation}
    \mathcal{R}[\eta;\lambda^{-1} \tilde{\varepsilon}_1,\lambda^{-1} \tilde{\varepsilon}_2] = \lambda^{-\Delta_n}\,\mathcal{R}[\eta; \tilde{\varepsilon}_1, \tilde{\varepsilon}_2]\,.
\end{equation}
This fixes $B_n( \tilde{\varepsilon}_1, \tilde{\varepsilon}_2) = b(n)\, \tilde{\varepsilon}_1^{\Delta_n} \tilde{\varepsilon}_2^{\Delta_n}$, where $b(n)$ is a constant independent of the geodesic radii, yielding
\begin{equation}
    \mathcal{R}_{(\mathbb{R}^2,\eta)}^{\tilde{\varepsilon}_1,\tilde{\varepsilon}_2}(w_1,\overbar{w}_1;w_2,\overbar{w}_2) = b(n)\,\biggl[\frac{ \tilde{\varepsilon}_1  \tilde{\varepsilon}_2}{(w_1-w_2)(\overbar{w}_1-\overbar{w}_2)}\biggr]^{\Delta_n}\,.
    \label{eq:final_sol_ward_identities}
\end{equation}
One can see this also from the fact that $\eta\rightarrow \lambda^2\eta$ is equivalent to a rescaling of the coordinates $(w,\overbar{w})\rightarrow \lambda (w,\overbar{w})$. This implies $(w_i,\overbar{w}_i)\rightarrow \lambda (w_i,\overbar{w}_i)$ so that \eqref{eq:final_sol_ward_identities} is invariant if the cut-offs are also scaled $\tilde{\varepsilon}_{i}\rightarrow \lambda\tilde{\varepsilon}_{i}$.

\end{appendix}

\bibliographystyle{JHEP}
\bibliography{bib}

\providecommand{\href}[2]{#2}\begingroup\raggedright\begin{thebibliography}{100}

\bibitem{Fulling:1972md}
S.A.~Fulling, \emph{{Nonuniqueness of canonical field quantization in
  Riemannian space-time}},
  \href{https://doi.org/10.1103/PhysRevD.7.2850}{\emph{Phys. Rev. D} {\bfseries
  7} (1973) 2850}.

\bibitem{Parker:1968mv}
L.~Parker, \emph{{Particle creation in expanding universes}},
  \href{https://doi.org/10.1103/PhysRevLett.21.562}{\emph{Phys. Rev. Lett.}
  {\bfseries 21} (1968) 562}.

\bibitem{Brandenberger:2012aj}
R.H.~Brandenberger and J.~Martin, \emph{{Trans-Planckian Issues for
  Inflationary Cosmology}},
  \href{https://doi.org/10.1088/0264-9381/30/11/113001}{\emph{Class. Quant.
  Grav.} {\bfseries 30} (2013) 113001}
  [\href{https://arxiv.org/abs/1211.6753}{{\ttfamily 1211.6753}}].

\bibitem{Dvali:2017eba}
G.~Dvali, C.~Gomez and S.~Zell, \emph{{Quantum Break-Time of de Sitter}},
  \href{https://doi.org/10.1088/1475-7516/2017/06/028}{\emph{JCAP} {\bfseries
  06} (2017) 028} [\href{https://arxiv.org/abs/1701.08776}{{\ttfamily
  1701.08776}}].

\bibitem{Agullo:2015qqa}
I.~Agullo and A.~Ashtekar, \emph{{Unitarity and ultraviolet regularity in
  cosmology}}, \href{https://doi.org/10.1103/PhysRevD.91.124010}{\emph{Phys.
  Rev. D} {\bfseries 91} (2015) 124010}
  [\href{https://arxiv.org/abs/1503.03407}{{\ttfamily 1503.03407}}].

\bibitem{Torre:1998eq}
C.G.~Torre and M.~Varadarajan, \emph{{Functional evolution of free quantum
  fields}}, \href{https://doi.org/10.1088/0264-9381/16/8/306}{\emph{Class.
  Quant. Grav.} {\bfseries 16} (1999) 2651}
  [\href{https://arxiv.org/abs/hep-th/9811222}{{\ttfamily hep-th/9811222}}].

\bibitem{Kuchar:1989wz}
K.~Kuchar, \emph{{Dirac Constraint Quantization of a Parametrized Field Theory
  by Anomaly - Free Operator Representations of Space-time Diffeomorphisms}},
  \href{https://doi.org/10.1103/PhysRevD.39.2263}{\emph{Phys. Rev. D}
  {\bfseries 39} (1989) 2263}.

\bibitem{Torre:1997zs}
C.G.~Torre and M.~Varadarajan, \emph{{Quantum fields at any time}},
  \href{https://doi.org/10.1103/PhysRevD.58.064007}{\emph{Phys. Rev. D}
  {\bfseries 58} (1998) 064007}
  [\href{https://arxiv.org/abs/hep-th/9707221}{{\ttfamily hep-th/9707221}}].

\bibitem{Erdmenger:2021wzc}
J.~Erdmenger, M.~Flory, M.~Gerbershagen, M.P.~Heller and A.-L.~Weigel,
  \emph{{Exact Gravity Duals for Simple Quantum Circuits}},
  \href{https://doi.org/10.21468/SciPostPhys.13.3.061}{\emph{SciPost Phys.}
  {\bfseries 13} (2022) 061}
  [\href{https://arxiv.org/abs/2112.12158}{{\ttfamily 2112.12158}}].

\bibitem{Erdmenger:2022lov}
J.~Erdmenger, A.-L.~Weigel, M.~Gerbershagen and M.P.~Heller, \emph{{From
  complexity geometry to holographic spacetime}},
  \href{https://doi.org/10.1103/PhysRevD.108.106020}{\emph{Phys. Rev. D}
  {\bfseries 108} (2023) 106020}
  [\href{https://arxiv.org/abs/2212.00043}{{\ttfamily 2212.00043}}].

\bibitem{Caputa:2018kdj}
P.~Caputa and J.M.~Magan, \emph{{Quantum Computation as Gravity}},
  \href{https://doi.org/10.1103/PhysRevLett.122.231302}{\emph{Phys. Rev. Lett.}
  {\bfseries 122} (2019) 231302}
  [\href{https://arxiv.org/abs/1807.04422}{{\ttfamily 1807.04422}}].

\bibitem{Erdmenger:2020sup}
J.~Erdmenger, M.~Gerbershagen and A.-L.~Weigel, \emph{{Complexity measures from
  geometric actions on Virasoro and Kac-Moody orbits}},
  \href{https://doi.org/10.1007/JHEP11(2020)003}{\emph{JHEP} {\bfseries 11}
  (2020) 003} [\href{https://arxiv.org/abs/2004.03619}{{\ttfamily
  2004.03619}}].

\bibitem{Flory:2020eot}
M.~Flory and M.P.~Heller, \emph{{Geometry of Complexity in Conformal Field
  Theory}}, \href{https://doi.org/10.1103/PhysRevResearch.2.043438}{\emph{Phys.
  Rev. Res.} {\bfseries 2} (2020) 043438}
  [\href{https://arxiv.org/abs/2005.02415}{{\ttfamily 2005.02415}}].

\bibitem{Flory:2020dja}
M.~Flory and M.P.~Heller, \emph{{Conformal field theory complexity from
  Euler-Arnold equations}},
  \href{https://doi.org/10.1007/JHEP12(2020)091}{\emph{JHEP} {\bfseries 12}
  (2020) 091} [\href{https://arxiv.org/abs/2007.11555}{{\ttfamily
  2007.11555}}].

\bibitem{Erdmenger:2024xmj}
J.~Erdmenger, J.~Kastikainen and T.~Schuhmann, \emph{{Towards complexity of
  primary-deformed Virasoro circuits}},
  \href{https://doi.org/10.1007/JHEP03(2025)127}{\emph{JHEP} {\bfseries 03}
  (2025) 127} [\href{https://arxiv.org/abs/2409.08319}{{\ttfamily
  2409.08319}}].

\bibitem{deBoer:2023lrd}
J.~de~Boer, V.~Godet, J.~Kastikainen and E.~Keski-Vakkuri, \emph{{Quantum
  information geometry of driven CFTs}},
  \href{https://doi.org/10.1007/JHEP09(2023)087}{\emph{JHEP} {\bfseries 09}
  (2023) 087} [\href{https://arxiv.org/abs/2306.00099}{{\ttfamily
  2306.00099}}].

\bibitem{Oblak:2016eij}
B.~Oblak, \emph{{BMS Particles in Three Dimensions}}, Ph.D. thesis, U.
  Brussels, Brussels U., 2016.
\newblock \href{https://arxiv.org/abs/1610.08526}{{\ttfamily 1610.08526}}.
\newblock 10.1007/978-3-319-61878-4.

\bibitem{Gendiar:2008udd}
A.~Gendiar, R.~Krcmar and T.~Nishino, \emph{{Spherical Deformation for
  One-Dimensional Quantum Systems}},
  \href{https://doi.org/10.1143/PTP.122.953}{\emph{Prog. Theor. Phys.}
  {\bfseries 122} (2009) 953}
  [\href{https://arxiv.org/abs/0810.0622}{{\ttfamily 0810.0622}}].

\bibitem{Hikihara:2011mtb}
T.~Hikihara and T.~Nishino, \emph{{Connecting distant ends of one-dimensional
  critical systems by a sine-square deformation}},
  \href{https://doi.org/10.1103/physrevb.83.060414}{\emph{Phys. Rev. B}
  {\bfseries 83} (2011) 060414}.

\bibitem{PhysRevA.83.052118}
A.~Gendiar, M.~Dani\ifmmode~\check{s}\else \v{s}\fi{}ka, Y.~Lee and T.~Nishino,
  \emph{Suppression of finite-size effects in one-dimensional correlated
  systems}, \href{https://doi.org/10.1103/PhysRevA.83.052118}{\emph{Phys. Rev.
  A} {\bfseries 83} (2011) 052118}.

\bibitem{Katsura:2011ss}
H.~Katsura, \emph{{Sine-square deformation of solvable spin chains and
  conformal field theories}},
  \href{https://doi.org/10.1088/1751-8113/45/11/115003}{\emph{J. Phys. A}
  {\bfseries 45} (2012) 115003}
  [\href{https://arxiv.org/abs/1110.2459}{{\ttfamily 1110.2459}}].

\bibitem{Wen:2016inm}
X.~Wen, S.~Ryu and A.W.W.~Ludwig, \emph{{Evolution operators in conformal field
  theories and conformal mappings: Entanglement Hamiltonian, the sine-square
  deformation, and others}},
  \href{https://doi.org/10.1103/PhysRevB.93.235119}{\emph{Phys. Rev. B}
  {\bfseries 93} (2016) 235119}
  [\href{https://arxiv.org/abs/1604.01085}{{\ttfamily 1604.01085}}].

\bibitem{Okunishi:2016zat}
K.~Okunishi, \emph{{Sine-square deformation and M{\"o}bius quantization of 2D
  conformal field theory}},
  \href{https://doi.org/10.1093/ptep/ptw060}{\emph{PTEP} {\bfseries 2016}
  (2016) 063A02} [\href{https://arxiv.org/abs/1603.09543}{{\ttfamily
  1603.09543}}].

\bibitem{Gawedzki:2017woc}
K.~Gawedzki, E.~Langmann and P.~Moosavi, \emph{{Finite-time universality in
  nonequilibrium CFT}},
  \href{https://doi.org/10.1007/s10955-018-2025-x}{\emph{J. Statist. Phys.}
  {\bfseries 172} (2018) 353}
  [\href{https://arxiv.org/abs/1712.00141}{{\ttfamily 1712.00141}}].

\bibitem{Langmann:2018skr}
E.~Langmann and P.~Moosavi, \emph{{Diffusive Heat Waves in Random Conformal
  Field Theory}},
  \href{https://doi.org/10.1103/PhysRevLett.122.020201}{\emph{Phys. Rev. Lett.}
  {\bfseries 122} (2019) 020201}
  [\href{https://arxiv.org/abs/1807.10239}{{\ttfamily 1807.10239}}].

\bibitem{MacCormack:2018rwq}
I.~MacCormack, A.~Liu, M.~Nozaki and S.~Ryu, \emph{{Holographic Duals of
  Inhomogeneous Systems: The Rainbow Chain and the Sine-Square Deformation
  Model}}, \href{https://doi.org/10.1088/1751-8121/ab3944}{\emph{J. Phys. A}
  {\bfseries 52} (2019) 505401}
  [\href{https://arxiv.org/abs/1812.10023}{{\ttfamily 1812.10023}}].

\bibitem{Moosavi:2019fas}
P.~Moosavi, \emph{{Inhomogeneous Conformal Field Theory Out of Equilibrium}},
  \href{https://doi.org/10.1007/s00023-021-01118-0}{\emph{Annales Henri
  Poincare} {\bfseries 25} (2024) 1083}
  [\href{https://arxiv.org/abs/1912.04821}{{\ttfamily 1912.04821}}].

\bibitem{Bernard:2019mqm}
D.~Bernard and P.~Le~Doussal, \emph{{Entanglement entropy growth in stochastic
  conformal field theory and the KPZ class}},
  \href{https://doi.org/10.1209/0295-5075/131/10007}{\emph{EPL} {\bfseries 131}
  (2020) 10007} [\href{https://arxiv.org/abs/1912.08458}{{\ttfamily
  1912.08458}}].

\bibitem{Lapierre:2020ftq}
B.~Lapierre and P.~Moosavi, \emph{{Geometric approach to inhomogeneous Floquet
  systems}}, \href{https://doi.org/10.1103/PhysRevB.103.224303}{\emph{Phys.
  Rev. B} {\bfseries 103} (2021) 224303}
  [\href{https://arxiv.org/abs/2010.11268}{{\ttfamily 2010.11268}}].

\bibitem{Han:2020kwp}
B.~Han and X.~Wen, \emph{{Classification of $SL_2$ deformed Floquet conformal
  field theories}},
  \href{https://doi.org/10.1103/PhysRevB.102.205125}{\emph{Phys. Rev. B}
  {\bfseries 102} (2020) 205125}
  [\href{https://arxiv.org/abs/2008.01123}{{\ttfamily 2008.01123}}].

\bibitem{Wen:2020wee}
X.~Wen, R.~Fan, A.~Vishwanath and Y.~Gu, \emph{{Periodically,
  quasiperiodically, and randomly driven conformal field theories}},
  \href{https://doi.org/10.1103/PhysRevResearch.3.023044}{\emph{Phys. Rev.
  Res.} {\bfseries 3} (2021) 023044}
  [\href{https://arxiv.org/abs/2006.10072}{{\ttfamily 2006.10072}}].

\bibitem{Caputa:2020mgb}
P.~Caputa and I.~MacCormack, \emph{{Geometry and Complexity of Path Integrals
  in Inhomogeneous CFTs}},
  \href{https://doi.org/10.1007/JHEP01(2021)027}{\emph{JHEP} {\bfseries 01}
  (2021) 027} [\href{https://arxiv.org/abs/2004.04698}{{\ttfamily
  2004.04698}}].

\bibitem{Goto:2023wai}
K.~Goto, M.~Nozaki, S.~Ryu, K.~Tamaoka and M.T.~Tan, \emph{{Scrambling and
  recovery of quantum information in inhomogeneous quenches in two-dimensional
  conformal field theories}},
  \href{https://doi.org/10.1103/PhysRevResearch.6.023001}{\emph{Phys. Rev.
  Res.} {\bfseries 6} (2024) 023001}
  [\href{https://arxiv.org/abs/2302.08009}{{\ttfamily 2302.08009}}].

\bibitem{Nozaki:2023fkx}
M.~Nozaki, K.~Tamaoka and M.T.~Tan, \emph{{Inhomogeneous quenches as state
  preparation in two-dimensional conformal field theories}},
  \href{https://doi.org/10.1103/PhysRevD.109.126014}{\emph{Phys. Rev. D}
  {\bfseries 109} (2024) 126014}
  [\href{https://arxiv.org/abs/2310.19376}{{\ttfamily 2310.19376}}].

\bibitem{Goto:2023yxb}
K.~Goto, T.~Guo, T.~Nosaka, M.~Nozaki, S.~Ryu and K.~Tamaoka, \emph{{Spatial
  deformation of many-body quantum chaotic systems and quantum information
  scrambling}}, \href{https://doi.org/10.1103/PhysRevB.109.054301}{\emph{Phys.
  Rev. B} {\bfseries 109} (2024) 054301}
  [\href{https://arxiv.org/abs/2305.01019}{{\ttfamily 2305.01019}}].

\bibitem{Das:2023xaw}
D.~Das, S.R.~Das, A.~Kundu and K.~Sengupta, \emph{{Exactly solvable floquet
  dynamics for conformal field theories in dimensions greater than two}},
  \href{https://doi.org/10.1007/JHEP09(2024)095}{\emph{JHEP} {\bfseries 09}
  (2024) 095} [\href{https://arxiv.org/abs/2311.13468}{{\ttfamily
  2311.13468}}].

\bibitem{Kudler-Flam:2023ahk}
J.~Kudler-Flam, M.~Nozaki, T.~Numasawa, S.~Ryu and M.T.~Tan, \emph{{Bridging
  two quantum quench problems {\textemdash} local joining quantum quench and
  M{\"o}bius quench {\textemdash} and their holographic dual descriptions}},
  \href{https://doi.org/10.1007/JHEP08(2024)213}{\emph{JHEP} {\bfseries 08}
  (2024) 213} [\href{https://arxiv.org/abs/2309.04665}{{\ttfamily
  2309.04665}}].

\bibitem{Liu:2023tiq}
X.~Liu, A.~McDonald, T.~Numasawa, B.~Lian and S.~Ryu, \emph{{Inhomogeneous
  Quantum Quenches of Conformal Field Theory with Boundaries}},
  \href{https://doi.org/10.1103/PhysRevLett.134.220404}{\emph{Phys. Rev. Lett.}
  {\bfseries 134} (2025) 220404}
  [\href{https://arxiv.org/abs/2309.04540}{{\ttfamily 2309.04540}}].

\bibitem{Mao:2024cnm}
W.~Mao, M.~Nozaki, K.~Tamaoka and M.T.~Tan, \emph{{Local operator quench
  induced by two-dimensional inhomogeneous and homogeneous CFT Hamiltonians}},
  \href{https://doi.org/10.1007/JHEP07(2024)200}{\emph{JHEP} {\bfseries 07}
  (2024) 200} [\href{https://arxiv.org/abs/2403.15851}{{\ttfamily
  2403.15851}}].

\bibitem{Lapierre:2024lga}
B.~Lapierre, T.~Numasawa, T.~Neupert and S.~Ryu, \emph{{Floquet engineered
  inhomogeneous quantum chaos in critical systems}},
  \href{https://arxiv.org/abs/2405.01642}{{\ttfamily 2405.01642}}.

\bibitem{Bernamonti:2024fgx}
A.~Bernamonti, F.~Galli and D.~Ge, \emph{{Boundary-induced transitions in
  M{\"o}bius quenches of holographic BCFT}},
  \href{https://doi.org/10.1007/JHEP06(2024)184}{\emph{JHEP} {\bfseries 06}
  (2024) 184} [\href{https://arxiv.org/abs/2402.16555}{{\ttfamily
  2402.16555}}].

\bibitem{Jiang:2024hgt}
H.~Jiang and M.~Mezei, \emph{{New horizons for inhomogeneous quenches and
  Floquet CFT}},  \href{https://arxiv.org/abs/2404.07884}{{\ttfamily
  2404.07884}}.

\bibitem{Das:2024lra}
J.~Das and A.~Kundu, \emph{{Flowery horizons {\&} bulk observers:
  sl$^{(q)}$(2,{\,}{\ensuremath{\mathbb{R}}}), drive in 2d holographic CFT}},
  \href{https://doi.org/10.1007/JHEP05(2025)035}{\emph{JHEP} {\bfseries 05}
  (2025) 035} [\href{https://arxiv.org/abs/2412.18536}{{\ttfamily
  2412.18536}}].

\bibitem{Goto:2021sqx}
K.~Goto, M.~Nozaki, S.~Ryu, K.~Tamaoka and M.T.~Tan, \emph{{Non-Equilibrating a
  Black Hole with Inhomogeneous Quantum Quench}},
  \href{https://arxiv.org/abs/2112.14388}{{\ttfamily 2112.14388}}.

\bibitem{Liska:2022vrd}
D.~Liska, V.~Gritsev, W.~Vleeshouwers and J.~Min{\'a}{\v{r}},
  \emph{{Holographic quantum scars}},
  \href{https://doi.org/10.21468/SciPostPhys.15.3.106}{\emph{SciPost Phys.}
  {\bfseries 15} (2023) 106}
  [\href{https://arxiv.org/abs/2212.05962}{{\ttfamily 2212.05962}}].

\bibitem{Li:2025rzl}
Z.~Li, Z.~Li and J.~Tian, \emph{{The Holography of the 2D inhomogeneously
  deformed CFT}},  \href{https://arxiv.org/abs/2502.11135}{{\ttfamily
  2502.11135}}.

\bibitem{Wen:2018vux}
X.~Wen and J.-Q.~Wu, \emph{{Quantum dynamics in sine-square deformed conformal
  field theory: Quench from uniform to nonuniform conformal field theory}},
  \href{https://doi.org/10.1103/PhysRevB.97.184309}{\emph{Phys. Rev. B}
  {\bfseries 97} (2018) 184309}
  [\href{https://arxiv.org/abs/1802.07765}{{\ttfamily 1802.07765}}].

\bibitem{Wen:2018agb}
X.~Wen and J.-Q.~Wu, \emph{{Floquet conformal field theory}},
  \href{https://arxiv.org/abs/1805.00031}{{\ttfamily 1805.00031}}.

\bibitem{Fan:2020orx}
R.~Fan, Y.~Gu, A.~Vishwanath and X.~Wen, \emph{{Floquet conformal field
  theories with generally deformed Hamiltonians}},
  \href{https://doi.org/10.21468/SciPostPhys.10.2.049}{\emph{SciPost Phys.}
  {\bfseries 10} (2021) 049}
  [\href{https://arxiv.org/abs/2011.09491}{{\ttfamily 2011.09491}}].

\bibitem{Tada:2014kza}
T.~Tada, \emph{{Sine-Square Deformation and its Relevance to String Theory}},
  \href{https://doi.org/10.1142/S0217732315500923}{\emph{Mod. Phys. Lett. A}
  {\bfseries 30} (2015) 1550092}
  [\href{https://arxiv.org/abs/1404.6343}{{\ttfamily 1404.6343}}].

\bibitem{Rodriguez-Laguna:2016roi}
J.~Rodr{\'\i}guez-Laguna, J.~Dubail, G.~Ram{\'\i}rez, P.~Calabrese and
  G.~Sierra, \emph{{More on the rainbow chain: entanglement, space-time
  geometry and thermal states}},
  \href{https://doi.org/10.1088/1751-8121/aa6268}{\emph{J. Phys. A} {\bfseries
  50} (2017) 164001} [\href{https://arxiv.org/abs/1611.08559}{{\ttfamily
  1611.08559}}].

\bibitem{dubail2017emergence}
J.~Dubail, J.-M.~St{\'e}phan and P.~Calabrese, \emph{Emergence of curved
  light-cones in a class of inhomogeneous luttinger liquids}, {\emph{SciPost
  Physics} {\bfseries 3} (2017) 019}.

\bibitem{Lapierre:2019rwj}
B.~Lapierre, K.~Choo, C.~Tauber, A.~Tiwari, T.~Neupert and R.~Chitra,
  \emph{{Emergent black hole dynamics in critical Floquet systems}},
  \href{https://doi.org/10.1103/PhysRevResearch.2.023085}{\emph{Phys. Rev.
  Res.} {\bfseries 2} (2020) 023085}
  [\href{https://arxiv.org/abs/1909.08618}{{\ttfamily 1909.08618}}].

\bibitem{Crawford:2021adf}
S.~Crawford, K.~Rejzner and B.~Vicedo, \emph{{Lorentzian 2D CFT from the pAQFT
  Perspective}},
  \href{https://doi.org/10.1007/s00023-022-01167-z}{\emph{Annales Henri
  Poincare} {\bfseries 23} (2022) 3525}
  [\href{https://arxiv.org/abs/2107.12347}{{\ttfamily 2107.12347}}].

\bibitem{Holzhey:1993kx}
C.F.E.~Holzhey, \emph{{A scene in the taming of the hole}},  other thesis,
  1993.

\bibitem{Holzhey:1994we}
C.~Holzhey, F.~Larsen and F.~Wilczek, \emph{{Geometric and renormalized entropy
  in conformal field theory}},
  \href{https://doi.org/10.1016/0550-3213(94)90402-2}{\emph{Nucl. Phys. B}
  {\bfseries 424} (1994) 443}
  [\href{https://arxiv.org/abs/hep-th/9403108}{{\ttfamily hep-th/9403108}}].

\bibitem{Fiola:1994ir}
T.M.~Fiola, J.~Preskill, A.~Strominger and S.P.~Trivedi, \emph{{Black hole
  thermodynamics and information loss in two-dimensions}},
  \href{https://doi.org/10.1103/PhysRevD.50.3987}{\emph{Phys. Rev. D}
  {\bfseries 50} (1994) 3987}
  [\href{https://arxiv.org/abs/hep-th/9403137}{{\ttfamily hep-th/9403137}}].

\bibitem{Ryu:2006bv}
S.~Ryu and T.~Takayanagi, \emph{{Holographic derivation of entanglement entropy
  from AdS/CFT}},
  \href{https://doi.org/10.1103/PhysRevLett.96.181602}{\emph{Phys. Rev. Lett.}
  {\bfseries 96} (2006) 181602}
  [\href{https://arxiv.org/abs/hep-th/0603001}{{\ttfamily hep-th/0603001}}].

\bibitem{Ryu:2006ef}
S.~Ryu and T.~Takayanagi, \emph{{Aspects of Holographic Entanglement Entropy}},
  \href{https://doi.org/10.1088/1126-6708/2006/08/045}{\emph{JHEP} {\bfseries
  08} (2006) 045} [\href{https://arxiv.org/abs/hep-th/0605073}{{\ttfamily
  hep-th/0605073}}].

\bibitem{Miyata:2024gvr}
A.~Miyata, M.~Nozaki, K.~Tamaoka and M.~Watanabe, \emph{{Hawking-Page and
  entanglement phase transition in 2d CFT on curved backgrounds}},
  \href{https://doi.org/10.1007/JHEP08(2024)190}{\emph{JHEP} {\bfseries 08}
  (2024) 190} [\href{https://arxiv.org/abs/2406.06121}{{\ttfamily
  2406.06121}}].

\bibitem{Bai:2024azk}
C.~Bai, A.~Miyata and M.~Nozaki, \emph{{Entanglement dynamics in 2d HCFTs on
  the curved background: the case of q-M{\"o}bius Hamiltonian}},
  \href{https://doi.org/10.1007/JHEP12(2024)208}{\emph{JHEP} {\bfseries 12}
  (2024) 208} [\href{https://arxiv.org/abs/2408.06594}{{\ttfamily
  2408.06594}}].

\bibitem{Calabrese:2004eu}
P.~Calabrese and J.L.~Cardy, \emph{{Entanglement entropy and quantum field
  theory}}, \href{https://doi.org/10.1088/1742-5468/2004/06/P06002}{\emph{J.
  Stat. Mech.} {\bfseries 0406} (2004) P06002}
  [\href{https://arxiv.org/abs/hep-th/0405152}{{\ttfamily hep-th/0405152}}].

\bibitem{Calabrese:2009qy}
P.~Calabrese and J.~Cardy, \emph{{Entanglement entropy and conformal field
  theory}}, \href{https://doi.org/10.1088/1751-8113/42/50/504005}{\emph{J.
  Phys. A} {\bfseries 42} (2009) 504005}
  [\href{https://arxiv.org/abs/0905.4013}{{\ttfamily 0905.4013}}].

\bibitem{Estienne:2025uhh}
B.~Estienne and J.~Lin, \emph{{Entanglement Entropy and Cauchy-Hadamard
  Renormalization}},  \href{https://arxiv.org/abs/2501.19014}{{\ttfamily
  2501.19014}}.

\bibitem{Christensen:1976vb}
S.M.~Christensen, \emph{{Vacuum Expectation Value of the Stress Tensor in an
  Arbitrary Curved Background: The Covariant Point Separation Method}},
  \href{https://doi.org/10.1103/PhysRevD.14.2490}{\emph{Phys. Rev. D}
  {\bfseries 14} (1976) 2490}.

\bibitem{davies1977quantum}
P.C.~Davies and S.A.~Fulling, \emph{Quantum vacuum energy in two dimensional
  space-times}, {\emph{Proceedings of the Royal Society of London. A.
  Mathematical and Physical Sciences} {\bfseries 354} (1977) 59}.

\bibitem{Christensen:1978yd}
S.M.~Christensen, \emph{{Regularization, Renormalization, and Covariant
  Geodesic Point Separation}},
  \href{https://doi.org/10.1103/PhysRevD.17.946}{\emph{Phys. Rev. D} {\bfseries
  17} (1978) 946}.

\bibitem{Wald:1978pj}
R.M.~Wald, \emph{{Trace Anomaly of a Conformally Invariant Quantum Field in
  Curved Space-Time}},
  \href{https://doi.org/10.1103/PhysRevD.17.1477}{\emph{Phys. Rev. D}
  {\bfseries 17} (1978) 1477}.

\bibitem{Polchinski:1986qf}
J.~Polchinski, \emph{{Vertex Operators in the Polyakov Path Integral}},
  \href{https://doi.org/10.1016/0550-3213(87)90389-0}{\emph{Nucl. Phys. B}
  {\bfseries 289} (1987) 465}.

\bibitem{Novotny:1998bw}
J.~Novotny and M.~Schnabl, \emph{{Point - splitting regularization of composite
  operators and anomalies}},
  \href{https://doi.org/10.1002/(SICI)1521-3978(200004)48:4<253::AID-PROP253>3.0.CO;2-3}{\emph{Fortsch.
  Phys.} {\bfseries 48} (2000) 253}
  [\href{https://arxiv.org/abs/hep-th/9803244}{{\ttfamily hep-th/9803244}}].

\bibitem{Cotler:2022weg}
J.~Cotler and A.~Strominger, \emph{{The Universe as a Quantum Encoder}},
  \href{https://arxiv.org/abs/2201.11658}{{\ttfamily 2201.11658}}.

\bibitem{Osborn:1991gm}
H.~Osborn, \emph{{Weyl consistency conditions and a local renormalization group
  equation for general renormalizable field theories}},
  \href{https://doi.org/10.1016/0550-3213(91)80030-P}{\emph{Nucl. Phys. B}
  {\bfseries 363} (1991) 486}.

\bibitem{Deser:1993yx}
S.~Deser and A.~Schwimmer, \emph{{Geometric classification of conformal
  anomalies in arbitrary dimensions}},
  \href{https://doi.org/10.1016/0370-2693(93)90934-A}{\emph{Phys. Lett. B}
  {\bfseries 309} (1993) 279}
  [\href{https://arxiv.org/abs/hep-th/9302047}{{\ttfamily hep-th/9302047}}].

\bibitem{Boulanger:2007st}
N.~Boulanger, \emph{{General solutions of the Wess-Zumino consistency condition
  for the Weyl anomalies}},
  \href{https://doi.org/10.1088/1126-6708/2007/07/069}{\emph{JHEP} {\bfseries
  07} (2007) 069} [\href{https://arxiv.org/abs/0704.2472}{{\ttfamily
  0704.2472}}].

\bibitem{Polyakov:1981rd}
A.M.~Polyakov, \emph{{Quantum Geometry of Bosonic Strings}},
  \href{https://doi.org/10.1016/0370-2693(81)90743-7}{\emph{Phys. Lett. B}
  {\bfseries 103} (1981) 207}.

\bibitem{Besken:2020snx}
M.~Be\c{s}ken, J.~De~Boer and G.~Mathys, \emph{{On local and integrated
  stress-tensor commutators}},
  \href{https://doi.org/10.1007/JHEP07(2021)148}{\emph{JHEP} {\bfseries 21}
  (2020) 148} [\href{https://arxiv.org/abs/2012.15724}{{\ttfamily
  2012.15724}}].

\bibitem{Fewster:2004nj}
C.J.~Fewster and S.~Hollands, \emph{{Quantum energy inequalities in
  two-dimensional conformal field theory}},
  \href{https://doi.org/10.1142/S0129055X05002406}{\emph{Rev. Math. Phys.}
  {\bfseries 17} (2005) 577}
  [\href{https://arxiv.org/abs/math-ph/0412028}{{\ttfamily math-ph/0412028}}].

\bibitem{Fewster:2018srj}
C.J.~Fewster and S.~Hollands, \emph{{Probability distributions for the stress
  tensor in conformal field theories}},
  \href{https://doi.org/10.1007/s11005-018-1124-6}{\emph{Lett. Math. Phys.}
  {\bfseries 109} (2019) 747}
  [\href{https://arxiv.org/abs/1805.04281}{{\ttfamily 1805.04281}}].

\bibitem{freistadt1955classical}
H.~Freistadt, \emph{Classical field theory in the hamilton-jacobi formalism},
  {\emph{Physical Review} {\bfseries 97} (1955) 1158}.

\bibitem{freistadt1956quantized}
H.~Freistadt, \emph{Quantized field theory in the hamilton-jacobi formalism},
  {\emph{Physical Review} {\bfseries 102} (1956) 274}.

\bibitem{Isler:1989ty}
K.~Isler, C.~Schmid and C.A.~Trugenberger, \emph{{Normal Ordering Prescription
  for the Energy Momentum Tensor of (1+1)-dimensional Fermions in Curved
  Space-times}},
  \href{https://doi.org/10.1016/0550-3213(90)90199-N}{\emph{Nucl. Phys. B}
  {\bfseries 339} (1990) 577}.

\bibitem{Capri:1992tb}
A.Z.~Capri and S.M.~Roy, \emph{{The Definition of time and quantum vacuum in
  (1+1)-dimensions}},
  \href{https://doi.org/10.1142/S0217732392002081}{\emph{Mod. Phys. Lett. A}
  {\bfseries 7} (1992) 2317}.

\bibitem{Massacand:1993cj}
C.M.~Massacand and C.~Schmid, \emph{{Particle production by tidal forces and
  the trace anomaly}},
  \href{https://doi.org/10.1006/aphy.1994.1046}{\emph{Annals Phys.} {\bfseries
  231} (1994) 363}.

\bibitem{Lamb:1994bz}
D.J.~Lamb and A.Z.~Capri, \emph{{Finite particle creation in (1+1)-dimensional
  compact in space}},
  \href{https://doi.org/10.1088/0264-9381/12/2/010}{\emph{Class. Quant. Grav.}
  {\bfseries 12} (1995) 413}
  [\href{https://arxiv.org/abs/hep-th/9412010}{{\ttfamily hep-th/9412010}}].

\bibitem{Capri:1995iw}
A.Z.~Capri, M.~Kobayashi and D.J.~Lamb, \emph{{Two observers calculate the
  trace anomaly}},
  \href{https://doi.org/10.1088/0264-9381/13/2/006}{\emph{Class. Quant. Grav.}
  {\bfseries 13} (1996) 179}
  [\href{https://arxiv.org/abs/hep-th/9508077}{{\ttfamily hep-th/9508077}}].

\bibitem{Page:1996ax}
D.N.~Page, \emph{{Stress tensors for instantaneous vacua in (1+1)-dimensions}},
  \href{https://doi.org/10.1088/0264-9381/14/11/007}{\emph{Class. Quant. Grav.}
  {\bfseries 14} (1997) 3041}
  [\href{https://arxiv.org/abs/gr-qc/9603005}{{\ttfamily gr-qc/9603005}}].

\bibitem{Hartman:2013qma}
T.~Hartman and J.~Maldacena, \emph{{Time Evolution of Entanglement Entropy from
  Black Hole Interiors}},
  \href{https://doi.org/10.1007/JHEP05(2013)014}{\emph{JHEP} {\bfseries 05}
  (2013) 014} [\href{https://arxiv.org/abs/1303.1080}{{\ttfamily 1303.1080}}].

\bibitem{Sivaramakrishnan:2017yng}
A.~Sivaramakrishnan, \emph{{Entanglement Entropy with a Time-dependent
  Hamiltonian}}, \href{https://doi.org/10.1103/PhysRevD.97.066003}{\emph{Phys.
  Rev. D} {\bfseries 97} (2018) 066003}
  [\href{https://arxiv.org/abs/1709.09776}{{\ttfamily 1709.09776}}].

\bibitem{Malvimat:2018cfe}
V.~Malvimat, S.~Mondal and G.~Sengupta, \emph{{Time Evolution of Entanglement
  Negativity from Black Hole Interiors}},
  \href{https://doi.org/10.1007/JHEP05(2019)183}{\emph{JHEP} {\bfseries 05}
  (2019) 183} [\href{https://arxiv.org/abs/1812.04424}{{\ttfamily
  1812.04424}}].

\bibitem{Shaposhnik:2022jzc}
L.~Shaposhnik, \emph{{Entanglement Entropy, Local IR/UV Connection and MPS in
  Weyl-deformed Geometries}},
  \href{https://arxiv.org/abs/2211.16430}{{\ttfamily 2211.16430}}.

\bibitem{Tonni:2017jom}
E.~Tonni, J.~Rodr{\'\i}guez-Laguna and G.~Sierra, \emph{{Entanglement
  hamiltonian and entanglement contour in inhomogeneous 1D critical systems}},
  \href{https://doi.org/10.1088/1742-5468/aab67d}{\emph{J. Stat. Mech.}
  {\bfseries 1804} (2018) 043105}
  [\href{https://arxiv.org/abs/1712.03557}{{\ttfamily 1712.03557}}].

\bibitem{Fefferman:2007rka}
C.~Fefferman and C.R.~Graham, \emph{{The ambient metric}}, {\emph{Ann. Math.
  Stud.} {\bfseries 178} (2011) 1}
  [\href{https://arxiv.org/abs/0710.0919}{{\ttfamily 0710.0919}}].

\bibitem{Imbimbo:1999bj}
C.~Imbimbo, A.~Schwimmer, S.~Theisen and S.~Yankielowicz,
  \emph{{Diffeomorphisms and holographic anomalies}},
  \href{https://doi.org/10.1088/0264-9381/17/5/322}{\emph{Class. Quant. Grav.}
  {\bfseries 17} (2000) 1129}
  [\href{https://arxiv.org/abs/hep-th/9910267}{{\ttfamily hep-th/9910267}}].

\bibitem{Brown:1986nw}
J.D.~Brown and M.~Henneaux, \emph{{Central Charges in the Canonical Realization
  of Asymptotic Symmetries: An Example from Three-Dimensional Gravity}},
  \href{https://doi.org/10.1007/BF01211590}{\emph{Commun. Math. Phys.}
  {\bfseries 104} (1986) 207}.

\bibitem{Penrose:1985bww}
R.~Penrose and W.~Rindler, \emph{{Spinors and Space-Time}}, Cambridge
  Monographs on Mathematical Physics, Cambridge Univ. Press, Cambridge, UK (4,
  2011),
  \href{https://doi.org/10.1017/CBO9780511564048}{10.1017/CBO9780511564048}.

\bibitem{Roberts:2012aq}
M.M.~Roberts, \emph{{Time evolution of entanglement entropy from a pulse}},
  \href{https://doi.org/10.1007/JHEP12(2012)027}{\emph{JHEP} {\bfseries 12}
  (2012) 027} [\href{https://arxiv.org/abs/1204.1982}{{\ttfamily 1204.1982}}].

\bibitem{Skenderis:2000in}
K.~Skenderis, \emph{{Asymptotically Anti-de Sitter space-times and their stress
  energy tensor}}, \href{https://doi.org/10.1142/S0217751X0100386X}{\emph{Int.
  J. Mod. Phys. A} {\bfseries 16} (2001) 740}
  [\href{https://arxiv.org/abs/hep-th/0010138}{{\ttfamily hep-th/0010138}}].

\bibitem{Ciambelli:2019bzz}
L.~Ciambelli and R.G.~Leigh, \emph{{Weyl Connections and their Role in
  Holography}}, \href{https://doi.org/10.1103/PhysRevD.101.086020}{\emph{Phys.
  Rev. D} {\bfseries 101} (2020) 086020}
  [\href{https://arxiv.org/abs/1905.04339}{{\ttfamily 1905.04339}}].

\bibitem{Ciambelli:2023ott}
L.~Ciambelli, A.~Delfante, R.~Ruzziconi and C.~Zwikel, \emph{{Symmetries and
  charges in Weyl-Fefferman-Graham gauge}},
  \href{https://doi.org/10.1103/PhysRevD.108.126003}{\emph{Phys. Rev. D}
  {\bfseries 108} (2023) 126003}
  [\href{https://arxiv.org/abs/2308.15480}{{\ttfamily 2308.15480}}].

\bibitem{Arenas-Henriquez:2024ypo}
G.~Arenas-Henriquez, F.~Diaz and D.~Rivera-Betancour, \emph{{Generalized
  Fefferman-Graham gauge and boundary Weyl structures}},
  \href{https://doi.org/10.1007/JHEP02(2025)007}{\emph{JHEP} {\bfseries 02}
  (2025) 007} [\href{https://arxiv.org/abs/2411.12513}{{\ttfamily
  2411.12513}}].

\bibitem{Brown:1992br}
J.D.~Brown and J.W.~York, Jr., \emph{{Quasilocal energy and conserved charges
  derived from the gravitational action}},
  \href{https://doi.org/10.1103/PhysRevD.47.1407}{\emph{Phys. Rev. D}
  {\bfseries 47} (1993) 1407}
  [\href{https://arxiv.org/abs/gr-qc/9209012}{{\ttfamily gr-qc/9209012}}].

\bibitem{Henningson:1998ey}
M.~Henningson and K.~Skenderis, \emph{{Holography and the Weyl anomaly}},
  \href{https://doi.org/10.1002/(SICI)1521-3978(20001)48:1/3<125::AID-PROP125>3.0.CO;2-B}{\emph{Fortsch.
  Phys.} {\bfseries 48} (2000) 125}
  [\href{https://arxiv.org/abs/hep-th/9812032}{{\ttfamily hep-th/9812032}}].

\bibitem{deHaro:2000vlm}
S.~de~Haro, S.N.~Solodukhin and K.~Skenderis, \emph{{Holographic reconstruction
  of space-time and renormalization in the AdS / CFT correspondence}},
  \href{https://doi.org/10.1007/s002200100381}{\emph{Commun. Math. Phys.}
  {\bfseries 217} (2001) 595}
  [\href{https://arxiv.org/abs/hep-th/0002230}{{\ttfamily hep-th/0002230}}].

\bibitem{Krasnov:2001cu}
K.~Krasnov, \emph{{On holomorphic factorization in asymptotically AdS 3-D
  gravity}}, \href{https://doi.org/10.1088/0264-9381/20/18/311}{\emph{Class.
  Quant. Grav.} {\bfseries 20} (2003) 4015}
  [\href{https://arxiv.org/abs/hep-th/0109198}{{\ttfamily hep-th/0109198}}].

\bibitem{Balasubramanian:1999zv}
V.~Balasubramanian and S.F.~Ross, \emph{{Holographic particle detection}},
  \href{https://doi.org/10.1103/PhysRevD.61.044007}{\emph{Phys. Rev. D}
  {\bfseries 61} (2000) 044007}
  [\href{https://arxiv.org/abs/hep-th/9906226}{{\ttfamily hep-th/9906226}}].

\bibitem{Lewkowycz:2013nqa}
A.~Lewkowycz and J.~Maldacena, \emph{{Generalized gravitational entropy}},
  \href{https://doi.org/10.1007/JHEP08(2013)090}{\emph{JHEP} {\bfseries 08}
  (2013) 090} [\href{https://arxiv.org/abs/1304.4926}{{\ttfamily 1304.4926}}].

\bibitem{Takayanagi:2011zk}
T.~Takayanagi, \emph{{Holographic Dual of BCFT}},
  \href{https://doi.org/10.1103/PhysRevLett.107.101602}{\emph{Phys. Rev. Lett.}
  {\bfseries 107} (2011) 101602}
  [\href{https://arxiv.org/abs/1105.5165}{{\ttfamily 1105.5165}}].

\bibitem{BarberoG:2016nkv}
J.F.~Barbero~G, J.~Margalef-Bentabol and E.J.S.~Villase{\~n}or,
  \emph{{Functional evolution of scalar fields in bounded one-dimensional
  regions}}, \href{https://doi.org/10.1088/1361-6382/aa5e8c}{\emph{Class.
  Quant. Grav.} {\bfseries 34} (2017) 065004}
  [\href{https://arxiv.org/abs/1611.09603}{{\ttfamily 1611.09603}}].

\bibitem{Moore:1970tmc}
G.T.~Moore, \emph{{Quantum Theory of the Electromagnetic Field in a
  Variable-Length One-Dimensional Cavity}},
  \href{https://doi.org/10.1063/1.1665432}{\emph{J. Math. Phys.} {\bfseries 11}
  (1970) 2679}.

\bibitem{Isler:1987ax}
K.~Isler, C.~Schmid and C.A.~Trugenberger, \emph{{Kinetic Normal Ordering and
  the (1+1)-dimensional U(1) Anomaly}},
  \href{https://doi.org/10.1016/0550-3213(88)90348-3}{\emph{Nucl. Phys. B}
  {\bfseries 301} (1988) 327}.

\bibitem{Jack:2013sha}
I.~Jack and H.~Osborn, \emph{{Constraints on RG Flow for Four Dimensional
  Quantum Field Theories}},
  \href{https://doi.org/10.1016/j.nuclphysb.2014.03.018}{\emph{Nucl. Phys. B}
  {\bfseries 883} (2014) 425}
  [\href{https://arxiv.org/abs/1312.0428}{{\ttfamily 1312.0428}}].

\bibitem{Anous:2016kss}
T.~Anous, T.~Hartman, A.~Rovai and J.~Sonner, \emph{{Black Hole Collapse in the
  1/c Expansion}}, \href{https://doi.org/10.1007/JHEP07(2016)123}{\emph{JHEP}
  {\bfseries 07} (2016) 123}
  [\href{https://arxiv.org/abs/1603.04856}{{\ttfamily 1603.04856}}].

\bibitem{Chapman:2021jbh}
S.~Chapman and G.~Policastro, \emph{{Quantum computational complexity from
  quantum information to black holes and back}},
  \href{https://doi.org/10.1140/epjc/s10052-022-10037-1}{\emph{Eur. Phys. J. C}
  {\bfseries 82} (2022) 128}
  [\href{https://arxiv.org/abs/2110.14672}{{\ttfamily 2110.14672}}].

\bibitem{Nandy:2024evd}
P.~Nandy, A.S.~Matsoukas-Roubeas, P.~Mart{\'\i}nez-Azcona, A.~Dymarsky and
  A.~del Campo, \emph{{Quantum dynamics in Krylov space: Methods and
  applications}},
  \href{https://doi.org/10.1016/j.physrep.2025.05.001}{\emph{Phys. Rept.}
  {\bfseries 1125-1128} (2025) 1}
  [\href{https://arxiv.org/abs/2405.09628}{{\ttfamily 2405.09628}}].

\bibitem{Baiguera:2025dkc}
S.~Baiguera, V.~Balasubramanian, P.~Caputa, S.~Chapman, J.~Haferkamp,
  M.P.~Heller et~al., \emph{{Quantum complexity in gravity, quantum field
  theory, and quantum information science}},
  \href{https://arxiv.org/abs/2503.10753}{{\ttfamily 2503.10753}}.

\bibitem{Rabinovici:2025otw}
E.~Rabinovici, A.~S{\'a}nchez-Garrido, R.~Shir and J.~Sonner, \emph{{Krylov
  Complexity}},  \href{https://arxiv.org/abs/2507.06286}{{\ttfamily
  2507.06286}}.

\bibitem{Balasubramanian:2022tpr}
V.~Balasubramanian, P.~Caputa, J.M.~Magan and Q.~Wu, \emph{{Quantum chaos and
  the complexity of spread of states}},
  \href{https://doi.org/10.1103/PhysRevD.106.046007}{\emph{Phys. Rev. D}
  {\bfseries 106} (2022) 046007}
  [\href{https://arxiv.org/abs/2202.06957}{{\ttfamily 2202.06957}}].

\bibitem{Rabinovici:2023yex}
E.~Rabinovici, A.~S{\'a}nchez-Garrido, R.~Shir and J.~Sonner, \emph{{A bulk
  manifestation of Krylov complexity}},
  \href{https://doi.org/10.1007/JHEP08(2023)213}{\emph{JHEP} {\bfseries 08}
  (2023) 213} [\href{https://arxiv.org/abs/2305.04355}{{\ttfamily
  2305.04355}}].

\bibitem{Heller:2024ldz}
M.P.~Heller, J.~Papalini and T.~Schuhmann, \emph{{Krylov spread complexity as
  holographic complexity beyond JT gravity}},
  \href{https://arxiv.org/abs/2412.17785}{{\ttfamily 2412.17785}}.

\bibitem{Caputa:2024sux}
P.~Caputa, B.~Chen, R.W.~McDonald, J.~Sim{\'o}n and B.~Strittmatter,
  \emph{{Spread Complexity Rate as Proper Momentum}},
  \href{https://arxiv.org/abs/2410.23334}{{\ttfamily 2410.23334}}.

\bibitem{He:2024pox}
P.-Z.~He, \emph{{Revisit the relationship between spread complexity rate and
  radial momentum}},  \href{https://arxiv.org/abs/2411.19172}{{\ttfamily
  2411.19172}}.

\bibitem{Fan:2024iop}
Z.-Y.~Fan, \emph{{Momentum-Krylov complexity correspondence}},
  \href{https://arxiv.org/abs/2411.04492}{{\ttfamily 2411.04492}}.

\bibitem{Giddings:2020dpb}
S.B.~Giddings, \emph{{Schr{\"o}dinger evolution of the Hawking state}},
  \href{https://doi.org/10.1103/PhysRevD.102.125022}{\emph{Phys. Rev. D}
  {\bfseries 102} (2020) 125022}
  [\href{https://arxiv.org/abs/2006.10834}{{\ttfamily 2006.10834}}].

\bibitem{Giddings:2021ipt}
S.B.~Giddings, \emph{{Schr{\"o}dinger evolution of two-dimensional black
  holes}}, \href{https://doi.org/10.1007/JHEP12(2021)025}{\emph{JHEP}
  {\bfseries 12} (2021) 025}
  [\href{https://arxiv.org/abs/2108.07824}{{\ttfamily 2108.07824}}].

\bibitem{Giddings:2022sss}
S.B.~Giddings and J.~Perkins, \emph{{Quantum evolution of the Hawking state for
  black holes}}, \href{https://doi.org/10.1103/PhysRevD.106.065011}{\emph{Phys.
  Rev. D} {\bfseries 106} (2022) 065011}
  [\href{https://arxiv.org/abs/2204.13126}{{\ttfamily 2204.13126}}].

\bibitem{Christopoulos:2021imm}
A.~Christopoulos, P.~Le~Doussal, D.~Bernard and A.~De~Luca, \emph{{Universal
  Out-of-Equilibrium Dynamics of 1D Critical Quantum Systems Perturbed by Noise
  Coupled to Energy}},
  \href{https://doi.org/10.1103/PhysRevX.13.011043}{\emph{Phys. Rev. X}
  {\bfseries 13} (2023) 011043}
  [\href{https://arxiv.org/abs/2110.15303}{{\ttfamily 2110.15303}}].

\bibitem{Katanaev:1986wk}
M.O.~Katanaev and I.V.~Volovich, \emph{{String Model with Dynamical Geometry
  and Torsion}},
  \href{https://doi.org/10.1016/0370-2693(86)90615-5}{\emph{Phys. Lett. B}
  {\bfseries 175} (1986) 413}
  [\href{https://arxiv.org/abs/hep-th/0209014}{{\ttfamily hep-th/0209014}}].

\bibitem{Strobl:1999wv}
T.~Strobl, \emph{{Gravity in two space-time dimensions}},  other thesis, 5,
  1999, [\href{https://arxiv.org/abs/hep-th/0011240}{{\ttfamily
  hep-th/0011240}}].

\bibitem{Wen:2024bzm}
X.~Wen, \emph{{Exactly solvable non-unitary time evolution in quantum critical
  systems I: effect of complex spacetime metrics}},
  \href{https://doi.org/10.1088/1742-5468/ad7c3d}{\emph{J. Stat. Mech.}
  {\bfseries 2024} (2024) 103103}
  [\href{https://arxiv.org/abs/2406.17059}{{\ttfamily 2406.17059}}].

\bibitem{Lapierre:2025zsg}
B.~Lapierre, P.~Pelliconi, S.~Ryu and J.~Sonner, \emph{{Driven Non-Unitary
  Dynamics of Quantum Critical Systems}},
  \href{https://arxiv.org/abs/2505.01508}{{\ttfamily 2505.01508}}.

\bibitem{Hollands:2019czd}
S.~Hollands, \emph{{Relative entropy for coherent states in chiral CFT}},
  \href{https://doi.org/10.1007/s11005-019-01238-z}{\emph{Lett. Math. Phys.}
  {\bfseries 110} (2020) 713}
  [\href{https://arxiv.org/abs/1903.07508}{{\ttfamily 1903.07508}}].

\bibitem{Panebianco:2019plp}
L.~Panebianco, \emph{{A formula for the relative entropy in chiral CFT}},
  \href{https://doi.org/10.1007/s11005-020-01296-8}{\emph{Lett. Math. Phys.}
  {\bfseries 110} (2020) 2363}
  [\href{https://arxiv.org/abs/1911.10136}{{\ttfamily 1911.10136}}].

\bibitem{flanders1973differentiation}
H.~Flanders, \emph{Differentiation under the integral sign}, {\emph{The
  American Mathematical Monthly} {\bfseries 80} (1973) 615}.

\bibitem{Reddiger:2019lqh}
M.~Reddiger and B.~Poirier, \emph{{The Differentiation Lemma and the Reynolds
  Transport Theorem for submanifolds with corners}},
  \href{https://doi.org/10.1142/S0219887823501372}{\emph{Int. J. Geom. Meth.
  Mod. Phys.} {\bfseries 20} (2023) 2350137}
  [\href{https://arxiv.org/abs/1906.03330}{{\ttfamily 1906.03330}}].

\bibitem{Adler:1976jx}
S.L.~Adler, J.~Lieberman and Y.J.~Ng, \emph{{Regularization of the Stress
  Energy Tensor for Vector and Scalar Particles Propagating in a General
  Background Metric}},
  \href{https://doi.org/10.1016/0003-4916(77)90313-X}{\emph{Annals Phys.}
  {\bfseries 106} (1977) 279}.

\bibitem{McAvity:1990we}
D.M.~McAvity and H.~Osborn, \emph{{A DeWitt expansion of the heat kernel for
  manifolds with a boundary}},
  \href{https://doi.org/10.1088/0264-9381/8/4/008}{\emph{Class. Quant. Grav.}
  {\bfseries 8} (1991) 603}.

\bibitem{Lazzarini:1990xid}
S.~Lazzarini, \emph{{SUR LES MODELES CONFORMES LAGRANGIENS BIDIMENSIONNELS}},
  Ph.D. thesis, Savoie U., 1990.

\bibitem{Grumiller:2016pqb}
D.~Grumiller and M.~Riegler, \emph{{Most general AdS$_{3}$ boundary
  conditions}}, \href{https://doi.org/10.1007/JHEP10(2016)023}{\emph{JHEP}
  {\bfseries 10} (2016) 023}
  [\href{https://arxiv.org/abs/1608.01308}{{\ttfamily 1608.01308}}].

\bibitem{Goddard:1986ee}
P.~Goddard, A.~Kent and D.I.~Olive, \emph{{Unitary Representations of the
  Virasoro and Supervirasoro Algebras}},
  \href{https://doi.org/10.1007/BF01464283}{\emph{Commun. Math. Phys.}
  {\bfseries 103} (1986) 105}.

\end{thebibliography}\endgroup

\end{document}